\documentclass{aa}  
\usepackage{graphicx}
\usepackage{hyperref}
\usepackage{txfonts}
\def\gmag{$G$\xspace}

\def\bpminrp{$G_{\rm BP}-G_{\rm RP}$\xspace}

\begin{document} 

\titlerunning{\textit{Gaia}~DR3: Validating the classification of variable YSO candidates} 

   \title{\textit{Gaia} Data Release 3}
   \subtitle{Validating the classification of variable young stellar object candidates}

   \author{G\'abor Marton
          \inst{\ref{konkoly},\ref{mtaexcellence}}
          \and
          P\'eter \'Abrah\'am\inst{\ref{konkoly},\ref{mtaexcellence},\ref{elte}}
          \and
          Lorenzo~Rimoldini\inst{\ref{inst-ch-eco}}
          \and
          Marc~Audard\inst{\ref{inst-ch-eco},\ref{inst-ch-obs}}
\and
          M\'aria Kun\inst{\ref{konkoly},\ref{mtaexcellence}}
          \and
          Zs\'ofia Nagy\inst{\ref{konkoly},\ref{mtaexcellence}}
          \and
          \'Agnes K\'osp\'al\inst{\ref{konkoly},\ref{elte},\ref{heidelberg},\ref{mtaexcellence}}
          \and
          L\'aszl\'o Szabados\inst{\ref{konkoly},\ref{mtaexcellence}}
          \and
Berry~Holl\inst{\ref{inst-ch-eco},\ref{inst-ch-obs}}
\and
Panagiotis~Gavras\inst{\ref{inst-es-rhea}}
\and
Nami~Mowlavi\inst{\ref{inst-ch-eco},\ref{inst-ch-obs}}
\and
Krzysztof~Nienartowicz\inst{\ref{inst-ch-sed}}
\and
Gr\'{e}gory~Jevardat~de~Fombelle\inst{\ref{inst-ch-eco}}
\and
Isabelle~Lecoeur-Ta\"ibi\inst{\ref{inst-ch-eco}}
\and
Lea~Karbevska\inst{\ref{inst-ch-eco},\ref{inst-fr}}
\and
Pedro~Garcia-Lario\inst{\ref{inst-es-esa}}
\and
Laurent~Eyer\inst{\ref{inst-ch-obs}}
          }

   \institute{Konkoly Observatory, Research Centre for Astronomy and Earth Sciences, E\"otv\"os Lor\'and Research Network (ELKH), H-1121 Budapest, Konkoly Thege Mikl\'os \'Ut 15-17., H-1121, Hungary\label{konkoly}\\
              \email{marton.gabor@csfk.org}
         \and
CSFK, MTA Centre of Excellence, Budapest, Konkoly Thege Mikl\'{o}s \'{U}t 15-17., H-1121, Hungary\label{mtaexcellence}
    \and
ELTE E\"otv\"os Lor\'and University, Institute of Physics, P\'azm\'any P\'eter s\'et\'any 1/A, H-1117 Budapest, Hungary\label{elte}
         \and
Max Planck Institute for Astronomy, K\"onigstuhl 17, D-69117 Heidelberg, Germany\label{heidelberg}
        \and
        Department of Astronomy, University of Geneva, Chemin d'Ecogia 16, 1290 Versoix, Switzerland\label{inst-ch-eco}
\and
Department of Astronomy, University of Geneva, Chemin Pegasi 51, 1290 Versoix, Switzerland\label{inst-ch-obs}
\and
RHEA for European Space Agency (ESA), Camino bajo del Castillo, s/n, Urbanizacion Villafranca del Castillo, Villanueva de la CaÃ±ada, 28692 Madrid, Spain\label{inst-es-rhea}
\and 
Sednai Sàrl, Geneva, Switzerland\label{inst-ch-sed}
\and 
Universit\'{e} de Caen Normandie, C\^{o}te de Nacre Boulevard Mar\'{e}chal Juin, 14032 Caen, France\label{inst-fr}
\and
European Space Agency (ESA), European Space Astronomy Centre (ESAC), Camino bajo del Castillo, s/n, Urbanizacion Villafranca del Castillo, Villanueva de la Cañada, 28692 Madrid, Spain\label{inst-es-esa}
}

   \date{Received -- --, 2022; accepted -- --, 2022}

 
  \abstract
   {The \textit{Gaia} third Data Release (DR3) presents the first catalogue of full-sky variable young stellar object (YSO) candidates observed by the \textit{Gaia} space telescope during the initial 34 months of science operations.}
   {Numerous types of variable stars were classified using photometric data collected by \textit{Gaia}. One of the new classes presented in the \textit{Gaia}~DR3 is the class of YSOs showing brightness variability. We analysed 79\,375 sources classified as YSO candidates in order to validate their young nature and investigate the completeness and purity of the sample.}
   {We cross-matched the \textit{Gaia}~DR3 YSO sample with numerous catalogues from the literature, including YSO catalogues based on optical and infrared data, as well as catalogues of extragalactic sources and Galactic variable stars. YSO catalogues were used to quantify the completeness of the \textit{Gaia}~DR3 YSO sample, while others were inspected to calculate the contamination.}
   {Among the 79\,375 potential YSO candidates published in the \textit{Gaia}~DR3 variable star catalogue, the majority of these objects are distributed along the line of sight of well-known star forming regions and the Galactic midplane. We find that the upper limit of contamination is 26.7\%, depending on the external catalogue used for the estimation, but find an average of $\sim10\%$ in general, while the completeness is at the percent level, taking into account that the \textit{Gaia}~DR3 YSO sample is based on sources that showed significant variability during the data-collection period. The number of sources in our sample that had not previously been catalogued as YSO candidates is $\sim$40\,000 objects.}
   {}


\maketitle
%

\section{Introduction}

\textit{Gaia} \citep{2016A&A...595A...1G} is a cornerstone mission of the European Space Agency. The mission is not only the most ambitious stellar (and extragalactic) astrometric project ever, but also one of the best transient discovery machines today. While other surveys, such as  YSOVAR \citep{2011ApJ...733...50M}, ASAS-SN \citep{2014ApJ...788...48S, 2017PASP..129j4502K}, and ZTF \citep{2019PASP..131a8003M}, provide photometric data and light curves for millions of sources, including young stellar objects (YSOs), \textit{Gaia} has the advantage of observing the whole sky. It collects photometric observations of some 1.8~billion stars down to a faint limit of 20.7 mag in the $G$~band and obtains low-resolution spectra down to $\sim$19~mag for an average of $\sim$80 epochs during the nominal five-year mission (which was completed and then extended until the end of 2022; hopefully it will be extended to the end of 2025 but this remains to be confirmed), although the cadence is highly dependent on the scanning law \citep{2016A&A...595A...1G}. \textit{Gaia} started monitoring the whole sky on 25~July 2014 and collects multi-epoch multi-band spectrophotometry and astrometric data for sources crossing its two fields of view (FoVs). A description of the \textit{Gaia} mission (spacecraft, instruments, survey, and measurement principles) as well as the structure and activities of the \textit{Gaia} Data Processing and Analysis Consortium (DPAC) can be found in \citet{2016A&A...595A...1G}. The \textit{Gaia} Data Release~3 (DR3) became public on 13~June 2022, and among many different products that have the potential to increase our fundamental understanding of the Galaxy it provides photometry in three passbands (\textit{Gaia} $G$, $G_{BP}$, and $G_{RP}$), five-parameter astrometry, and radial velocities (RVs) collected over the initial 34~months of observations; these are the most relevant quantities for our case. A summary of the \textit{Gaia}~DR3 contents and survey properties is provided in \citet{DR3-DPACP-185}.

One important research area, where collecting large amounts of data brings fundamental new results, is star formation. How the Sun ---and 
stars in general--- was born is identified as one of the most important questions of modern astronomy. Throughout their early evolution, stars show different features in their spectral energy distribution (SED). Initially, protostars are deeply embedded in their parental clouds, surrounded by dense dust and gas envelopes, which allow their detection only at sub-millimetre (mm) and far-infrared (FIR) wavelengths. At later stages when the envelope dissolves, young stars become apparent at optical wavelengths as well, which allows us to detect them with the \textit{Gaia} space telescope. However, at these evolutionary stages they still have circumstellar matter, their protoplanetary discs are still evolving, the stellar magnetosphere drives accretion columns, and accretion shocks are seen, all of which lead to variability in their emitted light, which can be studied to infer the underlying physics of star and planet formation.

\textit{Gaia} scans the sky repeatedly with successive observations following an irregular sequence in time ---based on the scanning law--- on a timescale from hours to months, and with a varying number of FoV transits depending on the celestial location \citep[e.g. see Appendix~A of][]{2017arXiv170203295E}. This allows the construction of light curves, which can be analysed for variability. 
More details about the general variability processing in \textit{Gaia}~DR3 are published in \citet{DR3-DPACP-162}. Variability in YSOs occurs on timescales that  span a wide range and depend on the physical processes \citep[see][]{2015ApJ...808...68H, 2022arXiv220311257F}. Circumbinary disc occultations last for  $\sim$100~days and can cause dimmings of  up to 4~mag. The EX~Lup-type outbursts occur on a timescale of couple of hundred days and can also cause brightenings of  2--4 mag, while FU~Ori-type outbursts last even longer and cause brightenings of even greater magnitude \citep[see e.g.][]{2014prpl.conf..387A}. These latter are therefore also a potential target of \textit{Gaia} observations. 

In this study, we focus on a new class introduced in the \textit{Gaia}~DR3 variability classification, the class of YSOs, for which the classification process resulted in a list of 79\,375 candidates. This output of the classification process was used in the current study. Our goal was threefold: (1) to validate their young nature, (2) to put constraints on the completeness of the catalogue, and (3) to check the purity and estimate the level of contamination. Having a large and reliable catalogue of variable YSOs can help further studies by providing a list of potential targets for more detailed analysis. Such analyses can be used to infer the underlying physics of the star and disc evolution, and their interaction.

\section{Data}\label{data}

\subsection{\textit{Gaia} data}
\textit{Gaia}~DR3 represents a significant improvement over Gaia DR2. The parallax precisions increased by 30 percent and proper motion precision improved by a factor of 2. Also, the systematic errors on the astrometry are lower by 30\%--40\% for the parallaxes, while those on the proper motions are lower by a factor $\sim$2.5. The longer temporal baseline of the observations also resulted in increased precision of the photometry and much better homogeneity across colour, magnitude, and celestial position. While \textit{Gaia} measures the position and brightness of the objects in its FoV, many parameters are provided that describe the quality of the data. Through the variability processing pipeline \citep{DR3-DPACP-162}, many other parameters are derived, which are helpful during the identification of variable sources and their verification.

One of the \textit{Gaia}~DR3 products is the supervised classification of variable sources \citep{DR3-DPACP-165}. Supervised classification was applied to classify several variability types using per-FoV epoch photometry in the three \textit{Gaia} bands. Among the 60~000 training sources, 5148 were YSOs and included subtypes such as T\,Tauri stars (classical, weak-lined, intermediate mass of types F to early G, and late G to early K types), FU\,Orionis, Herbig Ae/Be, and UX\,Orionis type stars. The time series were characterised by about two dozen features \citep[described in sect.~10.3.3 of][]{DR3-documentation}, which were used to train classifiers employing Random Forest \citep{Breiman.Random.Forest} and XGBoost \citep{Chen:2016:XST:2939672.2939785} algorithms.

The results of this supervised classification were verified in a post-processing phase, where the parameter space was analysed in order to maximise the purity of the sample and minimise its contamination. Several cuts were applied to variability parameters and astrometric values, such as parallax errors and proper motion, to exclude sources that have distances that are too small or too large to be taken into account as potential young stars \citep[see][for details]{DR3-DPACP-165}. 

\subsection{KYSO catalogue}
To help the identification of YSOs among the \textit{Gaia}~DR3 variable stars, we created a catalogue, the Konkoly Optical YSO (hereafter KYSO) catalogue\footnote{The KYSO table is only available in electronic form at the CDS via anonymous ftp to \hyperref[cdsarc.cds.unistra.fr]{cdsarc.cds.unistra.fr} (130.79.128.5) or via \hyperref[https://cdsarc.cds.unistra.fr/cgi-bin/qcat?J/A+A/]{https://cdsarc.cds.unistra.fr/cgi-bin/qcat?J/A+A/}.}, part of which served as a training sample for the supervised classification of \textit{Gaia}~DR3 variable stars \citep{DR3-DPACP-165} and was also used in the post-processing verification phase. The KYSO contains nearly 12\,000 objects, which are carefully selected young stars identified in the optical domain, and their young nature is mostly confirmed using spectroscopic data. More details about how the KYSO catalogue was compiled are described in Appendix~\ref{kysoappendix}. The KYSO catalogue is published as a Vizier table so that future classification studies can benefit from it. The KYSO catalogue was also used as a first step in our validation process and a detailed analysis was performed. We note that the KYSO catalogue has evolved since it was provided for training purposes for the supervised classification, as new sources have been added, and less reliable YSOs removed in order to increase the reliability of the catalogue.

\subsection{Other catalogues}
 We used several existing YSO catalogues from the literature based on optical and infrared surveys and observations, including data from \textit{Gaia}~DR2, infrared data from 2MASS, and the \textit{Spitzer} and \textit{WISE} surveys. These catalogues are the \citet{2019MNRAS.487.2522M} probabilistic YSO catalogue, based on \textit{Gaia}~DR2 optical data, 2MASS near-infrared (NIR) photometry, and \textit{WISE} mid-infrared (MIR) observations from the \href{https://wise2.ipac.caltech.edu/docs/release/allwise/expsup/}{AllWISE catalogue}, the SPICY (Spitzer/IRAC Candidate YSO) catalogue of \citet{2021ApJS..254...33K} based on MIR observations of  \textit{Spitzer,} and the \citet{2018A&A...619A.106G} study of Orion A based on the ESO--VISTA NIR survey. While the \textit{Gaia}~DR3 YSO candidate sample is based on optical variability, the listed catalogues from the literature used the IR excess in the SED as a signature of the young nature. We also cross-matched the \textit{Gaia}~DR3 YSO candidates with the SIMBAD database \citep{2000A&AS..143....9W} and other large catalogues listing extragalactic sources and other types of variable stars identified in the optical domain, as these sources can show colours and light-curve features that are similar to those of YSOs. Based on the results of the cross-matches we calculated an estimated completeness and purity for the \textit{Gaia}~DR3 YSO candidates.

It is important to note that  the \textit{Gaia} observations were of different angular resolution and sensitivity when compared to other observations in the literature. The 2MASS survey \citep{2006AJ....131.1163S} collected NIR data in the $J$ (1.2\,$\mu$m), $H$ (1.65\,$\mu$m), and $K_s$ (2.16\,$\mu$m) bands; the system PSF was $\sim2\farcs5$, and the average limiting magnitude was $\sim$14. The \textit{WISE} mission \citep{2010AJ....140.1868W} observed the whole sky with an angular resolution $6\farcs1$, $6\farcs4$, $6\farcs5$, and $12\farcs0$ at 3.4, 4.6, 12, and 22 $\mu$m, respectively. For the \textit{WISE} data, we used the AllWISE catalogue, which is >95\% complete for sources with W1<17.1 and W2<15.7~mag. These differences and the apparent motions of several objects make it difficult to simply cross-match such large catalogues and make it challenging to avoid source confusion. In the case of the AllWISE catalogue, 2MASS data are already included, where the position reconstruction was done using bright 2MASS point sources as the astrometric reference. For AllWISE, the proper motion of the reference stars in the 11~years separating the \textit{WISE} and 2MASS surveys has been integrated into the solutions to improve the absolute astrometric accuracy\footnote{\url{https://wise2.ipac.caltech.edu/docs/release/allwise/expsup/sec2\_1.html}}. The cross-match of the \textit{Gaia}~DR2 and the AllWISE catalogues was done by \citet{2019A&A...621A.144M}. In cases where we used 2MASS and AllWISE data for the validation, we relied on these catalogues, matched the \textit{Gaia}~DR3 YSO  sources to the DR2 positions, and inferred the data from the DR2xAllWISE (including 2MASS) table. In other cases, when the cross-match with such accuracy was not already available, we used the TOPCAT software \citep{2005ASPC..347...29T} with a search radius of 1$\arcsec$.




\section{Validation}\label{validation}

In this section, we consider three different aspects of the validation. In Section~\ref{youngnature}, we compare the parameters of the \textit{Gaia}~DR3 YSO sample to those of objects listed in different YSO catalogues from the literature in order to validate the young nature of the \textit{Gaia}~DR3 objects. These catalogues are based on different domains of the electromagnetic spectrum. The KYSO is based on optical data, the \citet{2018A&A...619A.106G} uses NIR data, the \citet{2021ApJS..254...33K} is based on MIR data, and finally the \citet{2019MNRAS.487.2522M} used optical, NIR, and MIR photometry. We also investigate the distance distributions of the \textit{Gaia}~DR3 YSOs on the all-sky and in individual star forming regions and compare them to literature values.

In Section~\ref{completeness}, we estimate the completeness of the \textit{Gaia}~DR3 YSO candidates by comparing the total number of the sources listed in a given YSO catalogue to the number of \textit{Gaia}~DR3 counterparts (meaning \textit{Gaia} actually observed these sources), to the number of counterparts that were part of the classification process (for sufficiently sampled signals that were considered as variable in \textit{Gaia}~DR3), and to the number of counterparts existing in the final \textit{Gaia}~DR3 YSO sample.

In Section~\ref{contamination}, we estimate the contamination level of the \textit{Gaia}~DR3 YSO sources by matching them to catalogues listing a specific type of objects, mostly identified in the optical domain to make the comparison reasonable.

\subsection{The young nature of the \textit{Gaia}~DR3 YSO candidates}\label{youngnature}

   \begin{figure}
   \centering
   \includegraphics[width=\hsize]{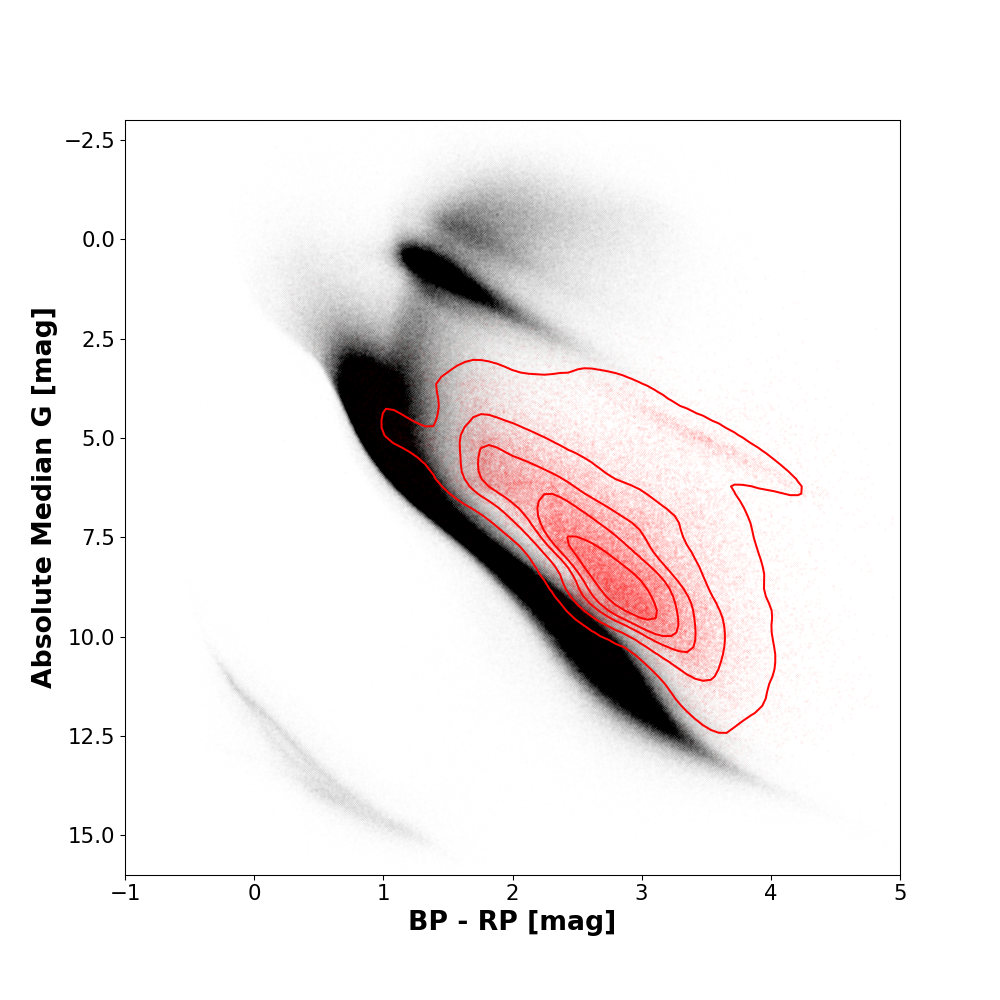}
      \caption{Observational HRD of \textit{Gaia}~DR3 YSOs (red dots) and reference sources based on 4.2~million \textit{Gaia} objects (black dots) selected based on their highly reliable parallax, sufficient S/N in both BP and RP bands, and the sufficiently high number of data points in their light curves. The \textit{Gaia}~DR3 YSOs occupy a specific region above the main sequence and below the giant branch. The contour levels are at 5\%, 25\%, 45\%, 65\%, and 85\% of the maximum density value. In the comparison with other catalogues, we use only the contours of the DR3 distribution for better visibility of the underlying data points. 
              }
         \label{dr3ysohrd}
   \end{figure} 

   \begin{figure}
   \centering
   \includegraphics[width=\hsize]{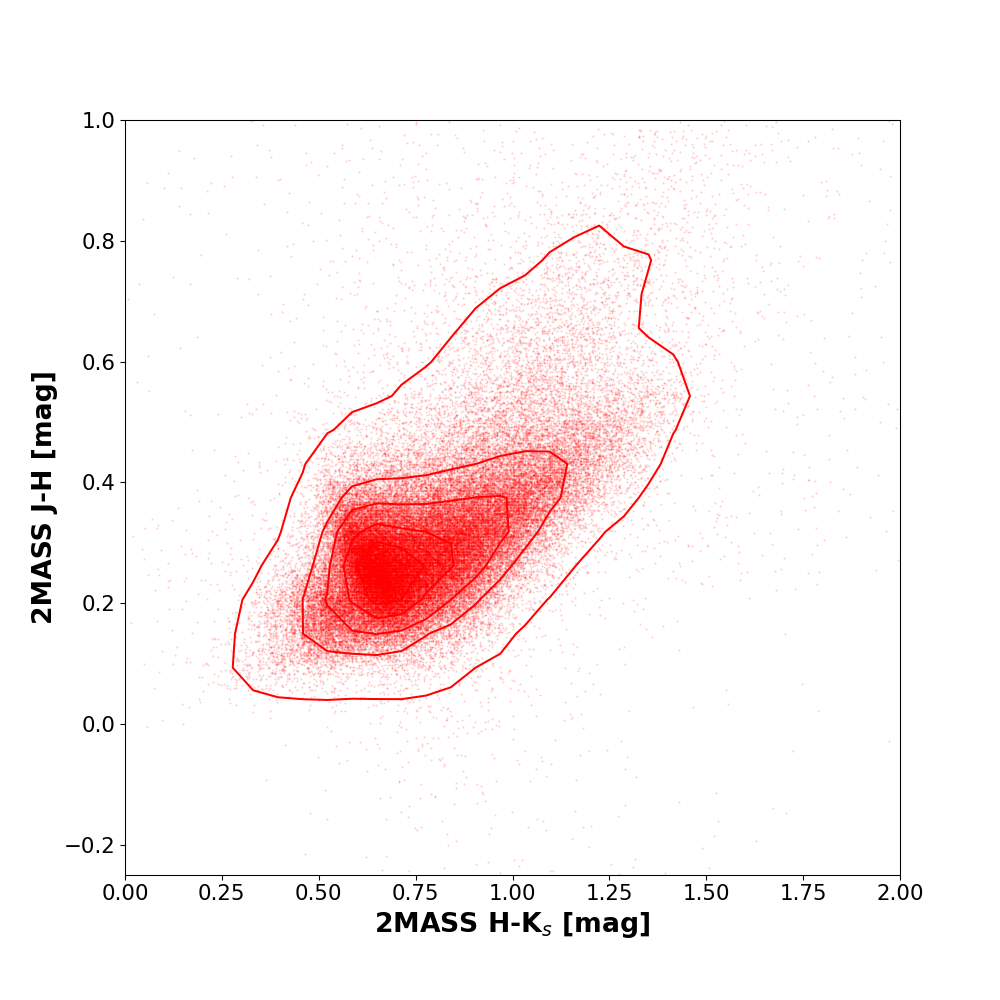}
      \caption{\textit{Gaia}~DR3 YSOs on the 2MASS colour--colour diagram. The contour levels are at 5\%, 25\%, 45\%, 65\%, and 85\% of the maximum density value. In the comparison with other catalogues, we use only the red contours of the DR3 distribution for better visibility of the underlying data points. The median, mean, standard deviation, and 5\% and 95\% quantiles of both colours are listed in Table~\ref{ysotable}.
              }
         \label{2masscolourdr3}
   \end{figure} 
\subsubsection{Cross-match with the KYSO catalogue}

As a first step of the validation of the \textit{Gaia}~DR3 YSO candidate sample, we cross-matched it with the KYSO catalogue. The KYSO catalogue provides a strong bias as the sources included are from various catalogues and from well-known star forming regions, while \textit{Gaia} provides a relatively homogeneous all-sky survey.

   \begin{figure*}
   \centering
   \includegraphics[width=\hsize]{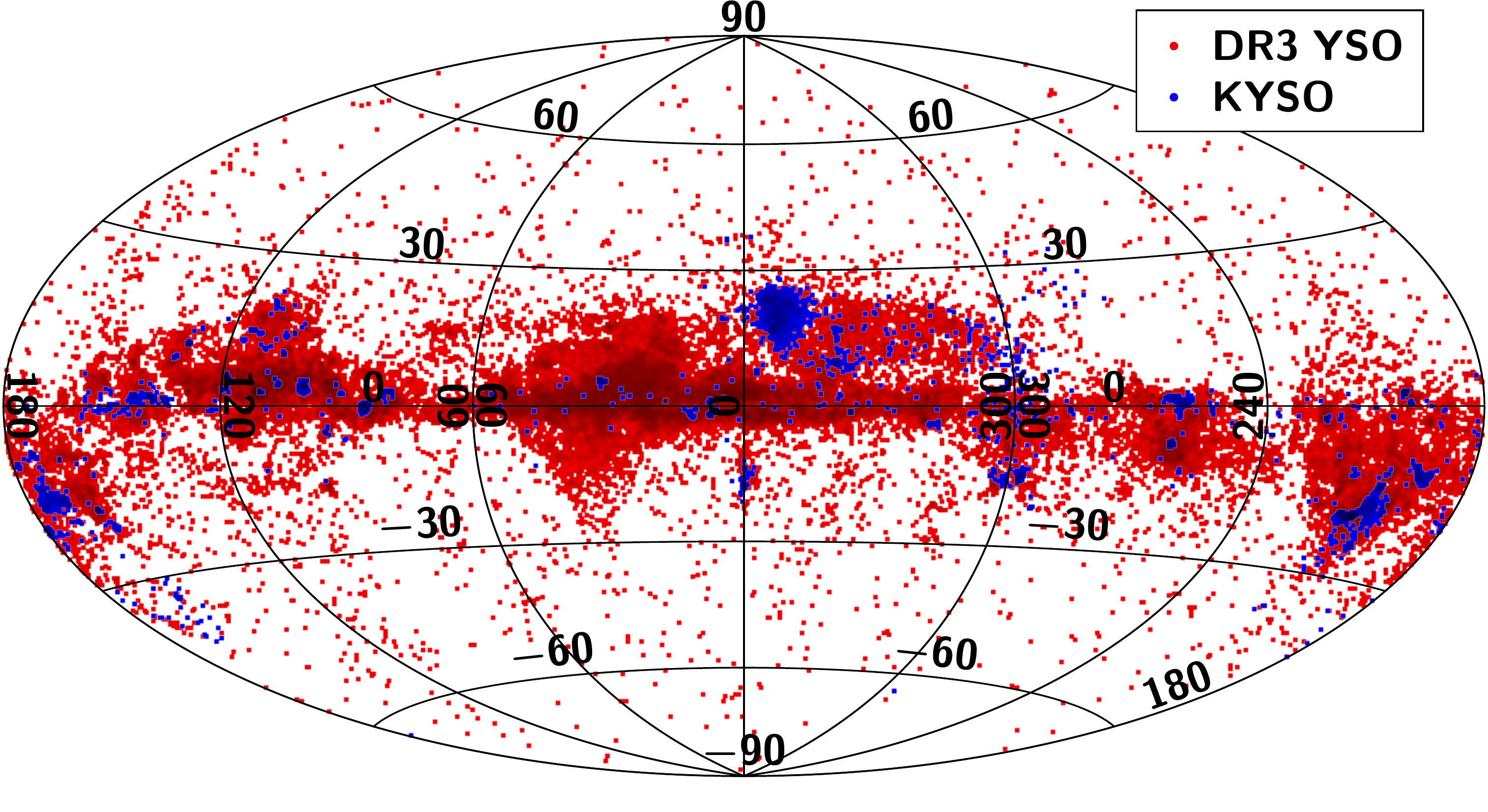}
      \caption{DR3 YSO sample (red dots) and the KYSO objects (blue dots) presented in Hammer-Aitoff projection in Galactic coordinates. The KYSO sample is heavily biased as it lists sources from well studied star forming regions, while \textit{Gaia} is an unbiased all-sky survey, and therefore \textit{Gaia}~DR3 YSO candidates can be seen in all directions. YSO candidates are seen at high Galactic latitudes ($|b|\geq 30\degr$), their number is 940, which is only 1.2\% of all the candidates. The median distance of these high-latitude sources is 363.6 pc, and 95\% of them are within a distance of 869.4 pc.
              }
         \label{aftercut}
   \end{figure*}   
   
In the case of a good classification, one would expect to see that YSOs occupy the same regions on the observational HRD and also on the various colour--colour diagrams. YSOs are seen in specific regions of the sky, and therefore we expect overdensities in the YSO candidate surface density distribution in the line of sight of the (1) regions where the training sample showed overdensities and (2) the known star forming regions and the Galactic midplane. Figure~\ref{aftercut} shows their distribution on the sky in Galactic coordinates, which clearly reflects the expected distribution.

\begin{figure}
   \centering
   \includegraphics[width=\hsize]{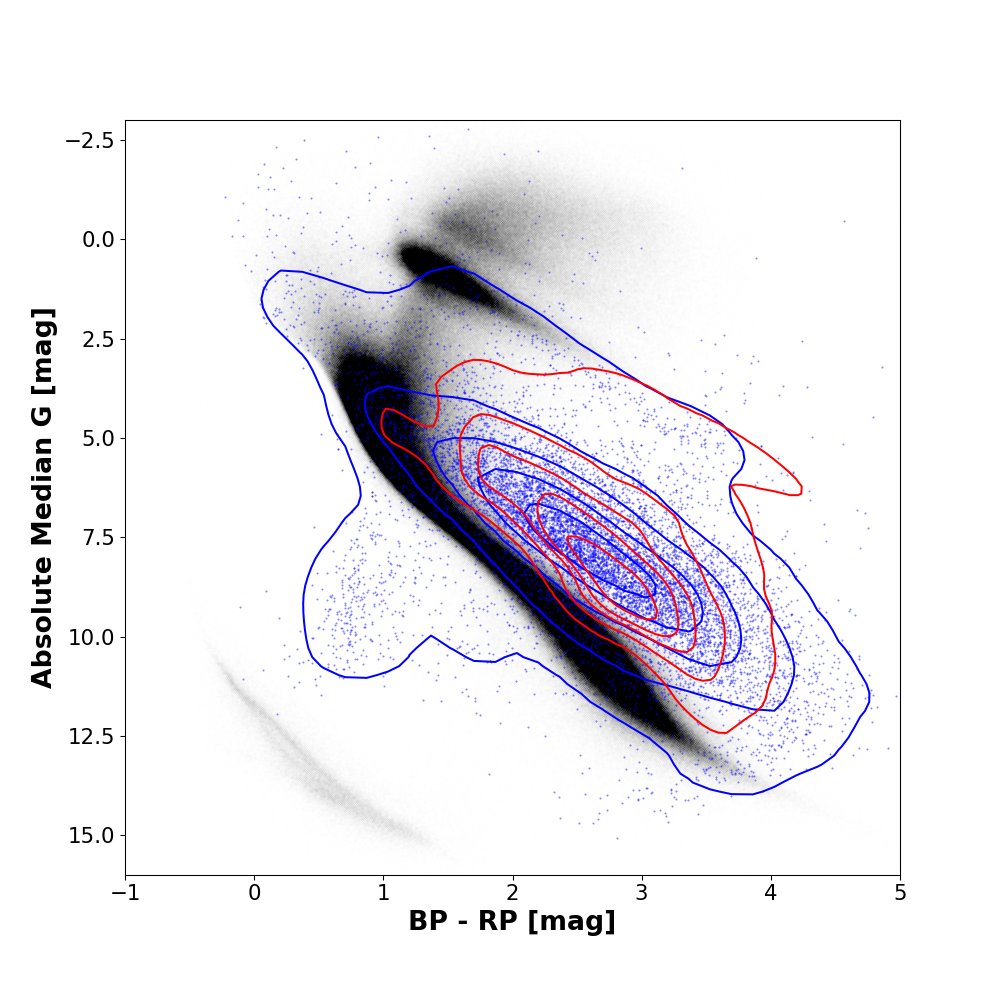}
      \caption{Observational HRD of \textit{Gaia}~DR3 YSOs (red contours), KYSOs (blue dots and blue contours), and reference sources (grey dots) as described in the caption of Fig.~\ref{dr3ysohrd}. Most of the KYSOs are located above the main sequence, but other parts of the HRD are also covered. The probability distributions of the \bpminrp colour and the $G_{\mathrm{Abs}}$ are shown in Figs.~\ref{bprpkyso} and \ref{absgkyso}. The main parameters of the distributions are listed in Table~\ref{ysotable}.
              }
         \label{kysohrd}
   \end{figure}

As a next step, we checked the location of the \textit{Gaia}~DR3 YSO sources and that of the KYSO objects on the observational Hertzsprung--Russell Diagram (HRD, often referred to as simply the colour--magnitude diagram). We note here that HRDs in this study are not corrected by Galactic extinction or reddening. Figure~\ref{kysohrd} shows how the KYSOs are distributed in the median \bpminrp colour versus \gmag band absolute magnitude diagram. As explained in \citet{2021AJ....161..147B}, a simple inversion of the parallax values does not give a precise distance, because the inversion can lead to bias. Therefore, the \gmag band absolute magnitudes were calculated using their distance values obtained from the \citet{2021AJ....161..147B} distance catalogue instead of the parallax listed in the \textit{Gaia}~DR3. In general, YSOs are mostly located above the main sequence, but depending on their mass and evolutionary stage, they can appear in other parts of the HRD as well (see Section~\ref{kysocompleteness}). The \textit{Gaia}~DR3 YSOs are located in a smaller region on the HRD, but highly overlapping with the KYSO catalogue, which shows that the classification process successfully identified those candidates that are located in the same place as the majority of the KYSOs. 
   
There are only a few sources located at the blue side of the HRD. The \bpminrp colour distribution is  shown in Fig.~\ref{bprpkyso}. While the histograms show similarities, it is seen that there are more KYSOs at the blue end of the diagram, where mostly high-mass YSOs and Ae/Be stars, are located. This means that the YSO classification was more sensitive to the redder, fainter, and lower-mass objects, because very few training objects were included below the main sequence, and also because of the strong competition with other classes at the blue end. 

We also analysed the \textit{Gaia} $G$ band absolute magnitude distribution of the KYSO sources and that of the sources classified as \textit{Gaia}~DR3 YSOs; the results are shown in~Fig.~\ref{absgkyso}. The distribution of KYSO sources is wider than that of the \textit{Gaia}~DR3 YSOs, but the two samples show a significant overlap. 

To see the IR excess distribution of the KYSOs and the \textit{Gaia}~DR3 YSOs, we cross-matched them with the 2MASS catalogue using a 1$\arcsec$ matching radius. Of the 79\,375 \textit{Gaia}~DR3 YSOs,  76\,879 (97\%) had a NIR counterpart, while 10\,613 (91\%) of the 11\,665 KYSOs had a match in the 2MASS database. 

\begin{figure}
   \centering
   \includegraphics[width=\hsize]{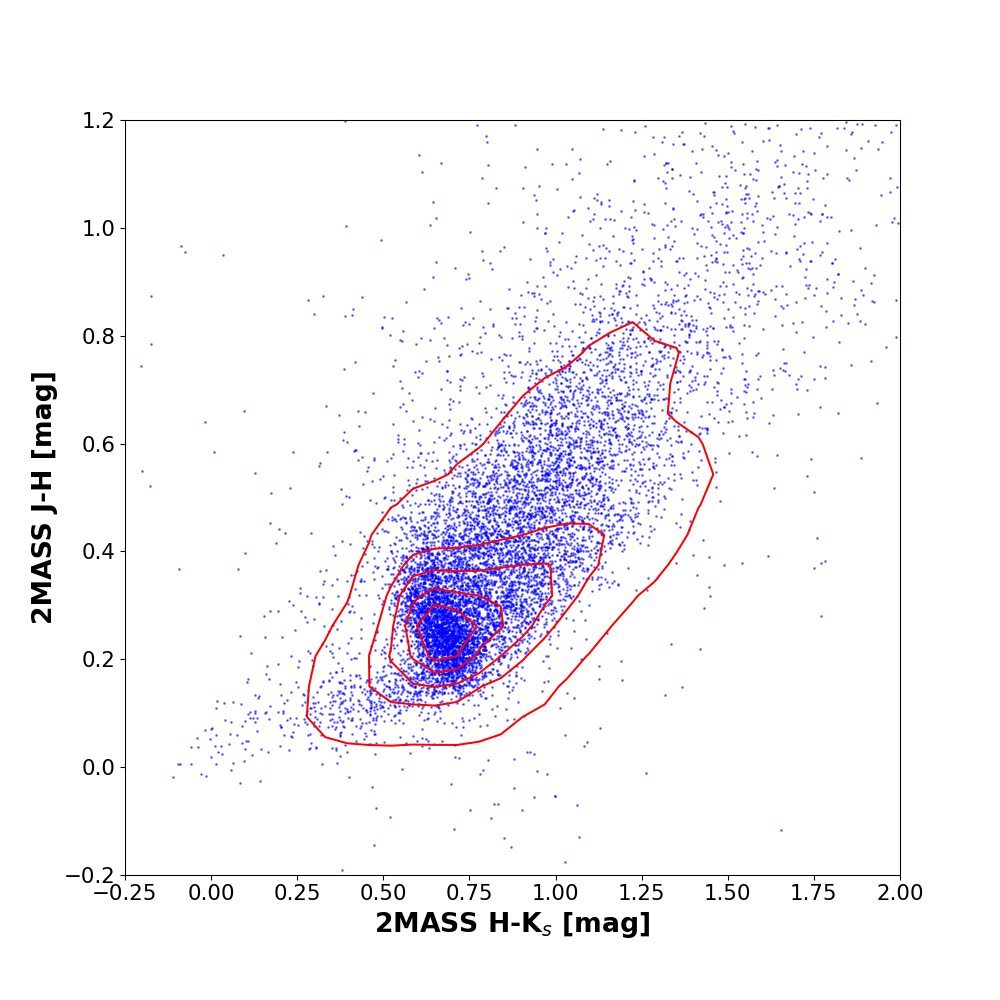}
      \caption{KYSO sources (blue dots) and \textit{Gaia}~DR3 YSOs (red contours) on the 2MASS colour--colour diagram. The median, mean, standard deviation, and 5\% and 95\% quantiles of both colours are listed in Table~\ref{ysotable}. The colour probability distributions are shown in Fig.~\ref{jhkyso} and ~\ref{hkkyso}. The main parameters of the distributions are listed in Table~\ref{ysotable}.
              }
         \label{jhhkkyso}
   \end{figure}

Figure~\ref{jhhkkyso} shows the distributions of the  \textit{Gaia}~DR3 YSO candidates and the KYSO sources on the 2MASS $J-H$ versus $H-K_s$ colour--colour diagram. Both samples occupy the same region on the diagram, and are mostly located between $0<J-H<1.5$ and $0<H-K_s<1$. 

Further details of the above-mentioned distributions are listed in Table~\ref{ysotable}, where we list the median and mean values, the standard deviation, and the threshold values below and above which 5\% of the samples are located. Figures showing the distributions are found in Appendix~\ref{distributions}. 

\begin{figure}
   \centering
   \includegraphics[width=\hsize]{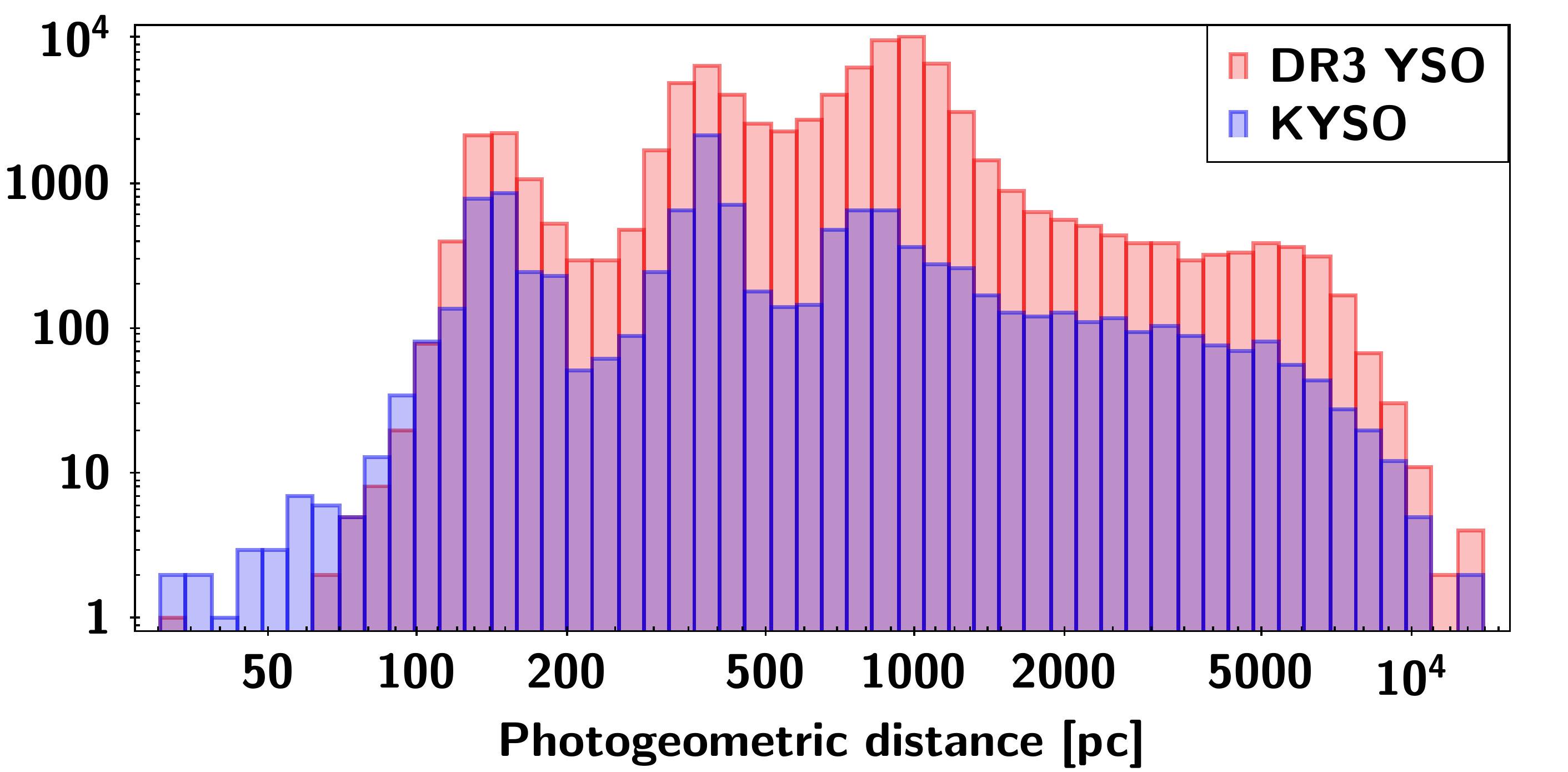}
      \caption{Distance distribution of KYSO sources (blue bars) and of sources classified as \textit{Gaia}~DR3 YSOs (red bars). The values show strong similarities, except that there are more KYSOs within 100~pc of the Sun. These nearby KYSO sources (87 have distances of smaller than 100~pc) are mostly members of the TW Hydrae association. There are also three strong peaks in the distance distribution. The first one, between $\sim$125 and $\sim$175~pc, corresponds to the Ophiuchus, Sco OB2 association, and Taurus regions. The second strong peak is seen at distances of between $\sim$300 and $\sim$500~pc. The vast majority of these sources are located in Orion. The third peak is between $\sim$300 and $\sim$1000~pc. These sources are also seen towards Orion and towards the Galactic midplane as well.
              }
         \label{kysodistance}
   \end{figure}

We also investigated the distance distribution of the KYSOs and \textit{Gaia}~DR3 YSOs based on the values of the median photogeometric distance posterior listed for \textit{Gaia} EDR3 by \citet{2021AJ....161..147B}. While the overall distributions of the distance values show strong similarities, the KYSO catalogue shows a small excess of very nearby stars (<100~pc); see Fig.~\ref{kysodistance}.

\begin{table}
\caption{Median, mean, standard deviation, and 5\% and 95\% quantiles of the $G_{\rm Abs}$, \bpminrp, $J-H$ and $H-K_s$ distributions of the \textit{Gaia}~DR3 YSO sample and other public catalogues used for the validation process. Each quantity is presented in magnitudes. Histograms for each sample and each quantity are presented in Appendix~\ref{distributions}.}
\label{ysotable}
\begin{tabular}{lrrrrr}
\hline
\hline
Parameter & Median & Mean & Stdev & 5\% & 95\% \\
\noalign{\vskip 2mm}
\hline
\multicolumn{1}{c}{}& \multicolumn{5}{c}{DR3 YSO candidate} \\
\hline
$G_{\rm Abs}$  & 7.623 & 7.506 & 1.854 & 4.384 & 10.316\\
\bpminrp  & 2.770 & 2.748 & 0.601 & 1.749 & 3.697\\
$J-H$  & 0.746 & 0.791 & 0.235 & 0.513 & 1.195\\
$H-K_s$  & 0.287 & 0.318 & 0.180 & 0.152 & 0.589\\
\hline
\multicolumn{1}{c}{}& \multicolumn{5}{c}{KYSO} \\
\hline
\noalign{\vskip 2mm}
$G_{\rm Abs}$  & 7.764 & 7.609 & 2.431 & 3.321 & 11.281 \\
\bpminrp  & 2.569 & 2.534 & 0.879 & 0.935 & 3.889 \\
$J-H$  & 0.796 & 0.881 & 0.368 & 0.495 & 1.549 \\
$H-K_s$  & 0.377 & 0.455 & 0.306 & 0.143 & 1.025 \\
\hline
\multicolumn{1}{c}{}& \multicolumn{5}{c}{G19 \citet{2019A&A...622A.149G}} \\
\hline
\noalign{\vskip 2mm}
$G_{\rm Abs}$  & 9.838 & 8.594 & 1.990 & 5.133 & 11.351 \\
\bpminrp  & 2.562 & 2.465 & 0.884 & 0.778 & 3.709 \\
$J-H$  & 0.951 & 1.203 & 0.745 & 0.568 & 2.778 \\
$H-K_s$  & 0.549 & 0.719 & 0.521 & 0.246 & 1.783 \\
\hline
\multicolumn{1}{c}{}& \multicolumn{5}{c}{SPICY \citet{2021ApJS..254...33K}} \\
\hline
\noalign{\vskip 2mm}
$G_{\rm Abs}$  & 6.228 & 6.024 & 1.657 & 2.947 & 8.386 \\
\bpminrp  & 2.754 & 2.777 & 0.752 & 1.580 & 4.066 \\
$J-H$  & 1.372 & 1.459 & 0.657 & 0.552 & 2.673 \\
$H-K_s$  & 0.899 & 1.008 & 0.606 & 0.287 & 2.166 \\
\hline
\multicolumn{1}{c}{}& \multicolumn{5}{c}{M19 \citet{2019MNRAS.487.2522M}} \\
\hline
\noalign{\vskip 2mm}
$G_{\rm Abs}$  & 6.182 & 6.546 & 2.191 & 3.608 & 10.705 \\
\bpminrp  & 2.709 & 2.466 & 0.811 & 1.173 & 3.491 \\
$J-H$  & 0.873 & 0.845 & 0.340 & 0.353 & 1.394 \\
$H-K_s$  & 0.384 & 0.383 & 0.217 & 0.100 & 0.758 \\
\hline
\end{tabular}
\end{table}

\subsubsection{Cross-match with the \citet{2019MNRAS.487.2522M} YSO catalogue}

In \citet{2019MNRAS.487.2522M} (referred to hereafter as the M19 catalogue), 101~million objects were classified from the DR2xAllWISE cross-match table into four main categories (evolved stars (E), main sequence stars (MS), extragalactic sources (EG), and YSOs (Y)) with the help of a Random Forest classifier, and class membership probabilities were assigned to each source. \citet{2014ApJ...791..131K} investigated the spurious detections in the \textit{WISE} W3 and W4 bands. Because M19 was heavily reliant on the \textit{WISE} data, the reliability of the W3 and W4 photometry values was also investigated. Another Random Forest classifier was built to decide whether the W3 and W4 band detections are spurious or reliable, and a probability value $R$ was given for each detection, as detailed in Section 2.7 in \citet{2019MNRAS.487.2522M}. If $R \geq 0.5$, one can assume that the W3 and W4 band photometry can be used and the $LY$ probability values are used from the M19 sample, which gives the probability that a source is a YSO; in the $LY$ acronym, the letter $Y$ refers to YSO, while the letter $L$ refers to the inclusion of longer wavelength W3 and W4 bands in the classification. In cases where $R < 0.5$, the $SY$ (where the letter $S$ refers to the shorter wavelength W1 and W2 bands; in these cases the \textit{WISE} W3 and W4 bands were not part of the classification process) probability was taken into account.

Based on the probability values, one can define a threshold above which one feels confident to accept the source as a reliably classified object. For analysing the properties of the \textit{Gaia}~DR3 sources and comparing it to the M19, we used YSOs from the M19 catalogue, where $R\geq0.5$ and $LY\geq0.95$, or $R<0.5$ and $SY\geq0.95$, meaning 259\,363 sources in total. 

\begin{figure}
   \centering
   \includegraphics[width=\hsize]{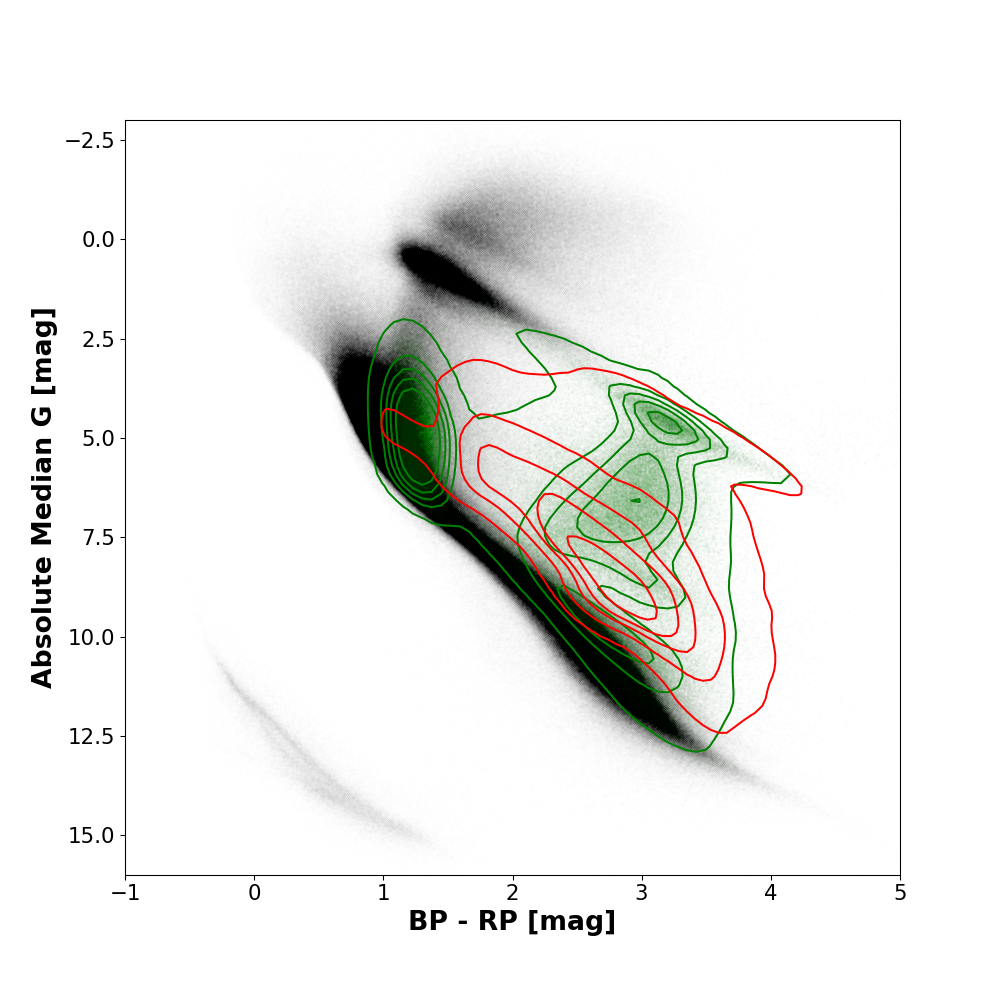}
      \caption{Same as Fig.~\ref{kysohrd} but for the \textit{Gaia}~DR3 (red contours) and M19  YSO candidates with $R\geq0.5$ and $LY\geq0.95$, or $R<0.5$ and $SY\geq0.95$ (green dots and contours). The probability distributions are shown in Figs.~\ref{bprpm19} and \ref{absgm19}. The main parameters of the distributions are listed in Table~\ref{ysotable}. Because the \textit{Gaia}~DR3 YSO distribution is narrower, for better visibility we plotted them on the top of the M19 distribution.
              }
         \label{hrdm19}
   \end{figure}

The observational HRD of the M19 and \textit{Gaia}~DR3 YSO candidates are shown in Fig.~\ref{hrdm19}. Similarly to the comparison with the KYSOs, the \textit{Gaia}~DR3 YSO candidates also show significant overlap with the distribution of the M19 sources, but the M19 distribution shows additional peaks at the bluer and brighter part of the HRD as well as close to the giant branch. The bright-blue tail of the distribution can be explained by the less accurate training sample used in the \cite{2019MNRAS.487.2522M} study, which allowed more evolved sources to be classified as YSOs with relatively high probability without the long-wavelength $W3$ and $W4$ measurements of the \textit{WISE} telescope. The peak close to the giant branch is also due to the uncertainty in the distance estimation of the source, and therefore these sources may appear more distant than they actually are. However, as listed in Table~\ref{ysotable}, 5\% of the \textit{Gaia}~DR3 YSO candidates are fainter than $\sim10.3$ mag, while in the M19 catalogue this quantile value is $\sim10.7$ mag. The 5\% quantile value at the bright end for the \textit{Gaia}~DR3 YSO candidates is $\sim4.4$ mag, while for the M19 YSOs, the same threshold is at $\sim3.6$ mag. Therefore, these differences cannot be considered significant.

\begin{figure}
   \centering
   \includegraphics[width=\hsize]{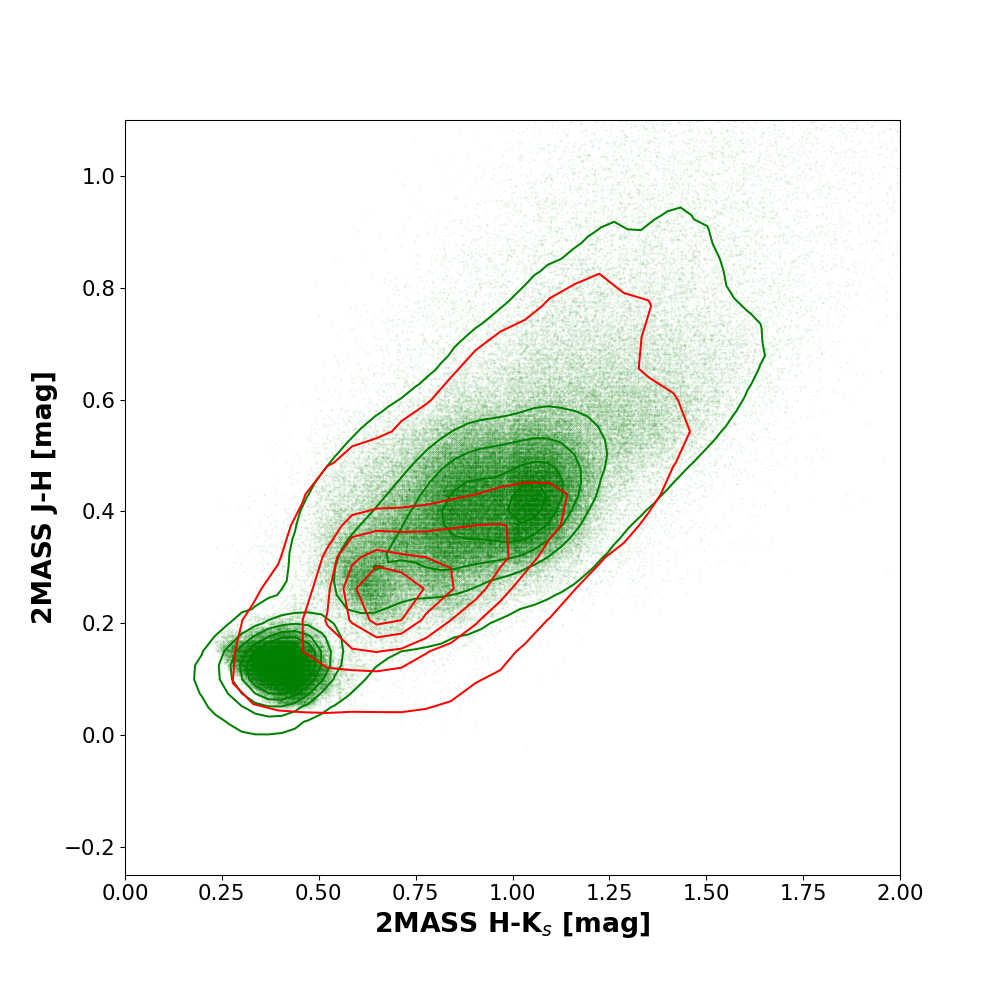}
      \caption{Same as Fig.~\ref{jhhkkyso}, but for the M19 sources (green dots and contours) and \textit{Gaia}~DR3 YSO candidates (red contours). The colour probability distributions are shown in Figs.~\ref{jhm19} and ~\ref{hkm19}, and the main parameters of the distributions are listed in Table~\ref{ysotable}.
              }
         \label{jhhkm19}
   \end{figure}

On the 2MASS J$-$H versus H$-$K$_s$ colour--colour diagram (Fig.~\ref{jhhkm19}), the \textit{Gaia}~DR3 YSOs are located in the same region as the M19 objects, but only show one maximum in the distribution, unlike the M19 sources, which show bimodality. As explained earlier, the method used for the M19 classification allowed more evolved objects to be classified as YSOs with relatively high probability when the long-wavelength \textit{WISE} measurements were not taken into account.

\subsubsection{Cross-match with the \citet{2021ApJS..254...33K} YSO catalogue}

Based on \textit{Spitzer} space telescope surveys of the Galactic midplane between $l \sim 255\degr$ and $110\degr$, including the GLIMPSE~I, II, and 3D, Vela--Carina, Cygnus X, and SMOG surveys (613~square degrees), \citet{2021ApJS..254...33K} published a probabilistic YSO catalogue, the SPICY catalog. These authors presented 117\,446 Spitzer/IRAC candidate YSOs.

\begin{figure}
   \centering
   \includegraphics[width=\hsize]{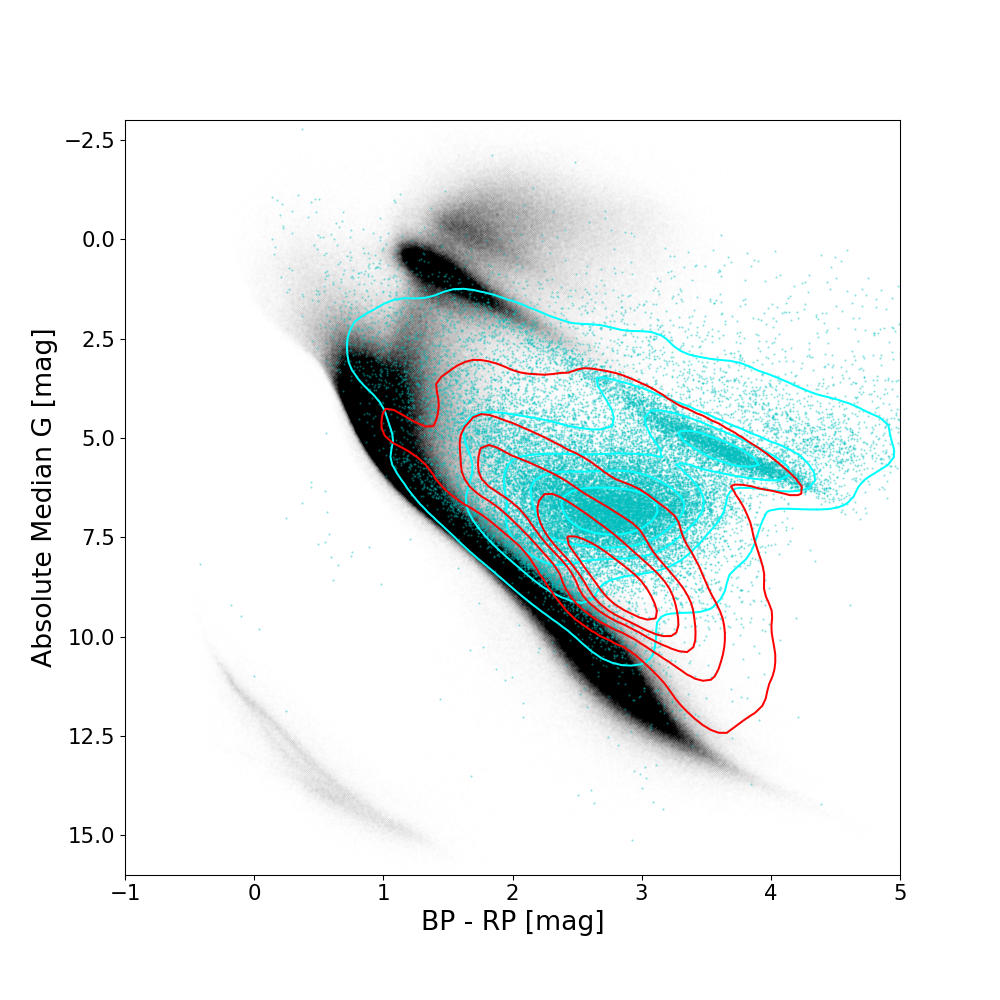}
      \caption{DR3 YSO candidates (red contours) and SPICY YSOs (cyan dots and contours) on the HRD. The \bpminrp and $G_{\rm Abs}$ distributions are shown in Figs.~\ref{bprpspicy} and ~\ref{absgspicy}, and the main parameters of the distributions are listed in Table~\ref{ysotable}. 
              }
         \label{hrdspicy}
   \end{figure}

Because \textit{Spitzer} observes the IR domain, it is an ideal tool for YSO discovery. Therefore, it is expected that more of the YSOs in the SPICY catalogue are in the earlier stages of evolution, showing more IR excess. Also, because the SPICY contains sources seen towards the Galactic midplane, where the amount of interstellar dust is very high and obscures the visible light, one can expect that only the brighter, higher-mass YSOs are seen with \textit{Gaia}. Figure~\ref{hrdspicy} shows the HRD for the \textit{Gaia}~DR3 YSO candidates and SPICY YSOs. The overlap is significant, but we see more \textit{Gaia}~DR3 YSO candidates in the 3 < \bpminrp < 4.5 and 5 < $G_{\mathrm{Abs}}$ < 10 region. These are objects  in nearby star forming regions and only a few of them are located in the region covered by the SPICY catalogue. While the median $G_{\mathrm{Abs}}$ brightness of the \textit{Gaia}~DR3 YSOs is 7.6 mag, the SPICY objects are brighter by 1 mag, and the median value is 6.2 magnitude. The median value of the \bpminrp is very similar, 2.7 for both samples.

\begin{figure}
   \centering
   \includegraphics[width=\hsize]{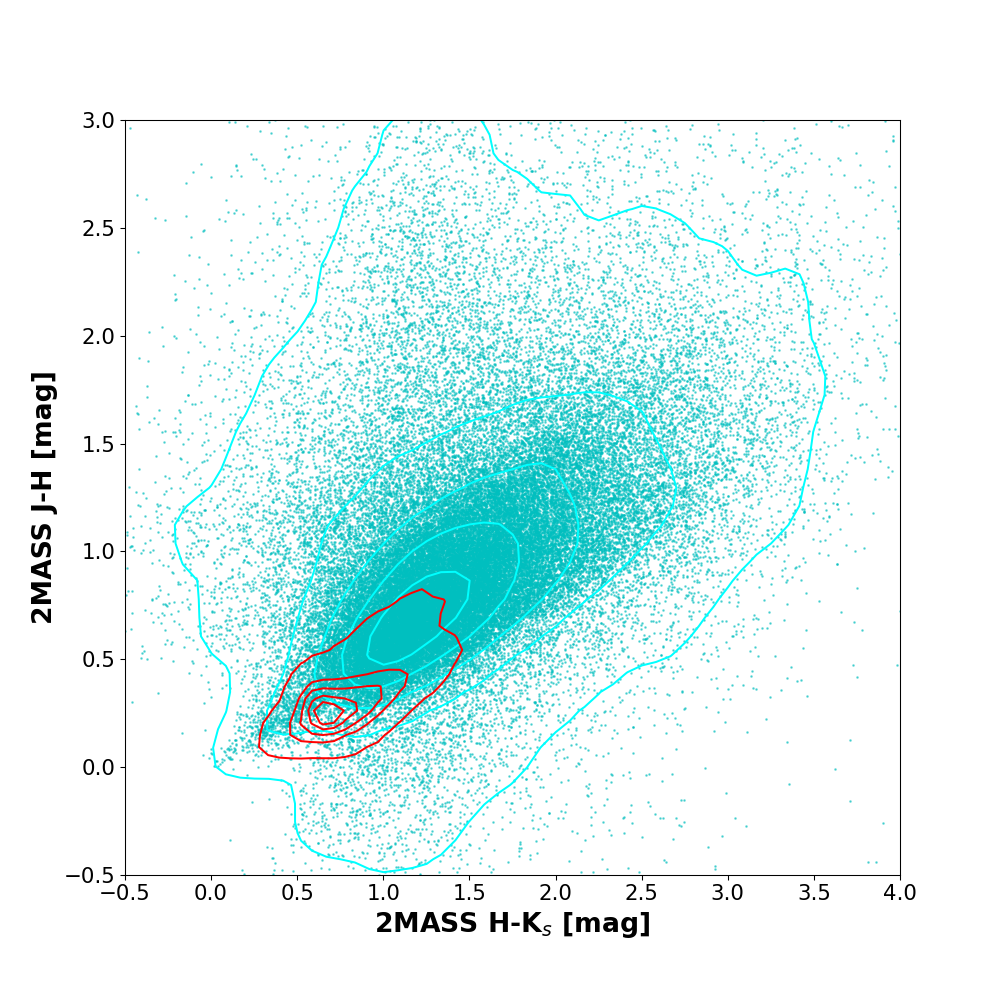}
      \caption{SPICY sources (cyan dots and contours) and \textit{Gaia}~DR3 YSOs (red contours) on the 2MASS colour--colour diagram. The \bpminrp and $G_{\rm Abs}$ distributions are shown in Figs.~\ref{jhspicy} and ~\ref{hkspicy}, and the main parameters of the distributions are listed in Table~\ref{ysotable}.
              }
         \label{jhhkspicy}
   \end{figure}

The 2MASS colour--colour distribution is shown in Fig.~\ref{jhhkspicy}. As expected, the SPICY objects cover a larger area on the diagram. While the median $J-H$ colour of the \textit{Gaia}~DR3 YSOs is 0.7 mag, this value is 1.4 mag for the SPICY objects. This difference is also seen in the median value of the $J-K_s$ colour distribution, as the median value for the \textit{Gaia}~DR3 YSO candidates is 0.3 magnitude while it is 0.9 for the SPICY objects.

\subsubsection{Cross-match with the \citet{2019A&A...622A.149G} paper}

\citet{2018A&A...619A.106G} used the \textit{Gaia}~DR2 distances of MIR-selected YSOs in the benchmark giant molecular cloud Orion A to infer its 3D shape and orientation based on the ESO--VISTA NIR survey. At a later stage, an updated source list was published in \citet{2019A&A...622A.149G}. We used this source list (hereafter G19) to match our sources to YSOs in the Orion A star forming region. 

\begin{figure}
   \centering
   \includegraphics[width=\hsize]{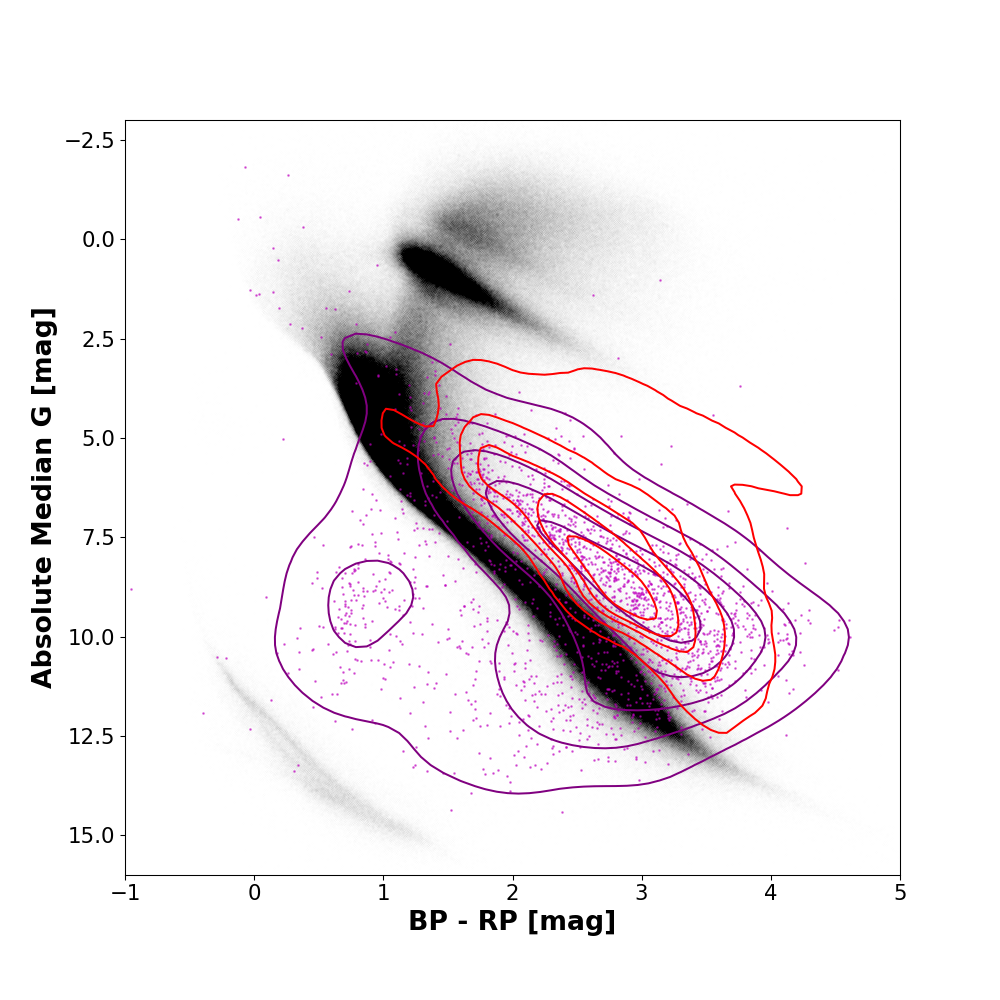}
      \caption{Same as Fig.~\ref{kysohrd} but for \textit{Gaia}~DR3 (red contours) and G19 (purple dots and contours) YSO candidates. Histograms of the two parameter distributions are shown in Figs.~\ref{absgg19} and \ref{bprpg19}. The main parameters of the distributions are listed in Table~\ref{ysotable}. The distribution shows strong similarities to that seen in Fig.~\ref{kysohrd}.
              }
         \label{hrdg19}
   \end{figure}

Again, as a first step, we checked the observational HRD and analysed the location of the objects shown in Fig.~\ref{hrdg19}. The G19 YSOs are seen not only above the main sequence, but also in the region between the main sequence and the location of the white dwarfs. This feature of their distribution is discussed in Section~\ref{kysocompleteness}. It is also clear from Fig.~\ref{hrdg19} that the G19 objects tend to have lower $G_{\mathrm{Abs}}$ values. As listed in Table~\ref{ysotable}, the median $G_{\mathrm{Abs}}$ of the G19 objects is 9.8 mag, while this value is 7.6 for the \textit{Gaia}~DR3 YSO candidates. Because the Orion A is a nearby star forming region with an average distance of $\sim420$ pc, it is expected that more of the fainter YSOs are seen with \textit{Gaia}.

\begin{figure}
   \centering
   \includegraphics[width=\hsize]{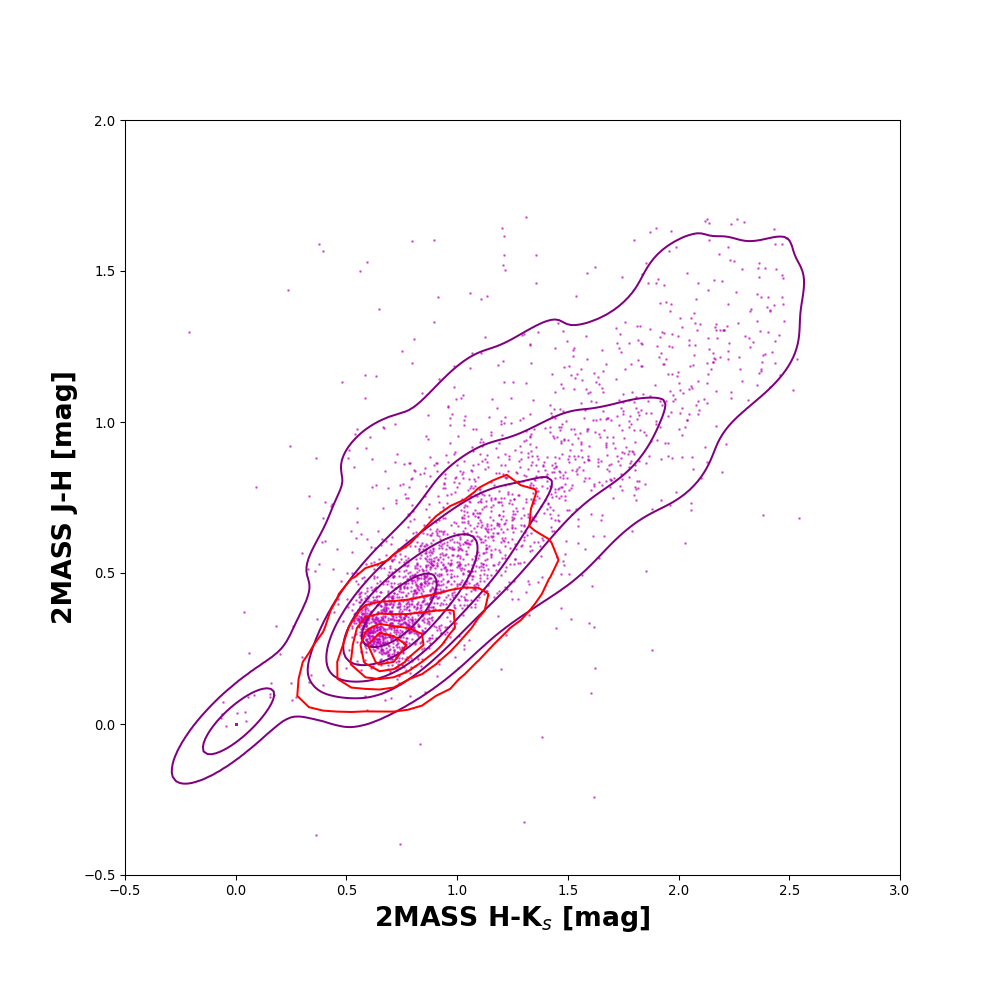}
      \caption{G19 sources (purple dots and contours) and \textit{Gaia}~DR3 YSOs (red contours) on the 2MASS colour--colour diagram. Probability distributions are shown in Figs.~\ref{jhg19} and \ref{hkg19}. The main parameters of the distributions are listed in Table~\ref{ysotable}.
              }
         \label{jhhkg19}
   \end{figure}

The distributions of G19 sources and \textit{Gaia}~DR3
YSOs on the 2MASS colour--colour diagram are presented in Fig.~\ref{jhhkg19}. The two samples show significant overlap, except that fewer G19 objects have J$-$H < 0.6 colour, but sources with larger IR excess are also detected. This can be also explained with the fact that the Orion A is a nearby system, and therefore those sources that do not emit the majority of their energy in the optical domain are still observable with \textit{Gaia}.

\begin{figure}
   \centering
   \includegraphics[width=\hsize]{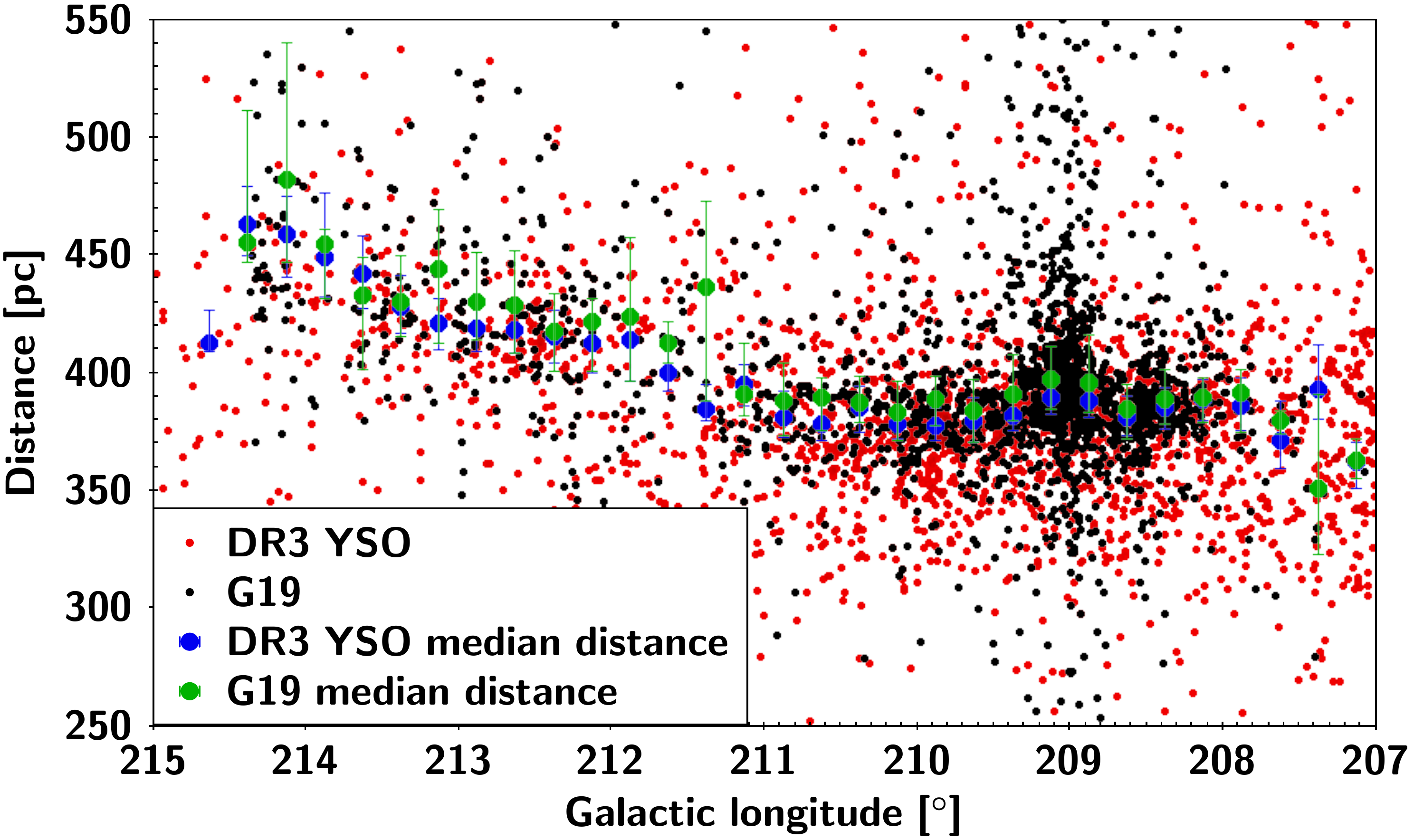}
      \caption{Distance of the \textit{Gaia}~DR3 YSO candidates (red dots) and the \citet{2018A&A...619A.106G} YSOs (black dots) as a function of Galactic longitude. Large dots represent the $med(D)$, which is median of the individual distance values in $0.25\degr$ bins, while error bars represent the $med(D-D_L)$ and $med(D-D_U)$, where $D_L$ and $D_U$ are the lower and upper limits of the individual distance values of the \textit{Gaia}~DR3 YSO candidates (blue dots) and of the \citet{2018A&A...619A.106G} YSOs (green dots).
              }
         \label{grossschedldistance}
   \end{figure}

\begin{figure}
   \centering
   \includegraphics[width=\hsize]{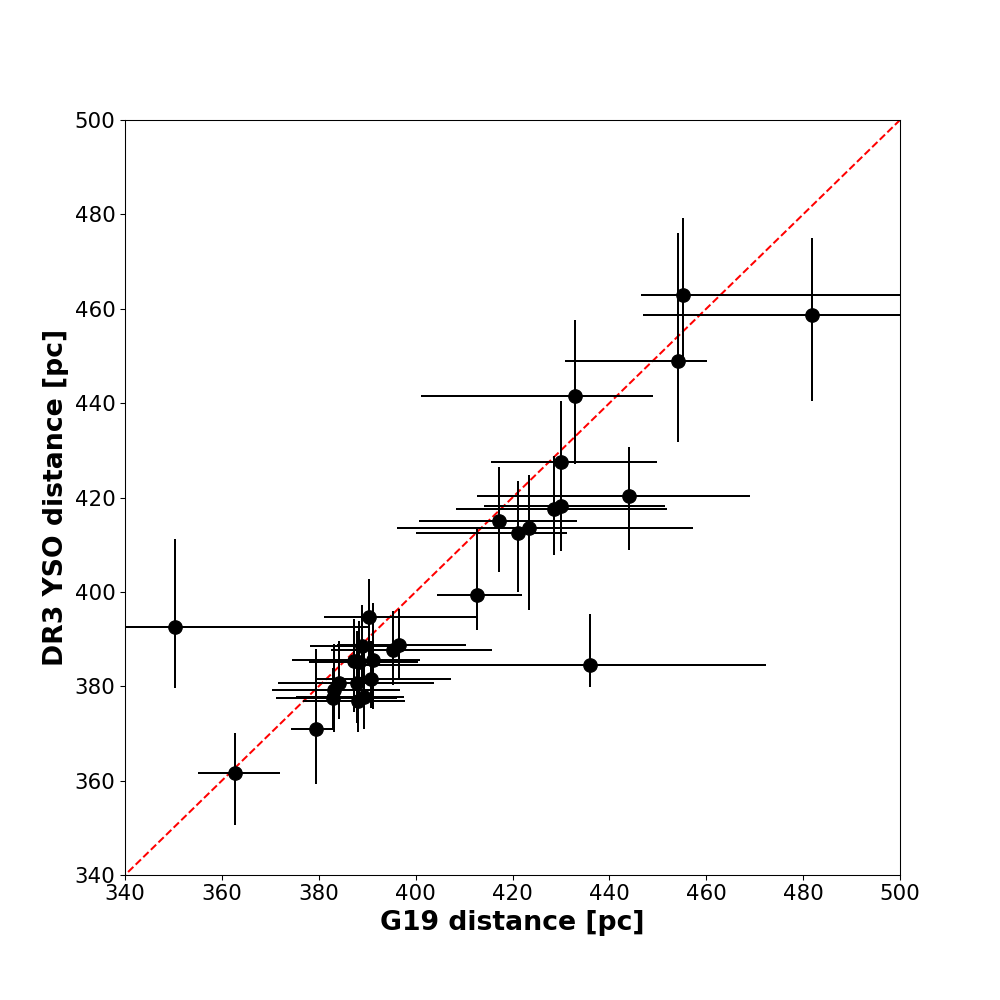}
      \caption{Distances of the DR3 YSO candidates versus those of the G19 YSOs in the star forming region Orion A. The error bars represent the same values as in Fig.\ref{grossschedldistance}.
              }
         \label{orionadistance}
   \end{figure}

We also tested the \textit{Gaia}~DR3 YSO sample against the 3D shape of the Orion~A cloud that was analysed in \citet{2018A&A...619A.106G}. Figures~\ref{grossschedldistance} and~\ref{orionadistance} clearly show that the median distance values for the \textit{Gaia}~DR3 YSO sample  as a function of Galactic longitude are in very good agreement with the findings of these latter authors, and we can confirm that the Orion A cloud is an elongated structure with its head part closer to us and its tail at a greater distance towards higher longitude values.

\subsubsection{Distances of star forming regions}
\begin{figure}
   \centering
   \includegraphics[width=\hsize]{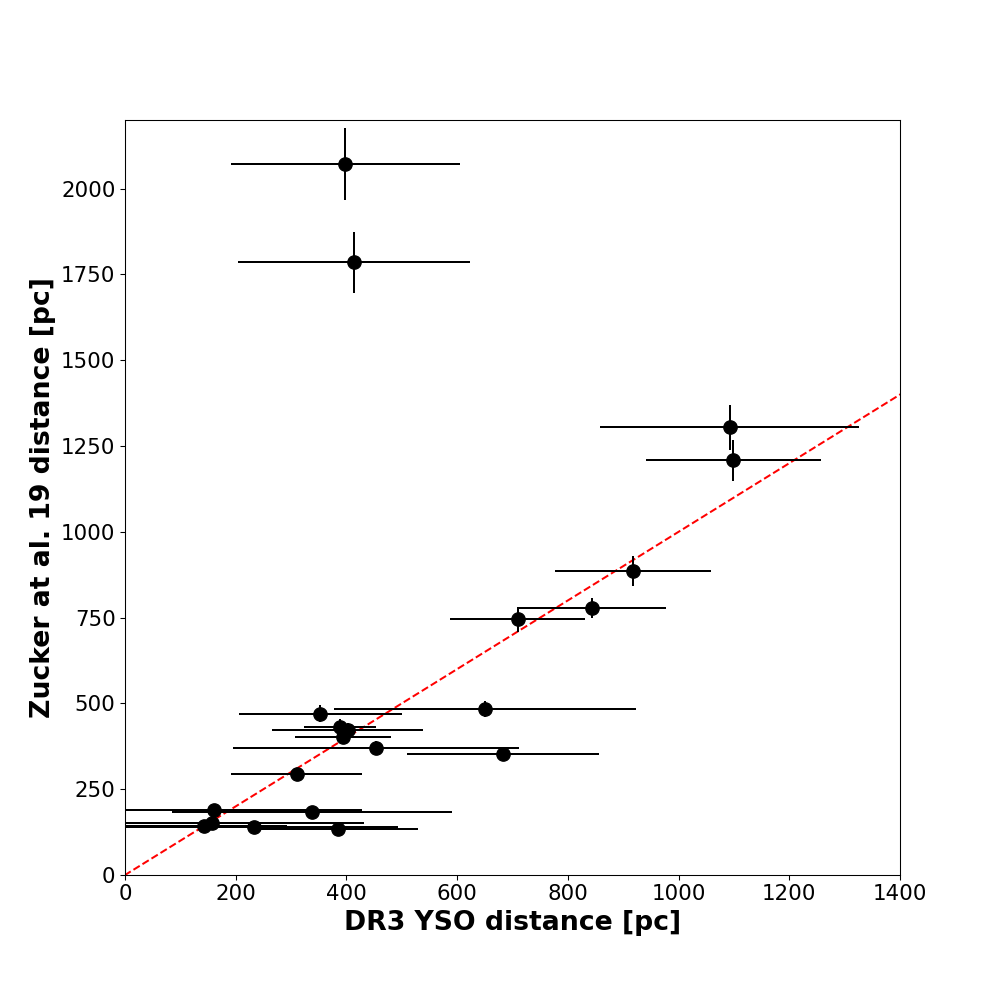}
      \caption{Median distance of \textit{Gaia}~DR3 YSOs in regions defined by \citet{2019ApJ...879..125Z} versus the distances reported by these latter authors based on \textit{Gaia}~DR2 data. Horizontal error bars represent the standard deviation of distances of \textit{Gaia}~DR3 YSOs in the given region. Vertical error bars are the systematic errors given in Table~1 of \citet{2019ApJ...879..125Z}. 
              }
         \label{zuckerdist}
   \end{figure}

\citet{2019ApJ...879..125Z} presented a uniform catalogue of accurate distances to local molecular clouds based on \textit{Gaia}~DR2. These authors used a sophisticated method to derive distances to 27 nearby star forming clouds, and reported upper and lower corner coordinates of regions in which they calculated average distances based on \textit{Gaia} astrometry and optical--NIR photometry. In this comparison, we simply calculated a median distance of the \textit{Gaia}~DR3 YSOs based on the distance values of \citet{2021AJ....161..147B}. The values are listed in Table~\ref{dr3distances}. As shown in Fig.~\ref{zuckerdist}, with the exception of two regions, the \textit{Gaia}~DR3 YSO results are in good agreement with the findings of \citet{2019ApJ...879..125Z}.

The two exceptions are the Gem~OB1 and the Maddalena regions. In the case of the Gem~OB1, the reported distance was 1786$\pm$89~pc, while the \textit{Gaia}~DR3 YSO sources showed a strong peak between 320 and 450~pc, and the farthest object was found to be at a distance of 1067.5~pc. In the direction of the Maddalena star forming region, we also see a peak in the distance distribution between 320 and 420~pc and the farthest object being at a distance of 1079.5~pc. For the case of Maddalena, it is known that despite its large mass, only low level star formation is happening in the cloud \citep[e.g. see][]{2015A&A...575A..79S}.

\begin{table*}\caption{Distances of star-forming clouds based on the \textit{Gaia}~DR3 YSO sample in comparison with those reported by \citet{2019ApJ...879..125Z}.}\label{dr3distances}
\tabcolsep=0.11cm
\begin{tabular}{lrrrrrrr}
\hline
\hline
Cloud name& DR3 median & DR3 parallax &B.-J.(2021) &B.-J.(2021)  & Zucker et al. & Zucker et al.  &DR3 YSO\\  
& parallax [mas]& MAD [mas] &med. dist. [pc]& dist. MAD [pc] & dist. [pc] &dist. err. [pc] & member count\\  
\hline
  AquilaS & 2.473&1.581&385.138 & 144.398 & 133 & 7 & 62\\
  California & 2.771& 0.633& 352.502 & 147.169 & 470 & 24 & 770\\
  Chamaeleon & 2.968& 1.733& 337.852 & 253.633 & 183 & 9 & 394\\
  CMaOB1 & 0.892& 0.188& 1098.853 & 158.152 & 1209 & 60 & 408\\
  CoronaAustralis & 6.351& 2.106& 156.923 & 275.613 & 151 & 8 & 28\\
  Crossbones & 1.051& 0.304& 918.044 & 140.597 & 886 & 44 & 176\\
  GemOB1 & 2.348& 0.931& 414.320 & 209.351 & 1786 & 89 & 64\\
  Lupus & 6.141& 2.233& 161.129 & 267.720 & 189 & 9 & 1059\\
  Maddalena & 2.351&0.835&398.265 & 207.223 & 2072 & 104 & 17\\
  MonOB1 & 1.373&0.175&709.091 & 121.870 & 745 & 37 & 577\\
  MonR2 & 1.146&0.264&844.402 & 132.447 & 778 & 39 & 473\\
  Ophiuchus & 6.978&1.435&141.990 & 150.515 & 144 & 7 & 791\\
  OrionA & 2.531& 0.267& 388.360 & 65.817 & 432 & 22 & 2072\\
  OrionB & 2.439&0.457&402.371 & 136.768 & 423 & 21 & 854\\
  OrionLam & 2.493& 0.374& 393.914 & 87.140 & 402 & 20 & 1075\\
  Perseus & 3.164& 0.496& 310.011 & 117.780 & 294 & 15 & 435\\
  Polaris & 1.420& 0.439& 682.982 & 174.088 & 352 & 18 & 15\\
  Rosette & 0.950& 0.581& 1092.109 & 233.491 & 1304 & 65 & 78\\
  SerpensAqr & 1.576& 0.470& 650.556 & 273.240 & 484 & 24 & 1136\\
  Taurus & 4.355& 2.535&232.237 & 260.722 & 141 & 7 & 432\\
  UrsaMajor & 2.154& 1.596& 453.741 & 258.613 & 371 & 19 & 12\\[2mm]
\hline\end{tabular}
\end{table*}

\subsubsection{Cross-match with the \textit{Gaia} Photometric Science Alerts}

The KYSO catalogue was also used to improve the classification of the \textit{Gaia} Photometric Science Alerts (GSA) \citep{2021A&A...652A..76H} with other YSO catalogues from the literature. Still, numerous objects remained unknown among the alerting sources, including possible YSOs. Therefore, we also checked which of the \textit{Gaia}~DR3 YSOs were also among the alerts. At the beginning of February 2022, 478 alerts appeared on the GSA Index website\footnote{\url{http://gsaweb.ast.cam.ac.uk/alerts/alertsindex}} classified as YSOs or mentioned in the comment section as possible YSOs from the total of 18\,976 alerts. A cross-match revealed that 8905 (46.9\%) of them are in the \textit{Gaia}~DR3 catalogue. The number of \textit{Gaia}~DR3 YSO candidates present in the alert list was found to be 159, which is 33.3\% of the possible YSOs alerts. Also, 41 sources were not identified as confirmed or possible YSOs. In the alert system, these are listed as mostly unknown, red stars in the direction of the Galactic midplane. Below, we present a list of YSO alerts that were or are being investigated with follow-up observations in more detail to infer the underlying physics of their brightness changes and in uncertain cases to confirm their YSO nature. The distance estimates for each source are based on the \textit{Gaia} EDR3 distance catalogue of \citet{2021AJ....161..147B}, while the probability value of being a YSO is from \citet{2019MNRAS.487.2522M}.

\begin{itemize}
\item Gaia22afv ($\alpha_{\rm J2000}$ = 04$^{\rm h}$ 03$^{\rm m}$ 38$\fs$86, $\delta_{\rm J2000}$ = 32$\degr$ 15$'$ 49$\farcs$93) is a candidate YSO, which triggered the \textit{Gaia} Alerts system on 2022 January~18 because of its brightening by $\sim$0.7~mag. Its distance is $263.7^{+6.5}_{-4.8}$~pc and the probability of being a YSO is 67.22\% 

\item Gaia18dlf ($\alpha_{\rm J2000}$ = 20$^{\rm h}$ 57$^{\rm m}$ 03$\fs$36, $\delta_{\rm J2000}$ = 43$\degr$ 41$'$ 44$\farcs$45) is a known YSO, which had a \textit{Gaia} alert on 2018 November~19 because of its long-term (on a timescale of a year) brightening by more than 1~mag. Its distance is $782^{+31}_{-34}$~pc. \citet{2011ApJS..193...25R} classified it as a flat-spectrum source based on the near- to mid-IR SED slope.

\item Gaia21egm (or V733~Cep, $\alpha_{\rm J2000}$ = 22$^{\rm h}$ 53$^{\rm m}$ 33$\fs$25, $\delta_{\rm J2000}$ = 62$\degr$ 32$'$ 23$\farcs$60) is a known FUor \citep{2007AJ....133.1000R}, which triggered a \textit{Gaia} alert on 2021 September~25 because of a drop in its brightness following a long-term fading. Its distance is $724.3^{+31.3}_{-26.8}$~pc. It exhibited a slow rise in its brightness by $\sim$4.5~mag between 1971 and 1993, after which it stayed at maximum brightness for several decades \citep{2010A&A...515A..24P}.

\item Gaia21arq ($\alpha_{\rm J2000}$ = 05$^{\rm h}$ 34$^{\rm m}$ 28$\fs$94, $\delta_{\rm J2000}$ = $-$05$\degr$ 08$'$ 38$\farcs$40) is a known YSO, which triggered a \textit{Gaia} alert on 2021 February~9, due to its brightening by 0.8~mag. Its distance is $354.6^{+17.7}_{-14.9}$~pc. It is a classical T~Tauri star, which was part of the H$\alpha$ survey of the Orion Nebula Cluster by \citet{2013ApJS..208...28S}.

\item Gaia18eap ($\alpha_{\rm J2000}$ = 02$^{\rm h}$ 34$^{\rm m}$ 34$\fs$63, $\delta_{\rm J2000}$ = 61$\degr$ 21$'$ 53$\farcs$64) is a YSO candidate, which had a \textit{Gaia} alert on 2018 December~29 because of its more than 1~mag brightening. The duration of the brightening event was about a year: the source returned to its original brightness by early 2020. Its distance is $980^{+248}_{-189}$~pc. It is classified as a flat-spectrum source \citep{2017ApJS..230....3S}.

\item Gaia21eox ($\alpha_{\rm J2000}$ = 00$^{\rm h}$ 04$^{\rm m}$ 13$\fs$63, $\delta_{\rm J2000}$ = 67$\degr$ 24$'$ 45$\farcs$50) is a YSO candidate at a distance of $977.02^{+56.72}_{-53.84}$~pc. It had a \textit{Gaia} alert on 2021 October~10 because of a 1~mag dimming episode, which ended by January 2022. Its probability of being a YSO is 98\%. 
\end{itemize}
As shown by the list above, the DR3 YSO candidate sample has the potential to contain eruptive YSOs, which are excellent targets for detailed observations aiming to better understand the fundamentals of the disc- and planet formation and evolution.

\subsection{Completeness}\label{completeness}

\subsubsection{Completeness based on the KYSOs}\label{kysocompleteness}

\begin{figure}
   \centering
   \includegraphics[width=\hsize]{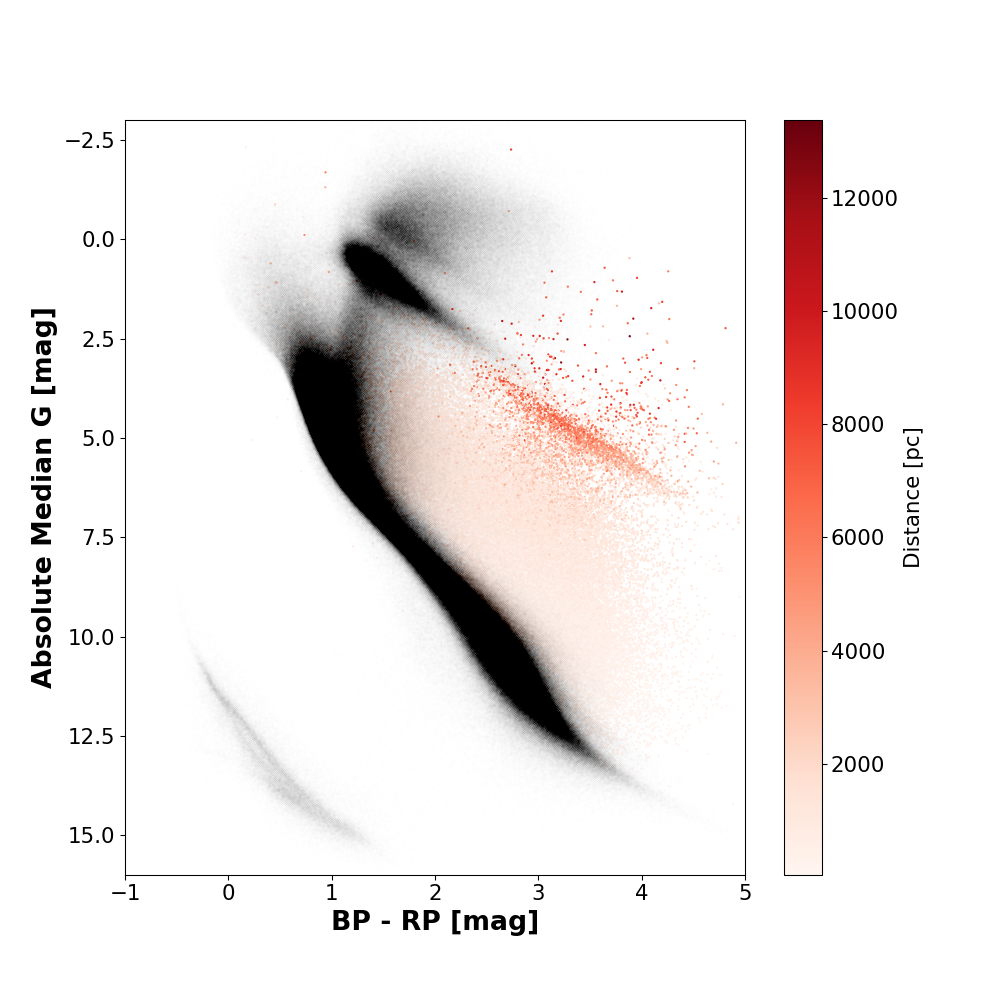}
      \caption{ KYSOs on the observational HRD. Colour coding corresponds to the distance of the individual objects. Grey dots represent the reference objects on the HRD. Many of the YSOs close to the giant branch or above it seem to be distant objects. Also, many of them have parallax errors that make the proper distance estimation problematic, and therefore their position on the diagram is more uncertain.
              }
         \label{distantkyso}
   \end{figure}

The KYSOs also occupy a region on the observational HRD close to or above the giant branch and below the main sequence, while the \textit{Gaia}~DR3 YSOs have a narrower distribution as seen in Fig.~\ref{kysohrd}. After visualising their distances, as shown in Fig.~\ref{distantkyso}, we realised that many of them are distant objects (plotted with dark blue or black colours) and also their $\varpi/\sigma_{\varpi}$ value is below 3, while for the \textit{Gaia}~DR3 YSOs we set a requirement for $\varpi/\sigma_{\varpi}\geq3$. This also means that the distance estimates of the KYSOs are less reliable in some cases and this can cause a less precise position on the colour--magnitude diagrams. Another feature in the distribution of the KYSOs on the observational HRD is an overdensity of sources with \bpminrp colour between 0 and 1~mag  and with an absolute median $G$ band brightness of between 11 and 5~mag, located between the main and white dwarf sequences. The vast majority (>95\%) of these sources are located in the Orion~star forming region and have M spectral type, and all are reported in the study of \citet{2013AJ....146...85H}. They are most likely YSOs for which we do not see their photosphere but scattered light from their protoplanetary discs. At high inclination, when the disc is seen edge-on, the sources appear to be fainter and bluer than they actually are \citep{2010A&A...521A..18G}. We note that this feature on the colour--magnitude diagram also appears in Fig.~\ref{hrdg19}, where the literature YSOs are also from the Orion.

\begin{figure}
   \centering
   \includegraphics[width=\hsize]{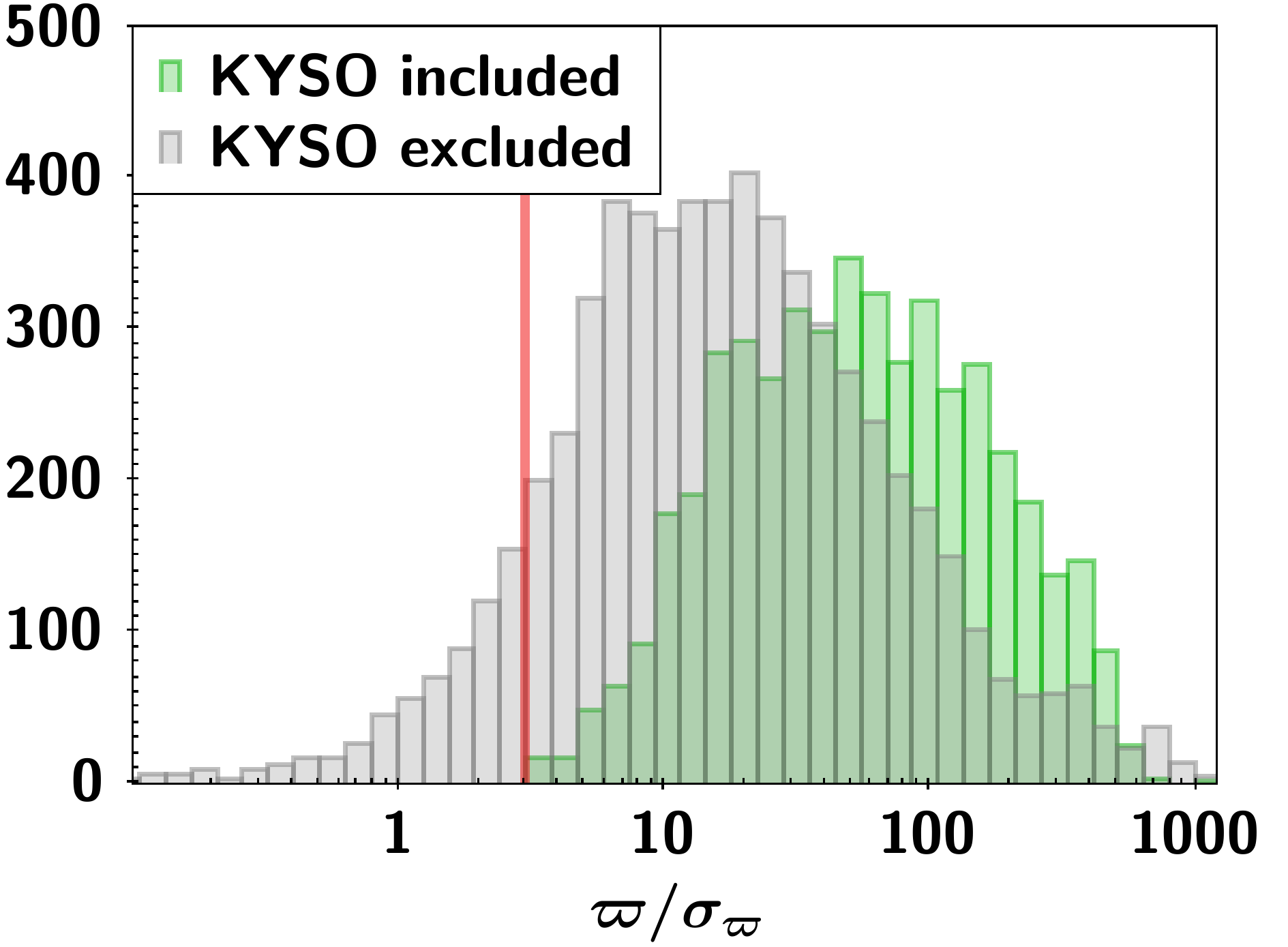}
      \caption{Distribution of parallax over its uncertainty $\varpi/\sigma_{\varpi}$  for  KYSO sources classified as \textit{Gaia}~DR3 YSOs (green bars) and those KYSOs that were excluded from the final \textit{Gaia}~DR3 YSO candidate list (grey bars). The vertical red line at $\varpi/\sigma_{\varpi}=3$ indicates the threshold below which we excluded all sources.
              }
         \label{varpi}
   \end{figure}

As a further step in our investigation, we checked the distribution of the parallax over its uncertainty ($\varpi/\sigma_{\varpi}$) for those sources that were included in the \textit{Gaia}~DR3 YSO sample from the KYSO catalogue and the excluded ones. The distribution of the values is shown in Fig.~\ref{varpi}. The excluded KYSO objects clearly tend to have lower values, meaning that their parallax is less reliable.

In total, 4656 KYSO sources were included in the final \textit{Gaia}~DR3 YSO sample, which means a $\sim$40\% completeness in this comparison. Of the KYSOs, 5\,286  were included in the classification process with the supervised classifier \citep{DR3-DPACP-165}, and 361 of them ended up in a class other than YSO. A quarter of the 11\,671 sources listed in the KYSO table were rejected based on data quality cuts applied in the classification verification phase to ensure that the \textit{Gaia}~DR3 YSO sample, while not complete, provides a reliable list of confirmed variable YSOs and potential YSO candidates. 

\subsubsection{Completeness based on \citet{2019MNRAS.487.2522M}}\label{m19completeness}
A 1$\arcsec$ radius was used to match the \textit{Gaia}~DR3 YSO source positions to the M19 catalogue, resulting in 40\,320 objects being in both the M19 and \textit{Gaia}~DR3 YSO samples, which means 51\% of the \textit{Gaia}~DR3 YSO candidates. While \textit{Gaia}~DR3 provides information on sources from all over the sky, the M19 catalogue was restricted to an area where the dust opacity value based on the Planck foreground maps \citep{2016A&A...594A..10P} was higher than $1.3\times 10^{-5}$. This means that YSOs were searched for on 25.4\% of the sky. Because of this cut, as many as 6\,037 \textit{Gaia}~DR3 YSO candidates cannot have a counterpart in the M19 sample. In the common area, 55\% of the sources have a match. 

In this completeness study, we used sources from the M19 catalogue that had YSO as their best label, meaning that the probability of being a YSO was higher than any of the other probabilities, which means that in cases where $R\geq0.5$, then $LY>LMS$, $LY>LEG$, and $LY>LE$, while if $R<0.5$, then $SY>SMS$, $SY>SEG$, and $SY>SE$, where $MS$, $EG$, and $E$ refer to the probability of being a main sequence source, an extragalactic object, and an evolved star, respectively. The number of sources in the M19 catalogue fulfilling these criteria is 15\,052\,388.
    
In this case, 6\,488 \textit{Gaia}~DR3 YSO candidates were found to have $R \geq 0.5$ and the highest probability label was $LY$ from the M19 catalogue, while 21\,897 candidates had $R < 0.5$ and best label $SY$. This means that, in total, 28\,385 \textit{Gaia}~DR3 YSO candidates were classified as possible YSOs according to the M19 catalogue. Compared to the total number of such sources in the M19 catalogue, the estimated completeness level is 0.19\%. 

If we compare to those M19 sources with $R \geq 0.5 $ and $ LY \geq0.95$ or $R < 0.5 $ and $ SY \geq0.95$ (259\,363 in total), the overlap with the \textit{Gaia}~DR3 YSO candidates is 6\,087 sources, and the completeness level is 2.35\%. The best labels (the object type with the highest probability for a given object) of possible evolved star, possible extragalactic source, and possible main sequence star were assigned to 9482 (23.5\%), 2143 (5.3\%), and 307 (0.8\%) sources, respectively. We cross-matched these sources with SIMBAD in order to investigate the completeness and contamination in more detail. Our findings are summarised in Table~\ref{m19simbad}. Based on the numbers, we find that 879 (65.5\%) out of the 1\,343 sources classified as evolved stars in the M19 are listed as some kind of YSO in SIMBAD, 142 (67\%) of the 212 M19 extragalactic sources that are in SIMBAD are also potential YSOs, and finally 146 (55.5\%) of the 264 SIMBAD counterparts of the M19 main sequence stars are also possibly YSOs.

\begin{table}\caption{SIMBAD classification of \textit{Gaia}~DR3 YSO candidates classified as possible contaminants in the M19 catalogue. The first row lists the total number of \textit{Gaia}~DR3 YSO candidates classified as either evolved star (E), extragalactic source (EG), or main sequence star (MS) in the M19. The second row shows the number of objects also found in SIMBAD. Rows (3)-(6) detail how many of these are classified in SIMBAD as some sort of YSO. The last row shows the number of sources that are not YSO related.}\label{m19simbad}
\begin{tabular}{lrrr}
\hline
\hline
\noalign{\vskip 2mm}
  \multicolumn{1}{c}{} &
  \multicolumn{1}{c}{M19 E} &
  \multicolumn{1}{c}{M19 EG} &
  \multicolumn{1}{c}{M19 MS} \\
\hline

  Total & 9482 & 2143 & 307\\
\\
  in SIMBAD & 1343 & 212 & 264\\
\\
  YSO & 278 & 24 & 41\\
  Candidate\_YSO & 392 & 112 & 54\\
  Orion V* & 54 & 2 & 12\\
  TTau* & 155 & 4 & 39\\
\\
  Probably not YSO & 464 & 70 & 118\\
\hline\end{tabular}
\end{table}

Combining the M19 classifications with the SIMBAD classifications results in 28\,385+1167=29\,552 YSOs. This means a small increase in completeness to 0.20\%.

\subsubsection{Completeness based on \citet{2021ApJS..254...33K}}

As the SPICY catalogue did not require data from the optical domain and is heavily concentrated on the Galactic midplane, one may expect a low fraction of matching sources. Using a 1$\arcsec$ radius, we find only 753 objects present in both the \textit{Gaia}~DR3 YSO sample and the SPICY catalogue. These can all be considered as possible YSOs, because the SPICY catalogue includes only sources that were classified as YSO with a probability of greater than 0.5 in at least one of the three classifiers used in their study.

SPICY also provided an estimated evolutionary class based on the IR slope of the SED. The age of a YSO is increasing in the following order: Class I, flat spectrum, Class II, and finally Class III. We found 8 Class I, 87 flat spectra, 598 Class II, 57 Class III objects, and 3 with uncertain class. In the SPICY catalogue, the numbers of objects in the respective classes are 15\,943, 23\,810, 59\,949, and 5\,352, meaning that 0.05\%, 0.4\%, 1\%, and 1.1\% of each class was found, which reflects that more evolved sources ---which we expect to be more visible in \textit{Gaia}--- are found more reliably. The detailed numbers are listed in Table~\ref{kuhntable}, which also shows that among the sources that were present in the variability analysis, the completeness is much higher as it was found to be 20.3\% considering all evolutionary classes, being the highest among Class II YSOs (23.36\%).

\begin{table*}\caption{Number of objects in the different samples from the SPICY catalogue. The meanings of the columns are as follows: (1) source type as reported in SPICY; (2) number of sources reported in SPICY; (3) number of sources from the SPICY catalogue also present in the \textit{Gaia}~DR3; (4) number of sources participating in the variability classification; (5) number of sources classified as YSO during variability classification; (6) fraction of SPICY sources classified as YSO, i.e. column (5) divided by column (2); (7) fraction of \textit{Gaia}~DR3 sources classified as YSO, i.e. column (5) divided by column (3); (8) fraction of those sources participating in the variability classification and classified as YSO, i.e. column (5) divided by column (4).}\label{kuhntable}
\begin{tabular}{lrrrrrrr}
\hline
\hline
\noalign{\vskip 2mm}
Source type & SPICY &\textit{Gaia}~DR3 & in var. &\textit{Gaia}~DR3 &\textit{Gaia}~DR3 YSO   &\textit{Gaia}~DR3 YSO  &\textit{Gaia}~DR3 YSO \\
 &  &  & class.  & YSO  & \% of SPICY & \% of \textit{Gaia}~DR3 & \% of var.\ class.\\
\hline
\noalign{\vskip 2mm}
  Total & 117\,446 & 37\,693 & 3\,693 & 751 & 0.64 & 1.99 & 20.30\\
  Class I & 15\,943 & 2\,529 & 130 & 8 & 0.05 & 0.32 & 6.15\\
  Flat spectrum & 23\,810 & 5\,480 & 514 & 87 & 0.37 & 1.59 & 16.93\\
  Class II & 59\,949 & 24\,486 & 2\,560& 598 & 1.00 & 2.44 & 23.36\\
  Class III & 5\,352 & 3\,413 & 462 & 57 & 1.07 & 1.67 & 12.34\\
\hline
\end{tabular}
\end{table*}

\subsubsection{Completeness based on \citet{2019A&A...622A.149G}}

Similarly to the SPICY catalogue, this study also reported evolutionary stages for the YSOs, and therefore we were able to estimate completeness as a function of age. The study lists 3117 sources and they are classified into several types as listed in the first column of Table~\ref{grosstable}. The number of sources classified into each type is listed in column~2. In order to obtain an estimate of the completeness of the \textit{Gaia}~DR3 YSO sample, we calculated three quantities. First, we checked how many of the different types of sources were visible to \textit{Gaia} based on \textit{Gaia}~DR3 (column~3), how many of them participated in the variability classification (column~4), and how many are among the \textit{Gaia}~DR3 YSOs (column~5). The ratio of columns~5 and~2 is listed in column~6. Column~7 is the ratio of columns~5 and~3, and column~8 is the ratio of columns~5 and~4.

The total completeness was found to be 19.9\%. As expected, the completeness is very low for Class~0 and~I sources, as they emit most of their radiation at IR wavelengths, but completeness grows with the evolutionary stage and for transition discs it reaches 38.3\%. 

\begin{table*}\caption{Number of objects in the different samples from the G19 catalogue. The columns list: (1) the source type as reported in G19; (2) the number of sources reported by G19; (3) the number of sources from the G19 catalogue also present in \textit{Gaia}~DR3; (4) the number of sources participating in the variability classification; (5) the number of sources classified as YSOs during variability classification; (6) the fraction of G19 sources classified as YSO, i.e. column (5) divided by column (2); (7) the fraction of \textit{Gaia}~DR3 sources classified as YSO, i.e. column (5) divided by column (3); (8) the fraction of those sources participating in the variability classification and classified as YSO; i.e. column (5) divided by column (4).} \label{grosstable}
\begin{tabular}{lrrrrrrr}
\hline
\hline
Source type & Gro{\ss}schedl &\textit{Gaia}~DR3 &in var. &\textit{Gaia}~DR3 &\textit{Gaia}~DR3 YSO   &\textit{Gaia}~DR3 YSO  &\textit{Gaia}~DR3 YSO  \\
&  et al.& & class. & YSO &  \% of G19 &\% of \textit{Gaia}~DR3 &\% of var. class.\\
\hline
  Total & 3117 & 2339 & 672& 621 & 19.9 & 26.5 & 92.4\\
  Class 0 & 60 & 0 & 0&0 & 0.0 & 0.0& 0.0\\
  Class I & 128 & 9 &1& 1 & 0.8 & 11.1&11.1\\
  Flat spectrum & 185 & 91 & 20& 13 & 7.0 & 14.3&65.0\\
  Class II/III PMS & 2012 & 1685 & 503& 473 & 23.5 & 28.1&94.0\\
  with disc &   &   &  &   &   &  &\\
  Anemic disc & 394 & 332 & 63& 57 & 14.5 & 17.2&90.5\\
  Transition disc & 201 & 195 & 83&77 & 38.3 & 39.5&92.8\\
  Galaxy & 39 & 7 & 1& 0 & 0.0 & 0.0&0.0\\
  Nebulosity, fuzzy & 45 & 2 &0& 0 & 0.0 & 0.0& 0.0\\
  contamination &  &  &&  &  & &\\
  MS star & 4 & 2 & 0&0 & 0.0 & 0.0& 0.0\\
  Class III w/o & 3 & 3 & 0&0 & 0.0 & 0.0& 0.0\\
  IR excess &  &  & & & & &\\
  Image artefact & 5 & 0 & 0&0 & 0.0 & 0.0& 0.0\\
  Galaxy candidate & 20 & 4 &0& 0 & 0.0 & 0.0& 0.0\\
  Uncertain YSO & 21 & 9 &0& 0 & 0.0 & 0.0& 0.0\\
  candidate &  &  & &  &  & &\\

\hline
\end{tabular}
\end{table*}

\subsubsection{Cross-match with the SIMBAD database}

The Object Type in SIMBAD is defined as a hierarchical classification, which emphasises the physical nature of the object rather than a peculiar emission in some region of the electromagnetic spectrum or the location in a peculiar cluster or external galaxy. Therefore, objects are only classified as peculiar emitters (in radio, IR, red, blue, UV, X-ray, or gamma ray) if nothing more about the nature of the object is known; that is, if it cannot be decided whether the object is a star, a multiple system, a nebula, or a galaxy \citep{2000A&AS..143....9W}. 

\begin{table}\caption{Number of sources associated with different SIMBAD objects types.}
\label{simbadtable}
\begin{tabular}{lrr}
\hline
\hline
\noalign{\vskip 2mm}
Simbad main\_type &  Number & \% \\[2mm]
\hline
\noalign{\vskip 2mm}
Candidate\_YSO  &       8683    &       42.29   \\
YSO     &       3975    &       19.36   \\
Star    &       2950    &       14.37   \\
TTau*   &       1898    &       9.24    \\
Em*     &       821     &       4.00    \\
Orion\_V*       &       773     &       3.77    \\
RotV*   &       389     &       1.89    \\
low-mass*       &       359     &       1.75    \\
V*      &       187     &       0.91    \\
Ae*     &       48      &       0.23    \\
Candidate\_LP*  &       48      &       0.23    \\
EB*     &       46      &       0.22    \\
**      &       33      &       0.16    \\
BYDra   &       25      &       0.12    \\
RGB*    &       25      &       0.12    \\
RSCVn   &       20      &       0.10    \\
SB*     &       20      &       0.10    \\
X       &       20      &       0.10    \\
Pec*    &       19      &       0.09    \\
Candidate\_TTau*        &       17      &       0.08    \\
NIR     &       15      &       0.07    \\
RRLyr   &       14      &       0.07    \\
EllipVar        &       13      &       0.06    \\
PulsV*  &       13      &       0.06    \\
Eruptive*       &       9       &       0.04    \\
MIR     &       8       &       0.04    \\
Candidate\_EB*  &       7       &       0.03    \\
LPV*    &       7       &       0.03    \\
Be*     &       6       &       0.03    \\
Candidate\_AGB* &       6       &       0.03    \\
Candidate\_brownD*      &       6       &       0.03    \\
HB*     &       6       &       0.03    \\
Irregular\_V*   &       6       &       0.03    \\
PM*     &       6       &       0.03    \\
brownD* &       5       &       0.02    \\
deltaCep        &       4       &       0.02    \\
Mira    &       4       &       0.02    \\
Planet  &       4       &       0.02    \\
PulsV*WVir      &       4       &       0.02    \\
V*?     &       4       &       0.02    \\
Candidate\_Ae*  &       3       &       0.01    \\
IR      &       3       &       0.01    \\
AGN\_Candidate  &       2       &       0.01    \\
Candidate\_Cepheid      &       2       &       0.01    \\
Candidate\_Mi*  &       2       &       0.01    \\
Candidate\_RRLyr        &       2       &       0.01    \\
Cepheid &       2       &       0.01    \\
BLLac\_Candidate        &       1       &       0.00    \\
denseCore       &       1       &       0.00    \\
FIR     &       1       &       0.00    \\
gammaDor        &       1       &       0.00    \\
Nova    &       1       &       0.00    \\
Planet? &       1       &       0.00    \\
PN      &       1       &       0.00    \\
post-AGB*       &       1       &       0.00    \\
PulsV*delSct    &       1       &       0.00    \\
Radio   &       1       &       0.00    \\
RotV*alf2CVn    &       1       &       0.00    \\
S*      &       1       &       0.00    \\[2mm]
\hline
\end{tabular}
\end{table}

The total number of objects we find in the SIMBAD database using a 1$\arcsec$ search radius is 20\,531. The first column in  Table~\ref{simbadtable} is the SIMBAD main\_type, the second column shows how many objects with the given SIMBAD main\_type were found in the \textit{Gaia}~DR3 YSO sample, while the last column is the percentage of the given main\_type according to the total number of 20\,531 associated SIMBAD objects. As one can see from Table~\ref{simbadtable}, there are multiple object types that we can use to estimate the completeness, because they can be considered as potential YSOs. We calculated the sum of objects belonging to the following types: Candidate\_YSO, YSO, TTau*, Candidate\_TTau*, Orion\_V*, Ae*, Candidate\_Ae*, and Be*. The total number of these objects is 15\,363, which means that 74.8\% of the total SIMBAD associations can be considered as potential YSOs. However, the number of sources listed in SIMBAD with the same main\_types is the following: 49\,945 YSO, 99\,097 Candidate\_YSO, 5\,079 TTau*, 317 Candidate\_TTau*, 2\,887 Orion\_V*, 147 Ae*, 67 Candidate\_Ae*, and 2\,215 Be*; in total, 159\,754 sources can be considered as possible YSOs. Based on these numbers, the estimated completeness using the SIMBAD catalogue is 9.6\%. This number is mostly determined by the large number of Candidate\_YSO type objects. If we do not take these into account, the estimated completeness is 11\%.

\subsubsection{Magnitude and extinction limitations}

The completeness in astronomy is mainly determined by the apparent brightness of the objects in the sky. This is especially true for YSOs, as in their early evolutionary stages most of their energy is emitted at wavelengths invisible to \textit{Gaia}. Moreover, they are located in obscure regions where the interstellar dust can modify their observed SEDs, resulting in lower apparent brightness at the visible wavelengths. 

\begin{figure}
   \centering
   \includegraphics[width=\hsize]{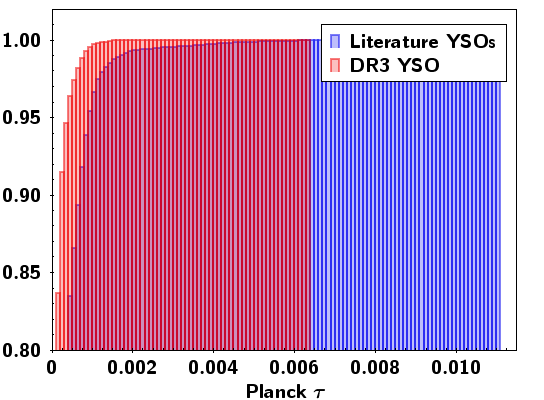}
      \caption{Cumulative distribution of the Planck $\tau$ values in the directions of the YSOs from the KYSO, SPICY, M19, and G19 catalogues (blue bars) and that of the \textit{Gaia}~DR3 YSO candidates (red bars)).
              }
         \label{cumulativetau}
   \end{figure}

Therefore, we also analysed the completeness of the \textit{Gaia}~DR3 YSO candidate sample as a function of the Planck dust opacity value \citep[$\tau$,][]{2016A&A...594A..10P}. As shown in Fig.~\ref{cumulativetau}, we had no objects in the \textit{Gaia}~DR3 YSO candidate sample above $\tau=0.0064$. We find that 99\% of the YSOs from the combined KYSO, SPICY, M19, and G19 catalogues are located above $\tau=0.0018,$ while 99\% of the \textit{Gaia}~DR3 YSO candidates are in regions where $\tau\geq0.0008$. 

\begin{figure}
   \centering
   \includegraphics[width=\hsize]{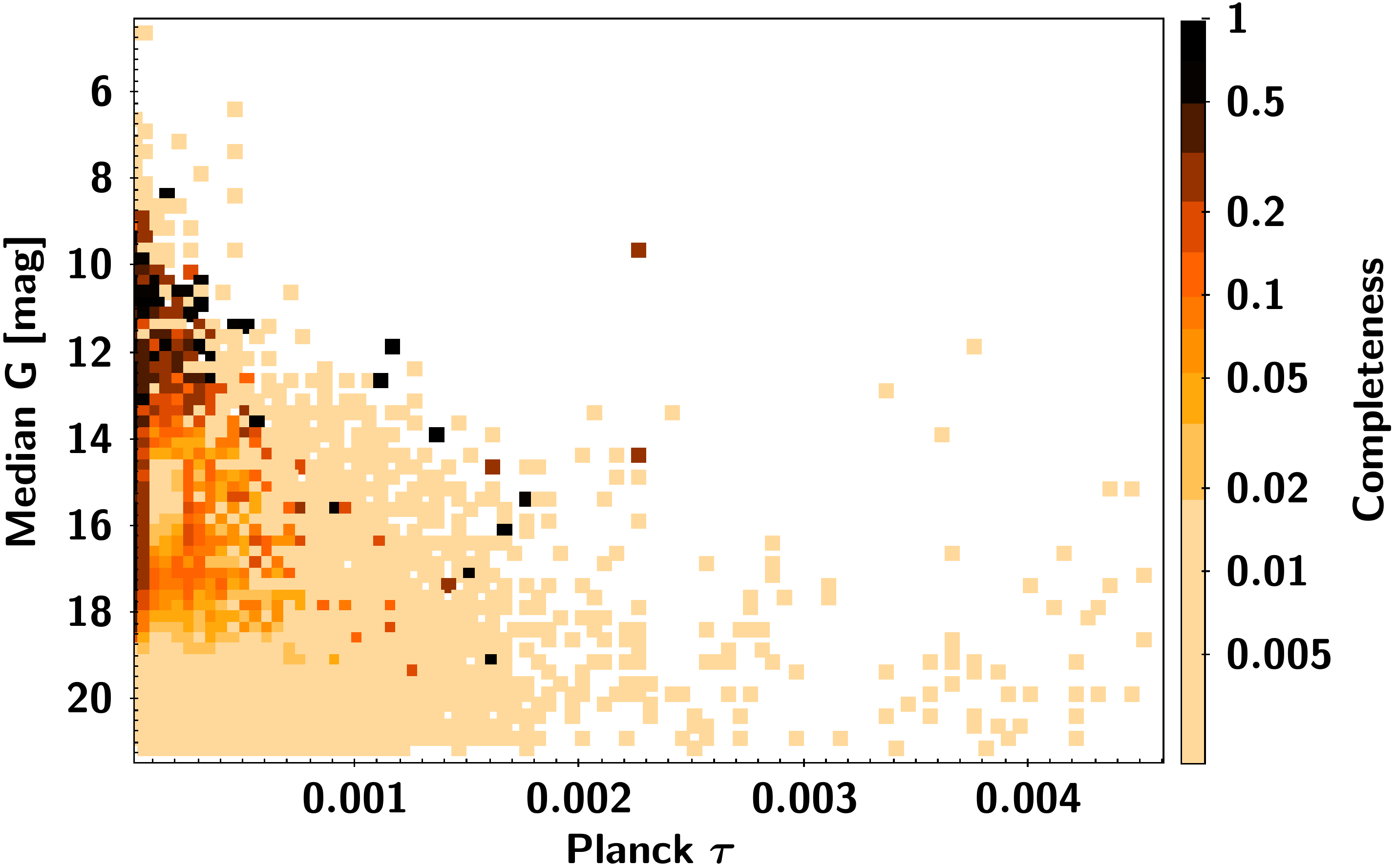}
      \caption{Completeness calculated based on objects from the KYSO, SPICY, M19, and G19 catalogues as a function of Planck dust opacity ($\tau$) and the absolute median \textit{Gaia} G band magnitude. Colour coding presents the fraction of objects that are in any of the mentioned catalogues and also in the \textit{Gaia}~DR3 YSO sample in $5\times10^{-5} \times 0.25$ mag size bins on a logarithmic scale.
              }
         \label{gtaucompleteness}
   \end{figure}

We also analysed the completeness as a function of the absolute G band magnitude and Planck $\tau$. As shown in Fig.~\ref{gtaucompleteness}, very faint YSOs were not recovered by the classification process and also sources at high $\tau$ values are missing, even if they are bright. On the other hand, sources brighter than 19 mag were found with a completeness of higher than 30\% in regions where $\tau$ is very low, but for all magnitude bins, the completeness decreases as $\tau$ increases.

\subsection{Contamination}\label{contamination}

In the previous sections, we mainly focus on the confirmation of the young nature of the sources and show that the \textit{Gaia}~DR3 YSO sample shows strong similarities to known YSOs in the colour--magnitude and colour--colour space, and that their distribution in the sky shows overdensities in the directions of the known star forming regions and towards the Galactic midplane. We also give estimations about the completeness of our catalogue, not only in general, but also as a function of evolutionary stage. In this subsection the SIMBAD and M19 catalogues are used to estimate the contamination rate of the \textit{Gaia}~DR3 YSO sample, complemented with various catalogues accessible through the VizieR service \citep{2000A&AS..143...23O}. Our goal is to find large catalogues listing sources classified into several object types that are potential contaminants of the \textit{Gaia}~DR3 YSO set. In all cases, we used a 1$\arcsec$ radius to find a counterpart to \textit{Gaia}~DR3 YSOs. 

\subsubsection{\citet{2019MNRAS.487.2522M}}
In the M19 catalogue, the number of sources that are likely not related to YSOs  after cross-matching them with SIMBAD and also do not have the best label of $SY$ or $LY$ (depending on the $R$ value) but are either main sequence stars, extragalactic sources, or evolved stars was found to be 652 (35.8\% of the 1819 sources listed in SIMBAD). If we assume that the contamination rate is 35.8\% among that sample of sources for which the best label is not YSO, then the expected number of such sources is 4272. This is 10.6\% of the total number of sources present in both the M19 and \textit{Gaia}~DR3 YSO samples and gives a lower limit of contamination for the whole \textit{Gaia}~DR3 YSO candidate list. However, a more realistic approach could be to estimate the upper limit of the contamination. As described in Sect.~\ref{m19completeness}, there are 40\,320 sources common in the M19 and \textit{Gaia}~DR3 YSO candidate catalogues and 29\,552 of them were identified as a possible YSO by either the M19 classification or the SIMBAD classification. Assuming that these classifications are realistic and that all the other sources are contaminant, 26.7\% of the \textit{Gaia}~DR3 YSO candidates are not young stars.

\subsubsection{SIMBAD}
In some cases, as a result of the cross-match with SIMBAD, the listed main type is too generic (like Star, Em*, low-mass*, etc.; see Table~\ref{simbadtable}). Therefore, in order to estimate the contamination, we took into account only the well-defined object types that definitely cannot be YSOs. The number of objects included  the following object types: BYDra, RGB*, RSCVn, SB*, RRLyr, Candidate\_AGB*, Candidate\_brownD*, HB*, brownD*, deltaCep, Mira, Planet, PulsV*WVir, AGN\_Candidate, Candidate\_Cepheid, Candidate\_Mi*, Candidate\_RRLyr, Cepheid, BLLac\_Candidate, gammaDor, Nova, Planet?, PN, post-AGB*, PulsV*delSct, Radio, RotV*alf2CVn, and S*. In total, these add up to 130 sources, 0.6\% of the \textit{Gaia}~DR3 YSO associated with sources in Simbad. We consider this number as a lower limit for the contamination. As an upper limit, we consider all types that cannot be YSOs as contaminants, which means all types except those in Sect.~\ref{m19completeness} (YSO, Candidate\_YSO, OrionV* and TTau*). Subtracting the number of these objects from the total number of 20\,531 objects common with SIMBAD leaves 5\,202 sources, meaning 25.3\% contamination.

\subsubsection{Cross-match with \citet{2022arXiv220701946G}}
\citet{DR3-DPACP-165} compared the DR3 variable YSO candidates with the catalogues of variable objects collected by \citet{2022arXiv220701946G} as part of the evaluation process of the machine-learning-based classification. As a result \citet{DR3-DPACP-165} found a contamination rate of 79.8\%. We  analysed the 5\,236 matching sources, including 1\,057 true positives (after excluding the trained YSO sources from the evaluation of completeness and contamination) and 4\,179 false positives. The majority of the contamination comes from two types of stars: RS CVn binaries (1542) and BY Dra-type stars (1435). 

About 99\% of the contaminants that have the label BY and 98\% of those with the label RS are from the Zwicky Transient Facility (ZTF) catalogue of \citet{2020ApJS..249...18C}. \citet{2022MNRAS.514.4932C} investigated this ZTF catalogue, paying special attention to the BY Dra-type objects, and collected additional photometric data to validate their nature. These authors found that most of the sources in their catalogue are rapid rotators, and are therefore very likely young stars for which a spin-down has not yet occurred.

To investigate the other main type of contamination, RS CVn-type stars, we did not find any paper that looks at ZTF data in a similar way to \citet{2022MNRAS.514.4932C}, but found that \citet{2022MNRAS.512.4835M} analysed the activity cycles in RS CVn-type stars. We analysed the reported rotational periods and found that while 90\% of the DR3 YSO candidates labelled RS have periods shorter than 9.41 days, 54\% of the \citet{2022MNRAS.512.4835M} RS CVn stars have periods longer than this threshold. The median period for those YSO candidates labelled RS is 3.67 days (MAD=2.27 days), while the median period for the \citet{2022MNRAS.512.4835M} objects is 11.26 days (MAD=8.68 days). Therefore, we conclude that a significant portion of the objects in our sample labelled RS CVn were likely misclassified by \citet{2020ApJS..249...18C} and are more likely YSOs.

Among the 1202 remaining sources that might contaminate the YSO sample, 487 are rotational variables (ROTs), which are defined as a generic class of spotted stars and are scattered all over the colour--magnitude diagram \citep{2022arXiv220701946G}, and therefore cannot be considered as contamination with certainty.

\subsubsection{Other catalogues}

In Table~\ref{viziertable}, we summarise the results of the cross-matches with large tables from the VizieR listing a total of 4\,828\,588 objects of specific types. We find 11~sources that might be galaxies, 16~sources that are listed as eclipsing or ellipsoidal binaries, 32~sources that are also listed as Cepheid/RR~Lyrae-type variables, 5~objects listed as Miras, and 1~star identified as a Mira-type variable. Also 36~sources are listed in the ASAS-SN catalogue of variable stars associated with objects that are certainly not YSO related. The sum of these numbers is 101 objects, which is 0.12\% of the variable \textit{Gaia}~DR3 YSO candidates. We also checked how many of our sources are located in the same regions as the catalogues used for the cross-match. To do so we used the HEALPix \citep{2005ApJ...622..759G} pixelisation of the sky with nside=256 resolution. This corresponds to $1.5978967\times10^{-5}$ sr pixel size. In some cases, the contamination in the areas covered by the different catalogues increases by an order of magnitude, but still remains at the percent level, except for the \citet{2019AJ....158...16S}, where the number of cross-matches suggests a contamination level of $\sim$15.6\%. in the overlapping area.

\begin{table*}\caption{Catalogues cross-matched with the \textit{Gaia}~DR3 YSO sample in order to quantify its contamination. Column~1 lists the titles of the corresponding papers. Column~2 lists the references to the papers. Column~3 gives the number of sources in the given catalogue. Column~4 is the number of \textit{Gaia}~DR3 YSO candidates that are found in the same HEALpixels as the sources of the given catalogue are located. Column~5 is the number of cross-matching \textit{Gaia}~DR3 YSOs found in the catalogue.}\label{viziertable}
\tabcolsep=0.05cm
\begin{tabular}{llrrr}
\hline
\hline
Paper title &  Reference & Nr.\ of entries & Nr.\ of \textit{Gaia}  & Nr.\ of matches \\
& & in the catalogue & DR3 YSOs in the & with \textit{Gaia} \\
& & &same region &DR3 YSOs\\[2mm]
\hline
HYPERLEDA I. Catalogue of galaxies & {\cite{2003A&A...412...45P}} & 983\,261 & 7\,600 &2 \\
6dF galaxy survey final redshift release & {\cite{2009MNRAS.399..683J}} & 124\,647 & 1\,867&0 \\
The Half Million Quasars (HMQ) catalogue & {\cite{2015PASA...32...10F}} & 510\,764 & 988&0 \\
OGLE fundamental-mode RR Lyrae in  & {\cite{2006ApJ...651..197C}} & 1\,913 & 334 &0 \\
Galactic Bulge RRab sample &&&\\
Galactic bulge eclipsing and ellipsoidal binaries& {\cite{2016AcA....66..405S}} & 450\,598 & 5\,677&16 \\
Identification of RR Lyrae stars from the DES &{\cite{2019AJ....158...16S}}&713\,816&192&30\\
JHKs photometry of VVV RR Lyrae stars &{\cite{2018ApJ...857...54D}}& 1\,892 & 965&0 \\
OGLE Galactic center Cepheids and RR Lyrae & {\cite{2017AcA....67..297S}} &40\,112 &5\,319&2\\
SDSS quasar catalogue, fourteenth data release & {\cite{2018A&A...613A..51P}}& 526\,356&345 &0 \\
General Catalogue of Galactic Carbon Stars &{\cite{2001BaltA..10....1A}} &6\,891 &2\,994& 0 \\
LCs re-analysis of Mira variables in ASAS &{\cite{2016ApJS..227....6V}}& 2\,875&363 &0 \\
Mira-like variables from the KELT survey &{\cite{2020ApJS..247...44A}}& 4\,132 &997 &5 \\
Mira Variables in the OGLE Bulge fields &{\cite{2005A&A...443..143G}}& 2\,691&345 &0 \\
Compiled catalogue  of stellar data of Miras & {\cite{2002A&A...384..925K}} &1\,103 & 78&0 \\
Extended catalogue  of NSVS red AGB &{\cite{2008OEJV...87....1U}}& 794 & 374&1 \\
variable stars &&&&\\
LEDA galaxies with DENIS measurements&{\cite{2005A&A...430..751P}}&753\,153&5\,917 &2\\
PSCz catalogue  &{\cite{2000MNRAS.317...55S}}&18\,351& 1\,272&7\\
ASAS-SN catalogue  of variable stars (*) &{\cite{2018MNRAS.477.3145J}}&666\,502& 55\,160& 36\\
Groups of galaxies in 2MASS survey &{\cite{2007ApJ...655..790C}}&18\,737&394 &0\\
\hline
\end{tabular}
\tablefoot{(*) The ASAS-SN catalogue provides a classification for each of their sources. Herein, we take into account  sources  that should not be considered as YSOs, such as CWB, EA, EB, EW, GCAS, and RCB: variables.}
\end{table*}

\section{Summary and conclusions}
We validate the first catalogue of variable YSO candidates observed during the first 34~months of operation of the \textit{Gaia} space telescope. As a result of the classification process, the catalogue lists 79\,375 YSO candidates as part of the third \textit{Gaia} Data Release (DR3). After analysing the different parameter distributions, such as colours, brightness, distance, and apparent positions on the sky, we conclude that the \textit{Gaia}~DR3 YSO sample contains sources that are very similar to confirmed YSOs, mainly including the lower-mass objects. By comparing the \textit{Gaia}~DR3 YSOs to catalogues listing evolutionary stages, we confirm that \textit{Gaia} is more sensitive to the more evolved Class II/III YSOs, which are visible in the optical bands covered by \textit{Gaia}. The estimated completeness at the very early stages of star formation, when most of the energy of   YSOs is emitted at FIR and MIR wavelengths, is close to zero, but it can range from a few percent to $\sim$40\% for objects with transitional discs, depending on the distance of the star forming regions they are located in.
The lower limit of contamination level is at least 0.6\% (according to SIMBAD), but the upper limit is 26.7\%. More realistically, the contamination is lower than the upper limit, but still around the 10\% level. Estimates of the contamination based on the cross-match with several other catalogues containing specific object types are even lower (below 0.2\%), but these objects are mostly located at higher galactic latitudes where only a small fraction of our sources are located. The estimated number of potentially new YSOs presented in the \textit{Gaia}~DR3 variable star catalogue is on the order of a few tens of thousands of objects, but below 44\,651. These objects were not classified as YSOs in the YSO catalogues we used for the validation and were not classified as a different type of object by the other catalogues used in this study.

\begin{acknowledgements}
We are thankful for our anonymous referee for all the comments and suggestion that helped us to improve the paper. This work has made use of data from the European Space Agency (ESA) mission \textit{Gaia} (\url{https://www.cosmos.esa.int/gaia}), processed by the \textit{Gaia} Data Processing and Analysis Consortium (DPAC, \url{https://www.cosmos.esa.int/web/gaia/dpac/consortium}). Funding for the DPAC has been provided by national institutions, some of which participate in the \textit{Gaia} Multilateral Agreement, which include, 
for Switzerland, the Swiss State Secretariat for Education, Research and Innovation through the ESA Prodex program, the `Mesures d'accompagnement', the `Activit\'{e}s Nationales Compl\'{e}mentaires', the Swiss National Science Foundation, and the Early Postdoc. Mobility fellowship; for Hungary, the ESA PRODEX contract nr.4000129910. G.M. acknowledges support from the European Union's Horizon 2020 research and innovation programme under grant agreement No. 101004141. This project has received funding from the European Research Council (ERC) under the European Union's Horizon 2020 research and innovation programme under grant agreement No 716155 (SACCRED). This research has made use of the SIMBAD database, operated at CDS, Strasbourg, France. This research has made use of the VizieR catalogue access tool, CDS, Strasbourg, France (DOI : 10.26093/cds/vizier). 
This work made use of software R~\citep{R-citation} and TOPCAT/STILTS \citep{2005ASPC..347...29T}.
\end{acknowledgements}

%
%
\bibliographystyle{aa} 
\bibliography{references}

\begin{thebibliography}{76}
\expandafter\ifx\csname natexlab\endcsname\relax\def\natexlab#1{#1}\fi

\bibitem[{{Alksnis} {et~al.}(2001){Alksnis}, {Balklavs}, {Dzervitis},
  {Eglitis}, {Paupers}, \& {Pundure}}]{2001BaltA..10....1A}
{Alksnis}, A., {Balklavs}, A., {Dzervitis}, U., {et~al.} 2001, Baltic
  Astronomy, 10, 1

\bibitem[{{Arnold} {et~al.}(2020){Arnold}, {McSwain}, {Pepper}, {Whitelock},
  {Hernitschek}, {James}, {Kuhn}, {Lund}, {Rodriguez}, {Siverd}, \&
  {Stassun}}]{2020ApJS..247...44A}
{Arnold}, R.~A., {McSwain}, M.~V., {Pepper}, J., {et~al.} 2020, \apjs, 247, 44

\bibitem[{{Audard} {et~al.}(2014){Audard}, {{\'A}brah{\'a}m}, {Dunham},
  {Green}, {Grosso}, {Hamaguchi}, {Kastner}, {K{\'o}sp{\'a}l}, {Lodato},
  {Romanova}, {Skinner}, {Vorobyov}, \& {Zhu}}]{2014prpl.conf..387A}
{Audard}, M., {{\'A}brah{\'a}m}, P., {Dunham}, M.~M., {et~al.} 2014, in
  Protostars and Planets VI, ed. H.~{Beuther}, R.~S. {Klessen}, C.~P.
  {Dullemond}, \& T.~{Henning}, 387

\bibitem[{{Bailer-Jones} {et~al.}(2021){Bailer-Jones}, {Rybizki}, {Fouesneau},
  {Demleitner}, \& {Andrae}}]{2021AJ....161..147B}
{Bailer-Jones}, C.~A.~L., {Rybizki}, J., {Fouesneau}, M., {Demleitner}, M., \&
  {Andrae}, R. 2021, \aj, 161, 147

\bibitem[{{Breiman}(2001)}]{Breiman.Random.Forest}
{Breiman}, L. 2001, \rm Machine Learning, 45, 5

\bibitem[{{Carpenter} {et~al.}(2001){Carpenter}, {Hillenbrand}, \&
  {Skrutskie}}]{2001AJ....121.3160C}
{Carpenter}, J.~M., {Hillenbrand}, L.~A., \& {Skrutskie}, M.~F. 2001, \aj, 121,
  3160

\bibitem[{{Chahal} {et~al.}(2022){Chahal}, {de Grijs}, {Kamath}, \&
  {Chen}}]{2022MNRAS.514.4932C}
{Chahal}, D., {de Grijs}, R., {Kamath}, D., \& {Chen}, X. 2022, \mnras, 514,
  4932

\bibitem[{Chen \& Guestrin(2016)}]{Chen:2016:XST:2939672.2939785}
Chen, T. \& Guestrin, C. 2016, in Proceedings of the 22nd ACM SIGKDD
  International Conference on Knowledge Discovery and Data Mining, KDD '16 (New
  York, NY, USA: ACM), 785--794

\bibitem[{{Chen} {et~al.}(2020){Chen}, {Wang}, {Deng}, {de Grijs}, {Yang}, \&
  {Tian}}]{2020ApJS..249...18C}
{Chen}, X., {Wang}, S., {Deng}, L., {et~al.} 2020, \apjs, 249, 18

\bibitem[{{Collinge} {et~al.}(2006){Collinge}, {Sumi}, \&
  {Fabrycky}}]{2006ApJ...651..197C}
{Collinge}, M.~J., {Sumi}, T., \& {Fabrycky}, D. 2006, \apj, 651, 197

\bibitem[{{Covey} {et~al.}(2010){Covey}, {Lada}, {Rom{\'a}n-Z{\'u}{\~n}iga},
  {Muench}, {Forbrich}, \& {Ascenso}}]{2010ApJ...722..971C}
{Covey}, K.~R., {Lada}, C.~J., {Rom{\'a}n-Z{\'u}{\~n}iga}, C., {et~al.} 2010,
  \apj, 722, 971

\bibitem[{{Crook} {et~al.}(2007){Crook}, {Huchra}, {Martimbeau}, {Masters},
  {Jarrett}, \& {Macri}}]{2007ApJ...655..790C}
{Crook}, A.~C., {Huchra}, J.~P., {Martimbeau}, N., {et~al.} 2007, \apj, 655,
  790

\bibitem[{{D{\'e}k{\'a}ny} {et~al.}(2018){D{\'e}k{\'a}ny}, {Hajdu}, {Grebel},
  {Catelan}, {Elorrieta}, {Eyheramendy}, {Majaess}, \&
  {Jord{\'a}n}}]{2018ApJ...857...54D}
{D{\'e}k{\'a}ny}, I., {Hajdu}, G., {Grebel}, E.~K., {et~al.} 2018, \apj, 857,
  54

\bibitem[{{Eyer} {et~al.}(2022){Eyer}, {Audard}, {Holl}, {Rimoldini},
  {Carnerero}, {Clementini}, {De Ridder}, {Distefano}, {Evans}, {Gavras},
  {Gomel}, {Lebzelter}, {Marton}, {Mowlavi}, {Panahi}, {Ripepi}, {Wyrzykowski},
  {Nienartowicz}, {Jevardat de Fombelle}, {Lecoeur-Taibi}, {Rohrbasser},
  {Riello}, {Garcia-Lario}, {Lanzafame}, {Mazeh}, {Raiteri}, {Zucker},
  {Abraham}, {Aerts}, {Aguado}, {Anderson}, {Bashi}, {Binnenfeld}, {Faigler},
  {Garofalo}, {Karbevska}, {Kospal}, {Kruszynska}, {Kun}, {Lanza}, {Leccia},
  {Marconi}, {Messina}, {Molinaro}, {Molnar}, {Muraveva}, {Musella}, {Nagy},
  {Pagano}, {Palaversa}, {Plachy}, {Rybicki}, {Shahaf}, {Szabados},
  {Szegedi-Elek}, {Trabucchi}, {Barblan}, \& {Roelens}}]{DR3-DPACP-162}
{Eyer}, L., {Audard}, M., {Holl}, B., {et~al.} 2022, arXiv e-prints,
  arXiv:2206.06416

\bibitem[{{Eyer} {et~al.}(2017){Eyer}, {Mowlavi}, {Evans}, {Nienartowicz},
  {Ordonez}, {Holl}, {Lecoeur-Taibi}, {Riello}, {Clementini}, {Cuypers}, {De
  Ridder}, {Lanzafame}, {Sarro}, {Charnas}, {Guy}, {Jevardat de Fombelle},
  {Rimoldini}, {S{\"u}veges}, {Mignard}, {Busso}, {De Angeli}, {van Leeuwen},
  {Dubath}, {Beck}, {Aguado}, {Debosscher}, {Distefano}, {Fuchs}, {Koubsky},
  {Lebzelter}, {Leccia}, {Lopez}, {Moitinho}, {Regibo}, {Ripepi}, {Roelens},
  {Szabados}, {Tingley}, {Votruba}, {Zucker}, {Aerts}, {Barblan},
  {Blanco-Cuaresma}, {Grenon}, {Jan}, {Lorenz}, {Miranda}, {Morgenthaler},
  {Ordenovic}, {Palaversa}, {Prsa}, {Ruiz-Fuertes}, {Anderson}, {Delgado},
  {Dzigan}, {Hudec}, {Jonckheere}, {Klagyivik}, {Kutka}, {Moniez}, {Nicoletti},
  {Park}, {Van Hemelryck}, {Varadi}, {Kochoska}, {Lanza}, {Marconi},
  {Marschalko}, {Messina}, {Musella}, {Pagano}, {Sadowski}, \&
  {Schultheis}}]{2017arXiv170203295E}
{Eyer}, L., {Mowlavi}, N., {Evans}, D.~W., {et~al.} 2017, arXiv e-prints,
  arXiv:1702.03295

\bibitem[{{Fischer} {et~al.}(2022){Fischer}, {Hillenbrand}, {Herczeg},
  {Johnstone}, {K{\'o}sp{\'a}l}, \& {Dunham}}]{2022arXiv220311257F}
{Fischer}, W.~J., {Hillenbrand}, L.~A., {Herczeg}, G.~J., {et~al.} 2022, arXiv
  e-prints, arXiv:2203.11257

\bibitem[{{Flesch}(2015)}]{2015PASA...32...10F}
{Flesch}, E.~W. 2015, \pasa, 32, e010

\bibitem[{{Gaia Collaboration} {et~al.}(2016){Gaia Collaboration}, {Prusti},
  {de Bruijne}, {Brown}, {Vallenari}, {Babusiaux}, {Bailer-Jones}, {Bastian},
  {Biermann}, {Evans}, {Eyer}, {Jansen}, {Jordi}, {Klioner}, {Lammers},
  {Lindegren}, {Luri}, {Mignard}, {Milligan}, {Panem}, {Poinsignon},
  {Pourbaix}, {Randich}, {Sarri}, {Sartoretti}, {Siddiqui}, {Soubiran},
  {Valette}, {van Leeuwen}, {Walton}, {Aerts}, {Arenou}, {Cropper}, {Drimmel},
  {H{\o}g}, {Katz}, {Lattanzi}, {O'Mullane}, {Grebel}, {Holland}, {Huc},
  {Passot}, {Bramante}, {Cacciari}, {Casta{\~n}eda}, {Chaoul}, {Cheek}, {De
  Angeli}, {Fabricius}, {Guerra}, {Hern{\'a}ndez}, {Jean-Antoine-Piccolo},
  {Masana}, {Messineo}, {Mowlavi}, {Nienartowicz}, {Ord{\'o}{\~n}ez-Blanco},
  {Panuzzo}, {Portell}, {Richards}, {Riello}, {Seabroke}, {Tanga},
  {Th{\'e}venin}, {Torra}, {Els}, {Gracia-Abril}, {Comoretto},
  {Garcia-Reinaldos}, {Lock}, {Mercier}, {Altmann}, {Andrae}, {Astraatmadja},
  {Bellas-Velidis}, {Benson}, {Berthier}, {Blomme}, {Busso}, {Carry},
  {Cellino}, {Clementini}, {Cowell}, {Creevey}, {Cuypers}, {Davidson}, {De
  Ridder}, {de Torres}, {Delchambre}, {Dell'Oro}, {Ducourant}, {Fr{\'e}mat},
  {Garc{\'\i}a-Torres}, {Gosset}, {Halbwachs}, {Hambly}, {Harrison}, {Hauser},
  {Hestroffer}, {Hodgkin}, {Huckle}, {Hutton}, {Jasniewicz}, {Jordan},
  {Kontizas}, {Korn}, {Lanzafame}, {Manteiga}, {Moitinho}, {Muinonen},
  {Osinde}, {Pancino}, {Pauwels}, {Petit}, {Recio-Blanco}, {Robin}, {Sarro},
  {Siopis}, {Smith}, {Smith}, {Sozzetti}, {Thuillot}, {van Reeven}, {Viala},
  {Abbas}, {Abreu Aramburu}, {Accart}, {Aguado}, {Allan}, {Allasia},
  {Altavilla}, {{\'A}lvarez}, {Alves}, {Anderson}, {Andrei}, {Anglada Varela},
  {Antiche}, {Antoja}, {Ant{\'o}n}, {Arcay}, {Atzei}, {Ayache}, {Bach},
  {Baker}, {Balaguer-N{\'u}{\~n}ez}, {Barache}, {Barata}, {Barbier}, {Barblan},
  {Baroni}, {Barrado y Navascu{\'e}s}, {Barros}, {Barstow}, {Becciani},
  {Bellazzini}, {Bellei}, {Bello Garc{\'\i}a}, {Belokurov}, {Bendjoya},
  {Berihuete}, {Bianchi}, {Bienaym{\'e}}, {Billebaud}, {Blagorodnova},
  {Blanco-Cuaresma}, {Boch}, {Bombrun}, {Borrachero}, {Bouquillon}, {Bourda},
  {Bouy}, {Bragaglia}, {Breddels}, {Brouillet}, {Br{\"u}semeister},
  {Bucciarelli}, {Budnik}, {Burgess}, {Burgon}, {Burlacu}, {Busonero}, {Buzzi},
  {Caffau}, {Cambras}, {Campbell}, {Cancelliere}, {Cantat-Gaudin}, {Carlucci},
  {Carrasco}, {Castellani}, {Charlot}, {Charnas}, {Charvet}, {Chassat},
  {Chiavassa}, {Clotet}, {Cocozza}, {Collins}, {Collins}, {Costigan}, {Crifo},
  {Cross}, {Crosta}, {Crowley}, {Dafonte}, {Damerdji}, {Dapergolas}, {David},
  {David}, {De Cat}, {de Felice}, {de Laverny}, {De Luise}, {De March}, {de
  Martino}, {de Souza}, {Debosscher}, {del Pozo}, {Delbo}, {Delgado},
  {Delgado}, {di Marco}, {Di Matteo}, {Diakite}, {Distefano}, {Dolding}, {Dos
  Anjos}, {Drazinos}, {Dur{\'a}n}, {Dzigan}, {Ecale}, {Edvardsson}, {Enke},
  {Erdmann}, {Escolar}, {Espina}, {Evans}, {Eynard Bontemps}, {Fabre},
  {Fabrizio}, {Faigler}, {Falc{\~a}o}, {Farr{\`a}s Casas}, {Faye}, {Federici},
  {Fedorets}, {Fern{\'a}ndez-Hern{\'a}ndez}, {Fernique}, {Fienga}, {Figueras},
  {Filippi}, {Findeisen}, {Fonti}, {Fouesneau}, {Fraile}, {Fraser}, {Fuchs},
  {Furnell}, {Gai}, {Galleti}, {Galluccio}, {Garabato}, {Garc{\'\i}a-Sedano},
  {Gar{\'e}}, {Garofalo}, {Garralda}, {Gavras}, {Gerssen}, {Geyer}, {Gilmore},
  {Girona}, {Giuffrida}, {Gomes}, {Gonz{\'a}lez-Marcos},
  {Gonz{\'a}lez-N{\'u}{\~n}ez}, {Gonz{\'a}lez-Vidal}, {Granvik}, {Guerrier},
  {Guillout}, {Guiraud}, {G{\'u}rpide}, {Guti{\'e}rrez-S{\'a}nchez}, {Guy},
  {Haigron}, {Hatzidimitriou}, {Haywood}, {Heiter}, {Helmi}, {Hobbs},
  {Hofmann}, {Holl}, {Holland}, {Hunt}, {Hypki}, {Icardi}, {Irwin}, {Jevardat
  de Fombelle}, {Jofr{\'e}}, {Jonker}, {Jorissen}, {Julbe}, {Karampelas},
  {Kochoska}, {Kohley}, {Kolenberg}, {Kontizas}, {Koposov}, {Kordopatis},
  {Koubsky}, {Kowalczyk}, {Krone-Martins}, {Kudryashova}, {Kull}, {Bachchan},
  {Lacoste-Seris}, {Lanza}, {Lavigne}, {Le Poncin-Lafitte}, {Lebreton},
  {Lebzelter}, {Leccia}, {Leclerc}, {Lecoeur-Taibi}, {Lemaitre}, {Lenhardt},
  {Leroux}, {Liao}, {Licata}, {Lindstr{\o}m}, {Lister}, {Livanou}, {Lobel},
  {L{\"o}ffler}, {L{\'o}pez}, {Lopez-Lozano}, {Lorenz}, {Loureiro},
  {MacDonald}, {Magalh{\~a}es Fernandes}, {Managau}, {Mann}, {Mantelet},
  {Marchal}, {Marchant}, {Marconi}, {Marie}, {Marinoni}, {Marrese},
  {Marschalk{\'o}}, {Marshall}, {Mart{\'\i}n-Fleitas}, {Martino}, {Mary},
  {Matijevi{\v{c}}}, {Mazeh}, {McMillan}, {Messina}, {Mestre}, {Michalik},
  {Millar}, {Miranda}, {Molina}, {Molinaro}, {Molinaro}, {Moln{\'a}r},
  {Moniez}, {Montegriffo}, {Monteiro}, {Mor}, {Mora}, {Morbidelli}, {Morel},
  {Morgenthaler}, {Morley}, {Morris}, {Mulone}, {Muraveva}, {Musella},
  {Narbonne}, {Nelemans}, {Nicastro}, {Noval}, {Ord{\'e}novic},
  {Ordieres-Mer{\'e}}, {Osborne}, {Pagani}, {Pagano}, {Pailler}, {Palacin},
  {Palaversa}, {Parsons}, {Paulsen}, {Pecoraro}, {Pedrosa}, {Pentik{\"a}inen},
  {Pereira}, {Pichon}, {Piersimoni}, {Pineau}, {Plachy}, {Plum}, {Poujoulet},
  {Pr{\v{s}}a}, {Pulone}, {Ragaini}, {Rago}, {Rambaux}, {Ramos-Lerate},
  {Ranalli}, {Rauw}, {Read}, {Regibo}, {Renk}, {Reyl{\'e}}, {Ribeiro},
  {Rimoldini}, {Ripepi}, {Riva}, {Rixon}, {Roelens}, {Romero-G{\'o}mez},
  {Rowell}, {Royer}, {Rudolph}, {Ruiz-Dern}, {Sadowski}, {Sagrist{\`a}
  Sell{\'e}s}, {Sahlmann}, {Salgado}, {Salguero}, {Sarasso}, {Savietto},
  {Schnorhk}, {Schultheis}, {Sciacca}, {Segol}, {Segovia}, {Segransan},
  {Serpell}, {Shih}, {Smareglia}, {Smart}, {Smith}, {Solano}, {Solitro},
  {Sordo}, {Soria Nieto}, {Souchay}, {Spagna}, {Spoto}, {Stampa}, {Steele},
  {Steidelm{\"u}ller}, {Stephenson}, {Stoev}, {Suess}, {S{\"u}veges}, {Surdej},
  {Szabados}, {Szegedi-Elek}, {Tapiador}, {Taris}, {Tauran}, {Taylor},
  {Teixeira}, {Terrett}, {Tingley}, {Trager}, {Turon}, {Ulla}, {Utrilla},
  {Valentini}, {van Elteren}, {Van Hemelryck}, {van Leeuwen}, {Varadi},
  {Vecchiato}, {Veljanoski}, {Via}, {Vicente}, {Vogt}, {Voss}, {Votruba},
  {Voutsinas}, {Walmsley}, {Weiler}, {Weingrill}, {Werner}, {Wevers},
  {Whitehead}, {Wyrzykowski}, {Yoldas}, {{\v{Z}}erjal}, {Zucker}, {Zurbach},
  {Zwitter}, {Alecu}, {Allen}, {Allende Prieto}, {Amorim},
  {Anglada-Escud{\'e}}, {Arsenijevic}, {Azaz}, {Balm}, {Beck}, {Bernstein},
  {Bigot}, {Bijaoui}, {Blasco}, {Bonfigli}, {Bono}, {Boudreault}, {Bressan},
  {Brown}, {Brunet}, {Bunclark}, {Buonanno}, {Butkevich}, {Carret}, {Carrion},
  {Chemin}, {Ch{\'e}reau}, {Corcione}, {Darmigny}, {de Boer}, {de Teodoro}, {de
  Zeeuw}, {Delle Luche}, {Domingues}, {Dubath}, {Fodor}, {Fr{\'e}zouls},
  {Fries}, {Fustes}, {Fyfe}, {Gallardo}, {Gallegos}, {Gardiol}, {Gebran},
  {Gomboc}, {G{\'o}mez}, {Grux}, {Gueguen}, {Heyrovsky}, {Hoar}, {Iannicola},
  {Isasi Parache}, {Janotto}, {Joliet}, {Jonckheere}, {Keil}, {Kim},
  {Klagyivik}, {Klar}, {Knude}, {Kochukhov}, {Kolka}, {Kos}, {Kutka}, {Lainey},
  {LeBouquin}, {Liu}, {Loreggia}, {Makarov}, {Marseille}, {Martayan},
  {Martinez-Rubi}, {Massart}, {Meynadier}, {Mignot}, {Munari}, {Nguyen},
  {Nordlander}, {Ocvirk}, {O'Flaherty}, {Olias Sanz}, {Ortiz}, {Osorio},
  {Oszkiewicz}, {Ouzounis}, {Palmer}, {Park}, {Pasquato}, {Peltzer}, {Peralta},
  {P{\'e}turaud}, {Pieniluoma}, {Pigozzi}, {Poels}, {Prat}, {Prod'homme},
  {Raison}, {Rebordao}, {Risquez}, {Rocca-Volmerange}, {Rosen}, {Ruiz-Fuertes},
  {Russo}, {Sembay}, {Serraller Vizcaino}, {Short}, {Siebert}, {Silva},
  {Sinachopoulos}, {Slezak}, {Soffel}, {Sosnowska}, {Strai{\v{z}}ys}, {ter
  Linden}, {Terrell}, {Theil}, {Tiede}, {Troisi}, {Tsalmantza}, {Tur},
  {Vaccari}, {Vachier}, {Valles}, {Van Hamme}, {Veltz}, {Virtanen}, {Wallut},
  {Wichmann}, {Wilkinson}, {Ziaeepour}, \& {Zschocke}}]{2016A&A...595A...1G}
{Gaia Collaboration}, {Prusti}, T., {de Bruijne}, J.~H.~J., {et~al.} 2016,
  \aap, 595, A1

\bibitem[{{Gaia Collaboration} {et~al.}(2022){Gaia Collaboration}, {Vallenari},
  {Brown}, {Prusti}, {de Bruijne}, {Arenou}, {Babusiaux}, {Biermann},
  {Creevey}, {Ducourant}, {Evans}, {Eyer}, {Guerra}, {Hutton}, {Jordi},
  {Klioner}, {Lammers}, {Lindegren}, {Luri}, {Mignard}, {Panem}, {Pourbaix},
  {Randich}, {Sartoretti}, {Soubiran}, {Tanga}, {Walton}, {Bailer-Jones},
  {Bastian}, {Drimmel}, {Jansen}, {Katz}, {Lattanzi}, {van Leeuwen}, {Bakker},
  {Cacciari}, {Casta{\~n}eda}, {De Angeli}, {Fabricius}, {Fouesneau},
  {Fr{\'e}mat}, {Galluccio}, {Guerrier}, {Heiter}, {Masana}, {Messineo},
  {Mowlavi}, {Nicolas}, {Nienartowicz}, {Pailler}, {Panuzzo}, {Riclet}, {Roux},
  {Seabroke}, {Sordo{\o}rcit}, {Th{\'e}venin}, {Gracia-Abril}, {Portell},
  {Teyssier}, {Altmann}, {Andrae}, {Audard}, {Bellas-Velidis}, {Benson},
  {Berthier}, {Blomme}, {Burgess}, {Busonero}, {Busso}, {C{\'a}novas}, {Carry},
  {Cellino}, {Cheek}, {Clementini}, {Damerdji}, {Davidson}, {de Teodoro},
  {Nu{\~n}ez Campos}, {Delchambre}, {Dell'Oro}, {Esquej},
  {Fern{\'a}ndez-Hern{\'a}ndez}, {Fraile}, {Garabato}, {Garc{\'\i}a-Lario},
  {Gosset}, {Haigron}, {Halbwachs}, {Hambly}, {Harrison}, {Hern{\'a}ndez},
  {Hestroffer}, {Hodgkin}, {Holl}, {Jan{\ss}en}, {Jevardat de Fombelle},
  {Jordan}, {Krone-Martins}, {Lanzafame}, {L{\"o}ffler}, {Marchal}, {Marrese},
  {Moitinho}, {Muinonen}, {Osborne}, {Pancino}, {Pauwels}, {Recio-Blanco},
  {Reyl{\'e}}, {Riello}, {Rimoldini}, {Roegiers}, {Rybizki}, {Sarro}, {Siopis},
  {Smith}, {Sozzetti}, {Utrilla}, {van Leeuwen}, {Abbas}, {{\'A}brah{\'a}m},
  {Abreu Aramburu}, {Aerts}, {Aguado}, {Ajaj}, {Aldea-Montero}, {Altavilla},
  {{\'A}lvarez}, {Alves}, {Anders}, {Anderson}, {Anglada Varela}, {Antoja},
  {Baines}, {Baker}, {Balaguer-N{\'u}{\~n}ez}, {Balbinot}, {Balog}, {Barache},
  {Barbato}, {Barros}, {Barstow}, {Bartolom{\'e}}, {Bassilana}, {Bauchet},
  {Becciani}, {Bellazzini}, {Berihuete}, {Bernet}, {Bertone}, {Bianchi},
  {Binnenfeld}, {Blanco-Cuaresma}, {Blazere}, {Boch}, {Bombrun}, {Bossini},
  {Bouquillon}, {Bragaglia}, {Bramante}, {Breedt}, {Bressan}, {Brouillet},
  {Brugaletta}, {Bucciarelli}, {Burlacu}, {Butkevich}, {Buzzi}, {Caffau},
  {Cancelliere}, {Cantat-Gaudin}, {Carballo}, {Carlucci}, {Carnerero},
  {Carrasco}, {Casamiquela}, {Castellani}, {Castro-Ginard}, {Chaoul},
  {Charlot}, {Chemin}, {Chiaramida}, {Chiavassa}, {Chornay}, {Comoretto},
  {Contursi}, {Cooper}, {Cornez}, {Cowell}, {Crifo}, {Cropper}, {Crosta},
  {Crowley}, {Dafonte}, {Dapergolas}, {David}, {David}, {de Laverny}, {De
  Luise}, {De March}, {De Ridder}, {de Souza}, {de Torres}, {del Peloso}, {del
  Pozo}, {Delbo}, {Delgado}, {Delisle}, {Demouchy}, {Dharmawardena}, {Di
  Matteo}, {Diakite}, {Diener}, {Distefano}, {Dolding}, {Edvardsson}, {Enke},
  {Fabre}, {Fabrizio}, {Faigler}, {Fedorets}, {Fernique}, {Fienga}, {Figueras},
  {Fournier}, {Fouron}, {Fragkoudi}, {Gai}, {Garcia-Gutierrez},
  {Garcia-Reinaldos}, {Garc{\'\i}a-Torres}, {Garofalo}, {Gavel}, {Gavras},
  {Gerlach}, {Geyer}, {Giacobbe}, {Gilmore}, {Girona}, {Giuffrida}, {Gomel},
  {Gomez}, {Gonz{\'a}lez-N{\'u}{\~n}ez}, {Gonz{\'a}lez-Santamar{\'\i}a},
  {Gonz{\'a}lez-Vidal}, {Granvik}, {Guillout}, {Guiraud},
  {Guti{\'e}rrez-S{\'a}nchez}, {Guy}, {Hatzidimitriou}, {Hauser}, {Haywood},
  {Helmer}, {Helmi}, {Sarmiento}, {Hidalgo}, {Hilger}, {H{\l}adczuk}, {Hobbs},
  {Holland}, {Huckle}, {Jardine}, {Jasniewicz}, {Jean-Antoine Piccolo},
  {Jim{\'e}nez-Arranz}, {Jorissen}, {Juaristi Campillo}, {Julbe}, {Karbevska},
  {Kervella}, {Khanna}, {Kontizas}, {Kordopatis}, {Korn}, {K{\'o}sp{\'a}l},
  {Kostrzewa-Rutkowska}, {Kruszy{\'n}ska}, {Kun}, {Laizeau}, {Lambert},
  {Lanza}, {Lasne}, {Le Campion}, {Lebreton}, {Lebzelter}, {Leccia}, {Leclerc},
  {Lecoeur-Taibi}, {Liao}, {Licata}, {Lindstr{\o}m}, {Lister}, {Livanou},
  {Lobel}, {Lorca}, {Loup}, {Madrero Pardo}, {Magdaleno Romeo}, {Managau},
  {Mann}, {Manteiga}, {Marchant}, {Marconi}, {Marcos}, {Marcos Santos},
  {Mar{\'\i}n Pina}, {Marinoni}, {Marocco}, {Marshall}, {Polo},
  {Mart{\'\i}n-Fleitas}, {Marton}, {Mary}, {Masip}, {Massari},
  {Mastrobuono-Battisti}, {Mazeh}, {McMillan}, {Messina}, {Michalik}, {Millar},
  {Mints}, {Molina}, {Molinaro}, {Moln{\'a}r}, {Monari}, {Mongui{\'o}},
  {Montegriffo}, {Montero}, {Mor}, {Mora}, {Morbidelli}, {Morel}, {Morris},
  {Muraveva}, {Murphy}, {Musella}, {Nagy}, {Noval}, {Oca{\~n}a}, {Ogden},
  {Ordenovic}, {Osinde}, {Pagani}, {Pagano}, {Palaversa}, {Palicio},
  {Pallas-Quintela}, {Panahi}, {Payne-Wardenaar}, {Pe{\~n}alosa Esteller},
  {Penttil{\"a}}, {Pichon}, {Piersimoni}, {Pineau}, {Plachy}, {Plum}, {Poggio},
  {Pr{\v{s}}a}, {Pulone}, {Racero}, {Ragaini}, {Rainer}, {Raiteri}, {Rambaux},
  {Ramos}, {Ramos-Lerate}, {Re Fiorentin}, {Regibo}, {Richards}, {Rios Diaz},
  {Ripepi}, {Riva}, {Rix}, {Rixon}, {Robichon}, {Robin}, {Robin}, {Roelens},
  {Rogues}, {Rohrbasser}, {Romero-G{\'o}mez}, {Rowell}, {Royer}, {Ruz Mieres},
  {Rybicki}, {Sadowski}, {S{\'a}ez N{\'u}{\~n}ez}, {Sagrist{\`a} Sell{\'e}s},
  {Sahlmann}, {Salguero}, {Samaras}, {Sanchez Gimenez}, {Sanna},
  {Santove{\~n}a}, {Sarasso}, {Schultheis}, {Sciacca}, {Segol}, {Segovia},
  {S{\'e}gransan}, {Semeux}, {Shahaf}, {Siddiqui}, {Siebert}, {Siltala},
  {Silvelo}, {Slezak}, {Slezak}, {Smart}, {Snaith}, {Solano}, {Solitro},
  {Souami}, {Souchay}, {Spagna}, {Spina}, {Spoto}, {Steele},
  {Steidelm{\"u}ller}, {Stephenson}, {S{\"u}veges}, {Surdej}, {Szabados},
  {Szegedi-Elek}, {Taris}, {Taylo}, {Teixeira}, {Tolomei}, {Tonello}, {Torra},
  {Torra}, {Torralba Elipe}, {Trabucchi}, {Tsounis}, {Turon}, {Ulla}, {Unger},
  {Vaillant}, {van Dillen}, {van Reeven}, {Vanel}, {Vecchiato}, {Viala},
  {Vicente}, {Voutsinas}, {Weiler}, {Wevers}, {Wyrzykowski}, {Yoldas}, {Yvard},
  {Zhao}, {Zorec}, {Zucker}, \& {Zwitter}}]{DR3-DPACP-185}
{Gaia Collaboration}, {Vallenari}, A., {Brown}, A.~G.~A., {et~al.} 2022, arXiv
  e-prints, arXiv:2208.00211

\bibitem[{{Gavras} {et~al.}(2022){Gavras}, {Rimoldini}, {Nienartowicz},
  {Jevardat de Fombelle}, {Holl}, {{\'A}brah{\'a}m}, {Audard}, {Carnerero},
  {Clementini}, {De Ridder}, {Distefano}, {Garcia-Lario}, {Garofalo},
  {K{\'o}sp{\'a}l}, {Kruszy{\'n}ska}, {Kun}, {Lecoeur-Ta{\"\i}bi}, {Marton},
  {Mazeh}, {Mowlavi}, {Raiteri}, {Ripepi}, {Szabados}, {Zucker}, \&
  {Eyer}}]{2022arXiv220701946G}
{Gavras}, P., {Rimoldini}, L., {Nienartowicz}, K., {et~al.} 2022, arXiv
  e-prints, arXiv:2207.01946

\bibitem[{{Getman} {et~al.}(2009){Getman}, {Feigelson}, {Luhman},
  {Sicilia-Aguilar}, {Wang}, \& {Garmire}}]{2009ApJ...699.1454G}
{Getman}, K.~V., {Feigelson}, E.~D., {Luhman}, K.~L., {et~al.} 2009, \apj, 699,
  1454

\bibitem[{{G{\'o}rski} {et~al.}(2005){G{\'o}rski}, {Hivon}, {Banday},
  {Wandelt}, {Hansen}, {Reinecke}, \& {Bartelmann}}]{2005ApJ...622..759G}
{G{\'o}rski}, K.~M., {Hivon}, E., {Banday}, A.~J., {et~al.} 2005, \apj, 622,
  759

\bibitem[{{Groenewegen} \& {Blommaert}(2005)}]{2005A&A...443..143G}
{Groenewegen}, M.~A.~T. \& {Blommaert}, J.~A.~D.~L. 2005, \aap, 443, 143

\bibitem[{{Gro{\ss}schedl} {et~al.}(2018){Gro{\ss}schedl}, {Alves}, {Meingast},
  {Ackerl}, {Ascenso}, {Bouy}, {Burkert}, {Forbrich}, {F{\"u}rnkranz},
  {Goodman}, {Hacar}, {Herbst-Kiss}, {Lada}, {Larreina}, {Leschinski},
  {Lombardi}, {Moitinho}, {Mortimer}, \& {Zari}}]{2018A&A...619A.106G}
{Gro{\ss}schedl}, J.~E., {Alves}, J., {Meingast}, S., {et~al.} 2018, \aap, 619,
  A106

\bibitem[{{Gro{\ss}schedl} {et~al.}(2019){Gro{\ss}schedl}, {Alves}, {Teixeira},
  {Bouy}, {Forbrich}, {Lada}, {Meingast}, {Hacar}, {Ascenso}, {Ackerl},
  {Hasenberger}, {K{\"o}hler}, {Kubiak}, {Larreina}, {Linhardt}, {Lombardi}, \&
  {M{\"o}ller}}]{2019A&A...622A.149G}
{Gro{\ss}schedl}, J.~E., {Alves}, J., {Teixeira}, P.~S., {et~al.} 2019, \aap,
  622, A149

\bibitem[{{Guarcello} {et~al.}(2010){Guarcello}, {Damiani}, {Micela}, {Peres},
  {Prisinzano}, \& {Sciortino}}]{2010A&A...521A..18G}
{Guarcello}, M.~G., {Damiani}, F., {Micela}, G., {et~al.} 2010, \aap, 521, A18

\bibitem[{{Herbig} \& {Bell}(1988)}]{1988cels.book.....H}
{Herbig}, G.~H. \& {Bell}, K.~R. 1988, {Third Catalog of Emission-Line Stars of
  the Orion Population : 3 : 1988}

\bibitem[{{Herbst} \& {Shevchenko}(1999)}]{1999AJ....118.1043H}
{Herbst}, W. \& {Shevchenko}, V.~S. 1999, \aj, 118, 1043

\bibitem[{{Hillenbrand} \& {Findeisen}(2015)}]{2015ApJ...808...68H}
{Hillenbrand}, L.~A. \& {Findeisen}, K.~P. 2015, \apj, 808, 68

\bibitem[{{Hillenbrand} {et~al.}(2013){Hillenbrand}, {Hoffer}, \&
  {Herczeg}}]{2013AJ....146...85H}
{Hillenbrand}, L.~A., {Hoffer}, A.~S., \& {Herczeg}, G.~J. 2013, \aj, 146, 85

\bibitem[{{Hodgkin} {et~al.}(2021){Hodgkin}, {Harrison}, {Breedt}, {Wevers},
  {Rixon}, {Delgado}, {Yoldas}, {Kostrzewa-Rutkowska}, {Wyrzykowski}, {van
  Leeuwen}, {Blagorodnova}, {Campbell}, {Eappachen}, {Fraser}, {Ihanec},
  {Koposov}, {Kruszy{\'n}ska}, {Marton}, {Rybicki}, {Brown}, {Burgess},
  {Busso}, {Cowell}, {De Angeli}, {Diener}, {Evans}, {Gilmore}, {Holland},
  {Jonker}, {van Leeuwen}, {Mignard}, {Osborne}, {Portell}, {Prusti},
  {Richards}, {Riello}, {Seabroke}, {Walton}, {{\'A}brah{\'a}m}, {Altavilla},
  {Baker}, {Bastian}, {O'Brien}, {de Bruijne}, {Butterley}, {Carrasco},
  {Casta{\~n}eda}, {Clark}, {Clementini}, {Copperwheat}, {Cropper},
  {Damljanovic}, {Davidson}, {Davis}, {Dennefeld}, {Dhillon}, {Dolding},
  {Dominik}, {Esquej}, {Eyer}, {Fabricius}, {Fridman}, {Froebrich}, {Garralda},
  {Gomboc}, {Gonz{\'a}lez-Vidal}, {Guerra}, {Hambly}, {Hardy}, {Holl},
  {Hourihane}, {Japelj}, {Kann}, {Kiss}, {Knigge}, {Kolb}, {Komossa},
  {K{\'o}sp{\'a}l}, {Kov{\'a}cs}, {Kun}, {Leto}, {Lewis}, {Littlefair},
  {Mahabal}, {Mundell}, {Nagy}, {Padeletti}, {Palaversa}, {Pigulski},
  {Pretorius}, {van Reeven}, {Ribeiro}, {Roelens}, {Rowell}, {Schartel},
  {Scholz}, {Schwope}, {Sip{\H{o}}cz}, {Smartt}, {Smith}, {Serraller},
  {Steeghs}, {Sullivan}, {Szabados}, {Szegedi-Elek}, {Tisserand}, {Tomasella},
  {van Velzen}, {Whitelock}, {Wilson}, \& {Young}}]{2021A&A...652A..76H}
{Hodgkin}, S.~T., {Harrison}, D.~L., {Breedt}, E., {et~al.} 2021, \aap, 652,
  A76

\bibitem[{{Jayasinghe} {et~al.}(2018){Jayasinghe}, {Kochanek}, {Stanek},
  {Shappee}, {Holoien}, {Thompson}, {Prieto}, {Dong}, {Pawlak}, {Shields},
  {Pojmanski}, {Otero}, {Britt}, \& {Will}}]{2018MNRAS.477.3145J}
{Jayasinghe}, T., {Kochanek}, C.~S., {Stanek}, K.~Z., {et~al.} 2018, \mnras,
  477, 3145

\bibitem[{{Jones} {et~al.}(2009){Jones}, {Read}, {Saunders}, {Colless},
  {Jarrett}, {Parker}, {Fairall}, {Mauch}, {Sadler}, {Watson}, {Burton},
  {Campbell}, {Cass}, {Croom}, {Dawe}, {Fiegert}, {Frankcombe}, {Hartley},
  {Huchra}, {James}, {Kirby}, {Lahav}, {Lucey}, {Mamon}, {Moore}, {Peterson},
  {Prior}, {Proust}, {Russell}, {Safouris}, {Wakamatsu}, {Westra}, \&
  {Williams}}]{2009MNRAS.399..683J}
{Jones}, D.~H., {Read}, M.~A., {Saunders}, W., {et~al.} 2009, \mnras, 399, 683

\bibitem[{{Kharchenko} {et~al.}(2002){Kharchenko}, {Kilpio}, {Malkov}, \&
  {Schilbach}}]{2002A&A...384..925K}
{Kharchenko}, N., {Kilpio}, E., {Malkov}, O., \& {Schilbach}, E. 2002, \aap,
  384, 925

\bibitem[{{Kochanek} {et~al.}(2017){Kochanek}, {Shappee}, {Stanek}, {Holoien},
  {Thompson}, {Prieto}, {Dong}, {Shields}, {Will}, {Britt}, {Perzanowski}, \&
  {Pojma{\'n}ski}}]{2017PASP..129j4502K}
{Kochanek}, C.~S., {Shappee}, B.~J., {Stanek}, K.~Z., {et~al.} 2017, \pasp,
  129, 104502

\bibitem[{{Koenig} \& {Leisawitz}(2014)}]{2014ApJ...791..131K}
{Koenig}, X.~P. \& {Leisawitz}, D.~T. 2014, \apj, 791, 131

\bibitem[{{Kuhn} {et~al.}(2021){Kuhn}, {de Souza}, {Krone-Martins},
  {Castro-Ginard}, {Ishida}, {Povich}, {Hillenbrand}, \& {COIN
  Collaboration}}]{2021ApJS..254...33K}
{Kuhn}, M.~A., {de Souza}, R.~S., {Krone-Martins}, A., {et~al.} 2021, \apjs,
  254, 33

\bibitem[{{Mamajek} {et~al.}(2002){Mamajek}, {Meyer}, \&
  {Liebert}}]{2002AJ....124.1670M}
{Mamajek}, E.~E., {Meyer}, M.~R., \& {Liebert}, J. 2002, \aj, 124, 1670

\bibitem[{{Marrese} {et~al.}(2019){Marrese}, {Marinoni}, {Fabrizio}, \&
  {Altavilla}}]{2019A&A...621A.144M}
{Marrese}, P.~M., {Marinoni}, S., {Fabrizio}, M., \& {Altavilla}, G. 2019,
  \aap, 621, A144

\bibitem[{{Mart{\'\i}nez} {et~al.}(2022){Mart{\'\i}nez}, {Mauas}, \&
  {Buccino}}]{2022MNRAS.512.4835M}
{Mart{\'\i}nez}, C.~I., {Mauas}, P.~J.~D., \& {Buccino}, A.~P. 2022, \mnras,
  512, 4835

\bibitem[{{Marton} {et~al.}(2019){Marton}, {{\'A}brah{\'a}m}, {Szegedi-Elek},
  {Varga}, {Kun}, {K{\'o}sp{\'a}l}, {Varga-Vereb{\'e}lyi}, {Hodgkin},
  {Szabados}, {Beck}, \& {Kiss}}]{2019MNRAS.487.2522M}
{Marton}, G., {{\'A}brah{\'a}m}, P., {Szegedi-Elek}, E., {et~al.} 2019, \mnras,
  487, 2522

\bibitem[{{Masci} {et~al.}(2019){Masci}, {Laher}, {Rusholme}, {Shupe}, {Groom},
  {Surace}, {Jackson}, {Monkewitz}, {Beck}, {Flynn}, {Terek}, {Landry},
  {Hacopians}, {Desai}, {Howell}, {Brooke}, {Imel}, {Wachter}, {Ye}, {Lin},
  {Cenko}, {Cunningham}, {Rebbapragada}, {Bue}, {Miller}, {Mahabal}, {Bellm},
  {Patterson}, {Juri{\'c}}, {Golkhou}, {Ofek}, {Walters}, {Graham}, {Kasliwal},
  {Dekany}, {Kupfer}, {Burdge}, {Cannella}, {Barlow}, {Van Sistine}, {Giomi},
  {Fremling}, {Blagorodnova}, {Levitan}, {Riddle}, {Smith}, {Helou}, {Prince},
  \& {Kulkarni}}]{2019PASP..131a8003M}
{Masci}, F.~J., {Laher}, R.~R., {Rusholme}, B., {et~al.} 2019, \pasp, 131,
  018003

\bibitem[{{Morales-Calder{\'o}n} {et~al.}(2011){Morales-Calder{\'o}n},
  {Stauffer}, {Hillenbrand}, {Gutermuth}, {Song}, {Rebull}, {Plavchan},
  {Carpenter}, {Whitney}, {Covey}, {Alves de Oliveira}, {Winston},
  {McCaughrean}, {Bouvier}, {Guieu}, {Vrba}, {Holtzman}, {Marchis}, {Hora},
  {Wasserman}, {Terebey}, {Megeath}, {Guinan}, {Forbrich}, {Hu{\'e}lamo},
  {Riviere-Marichalar}, {Barrado}, {Stapelfeldt}, {Hern{\'a}ndez}, {Allen},
  {Ardila}, {Bayo}, {Favata}, {James}, {Werner}, \&
  {Wood}}]{2011ApJ...733...50M}
{Morales-Calder{\'o}n}, M., {Stauffer}, J.~R., {Hillenbrand}, L.~A., {et~al.}
  2011, \apj, 733, 50

\bibitem[{{Ochsenbein} {et~al.}(2000){Ochsenbein}, {Bauer}, \&
  {Marcout}}]{2000A&AS..143...23O}
{Ochsenbein}, F., {Bauer}, P., \& {Marcout}, J. 2000, \aaps, 143, 23

\bibitem[{{P{\^a}ris} {et~al.}(2018){P{\^a}ris}, {Petitjean}, {Aubourg},
  {Myers}, {Streblyanska}, {Lyke}, {Anderson}, {Armengaud}, {Bautista},
  {Blanton}, {Blomqvist}, {Brinkmann}, {Brownstein}, {Brandt}, {Burtin},
  {Dawson}, {de la Torre}, {Georgakakis}, {Gil-Mar{\'\i}n}, {Green}, {Hall},
  {Kneib}, {LaMassa}, {Le Goff}, {MacLeod}, {Mariappan}, {McGreer}, {Merloni},
  {Noterdaeme}, {Palanque-Delabrouille}, {Percival}, {Ross}, {Rossi},
  {Schneider}, {Seo}, {Tojeiro}, {Weaver}, {Weijmans}, {Y{\`e}che}, {Zarrouk},
  \& {Zhao}}]{2018A&A...613A..51P}
{P{\^a}ris}, I., {Petitjean}, P., {Aubourg}, {\'E}., {et~al.} 2018, \aap, 613,
  A51

\bibitem[{{Paturel} {et~al.}(2003){Paturel}, {Petit}, {Prugniel}, {Theureau},
  {Rousseau}, {Brouty}, {Dubois}, \& {Cambr{\'e}sy}}]{2003A&A...412...45P}
{Paturel}, G., {Petit}, C., {Prugniel}, P., {et~al.} 2003, \aap, 412, 45

\bibitem[{{Paturel} {et~al.}(2005){Paturel}, {Vauglin}, {Petit},
  {Borsenberger}, {Epchtein}, {Fouqu{\'e}}, \& {Mamon}}]{2005A&A...430..751P}
{Paturel}, G., {Vauglin}, I., {Petit}, C., {et~al.} 2005, \aap, 430, 751

\bibitem[{{Peneva} {et~al.}(2010){Peneva}, {Semkov}, {Munari}, \&
  {Birkle}}]{2010A&A...515A..24P}
{Peneva}, S.~P., {Semkov}, E.~H., {Munari}, U., \& {Birkle}, K. 2010, \aap,
  515, A24

\bibitem[{{Planck Collaboration} {et~al.}(2016){Planck Collaboration}, {Adam},
  {Ade}, {Aghanim}, {Alves}, {Arnaud}, {Ashdown}, {Aumont}, {Baccigalupi},
  {Banday}, {Barreiro}, {Bartlett}, {Bartolo}, {Battaner}, {Benabed},
  {Beno{\^\i}t}, {Benoit-L{\'e}vy}, {Bernard}, {Bersanelli}, {Bielewicz},
  {Bock}, {Bonaldi}, {Bonavera}, {Bond}, {Borrill}, {Bouchet}, {Boulanger},
  {Bucher}, {Burigana}, {Butler}, {Calabrese}, {Cardoso}, {Catalano},
  {Challinor}, {Chamballu}, {Chary}, {Chiang}, {Christensen}, {Clements},
  {Colombi}, {Colombo}, {Combet}, {Couchot}, {Coulais}, {Crill}, {Curto},
  {Cuttaia}, {Danese}, {Davies}, {Davis}, {de Bernardis}, {de Rosa}, {de
  Zotti}, {Delabrouille}, {D{\'e}sert}, {Dickinson}, {Diego}, {Dole},
  {Donzelli}, {Dor{\'e}}, {Douspis}, {Ducout}, {Dupac}, {Efstathiou}, {Elsner},
  {En{\ss}lin}, {Eriksen}, {Falgarone}, {Fergusson}, {Finelli}, {Forni},
  {Frailis}, {Fraisse}, {Franceschi}, {Frejsel}, {Galeotta}, {Galli}, {Ganga},
  {Ghosh}, {Giard}, {Giraud-H{\'e}raud}, {Gjerl{\o}w}, {Gonz{\'a}lez-Nuevo},
  {G{\'o}rski}, {Gratton}, {Gregorio}, {Gruppuso}, {Gudmundsson}, {Hansen},
  {Hanson}, {Harrison}, {Helou}, {Henrot-Versill{\'e}},
  {Hern{\'a}ndez-Monteagudo}, {Herranz}, {Hildebrandt}, {Hivon}, {Hobson},
  {Holmes}, {Hornstrup}, {Hovest}, {Huffenberger}, {Hurier}, {Jaffe}, {Jaffe},
  {Jones}, {Juvela}, {Keih{\"a}nen}, {Keskitalo}, {Kisner}, {Kneissl},
  {Knoche}, {Kunz}, {Kurki-Suonio}, {Lagache}, {L{\"a}hteenm{\"a}ki},
  {Lamarre}, {Lasenby}, {Lattanzi}, {Lawrence}, {Le Jeune}, {Leahy},
  {Leonardi}, {Lesgourgues}, {Levrier}, {Liguori}, {Lilje}, {Linden-V{\o}rnle},
  {L{\'o}pez-Caniego}, {Lubin}, {Mac{\'\i}as-P{\'e}rez}, {Maggio}, {Maino},
  {Mandolesi}, {Mangilli}, {Maris}, {Marshall}, {Martin},
  {Mart{\'\i}nez-Gonz{\'a}lez}, {Masi}, {Matarrese}, {McGehee}, {Meinhold},
  {Melchiorri}, {Mendes}, {Mennella}, {Migliaccio}, {Mitra},
  {Miville-Desch{\^e}nes}, {Moneti}, {Montier}, {Morgante}, {Mortlock}, {Moss},
  {Munshi}, {Murphy}, {Naselsky}, {Nati}, {Natoli}, {Netterfield},
  {N{\o}rgaard-Nielsen}, {Noviello}, {Novikov}, {Novikov}, {Orlando},
  {Oxborrow}, {Paci}, {Pagano}, {Pajot}, {Paladini}, {Paoletti}, {Partridge},
  {Pasian}, {Patanchon}, {Pearson}, {Perdereau}, {Perotto}, {Perrotta},
  {Pettorino}, {Piacentini}, {Piat}, {Pierpaoli}, {Pietrobon}, {Plaszczynski},
  {Pointecouteau}, {Polenta}, {Pratt}, {Pr{\'e}zeau}, {Prunet}, {Puget},
  {Rachen}, {Reach}, {Rebolo}, {Reinecke}, {Remazeilles}, {Renault}, {Renzi},
  {Ristorcelli}, {Rocha}, {Rosset}, {Rossetti}, {Roudier},
  {Rubi{\~n}o-Mart{\'\i}n}, {Rusholme}, {Sandri}, {Santos}, {Savelainen},
  {Savini}, {Scott}, {Seiffert}, {Shellard}, {Spencer}, {Stolyarov}, {Stompor},
  {Strong}, {Sudiwala}, {Sunyaev}, {Sutton}, {Suur-Uski}, {Sygnet}, {Tauber},
  {Terenzi}, {Toffolatti}, {Tomasi}, {Tristram}, {Tucci}, {Tuovinen}, {Umana},
  {Valenziano}, {Valiviita}, {Van Tent}, {Vielva}, {Villa}, {Wade}, {Wandelt},
  {Wehus}, {Wilkinson}, {Yvon}, {Zacchei}, \& {Zonca}}]{2016A&A...594A..10P}
{Planck Collaboration}, {Adam}, R., {Ade}, P.~A.~R., {et~al.} 2016, \aap, 594,
  A10

\bibitem[{{R Core Team}(2018)}]{R-citation}
{R Core Team}. 2018, R: A Language and Environment for Statistical Computing, R
  Foundation for Statistical Computing, Vienna, Austria

\bibitem[{{Rebull} {et~al.}(2011){Rebull}, {Guieu}, {Stauffer}, {Hillenbrand},
  {Noriega-Crespo}, {Stapelfeldt}, {Carey}, {Carpenter}, {Cole}, {Padgett},
  {Strom}, \& {Wolff}}]{2011ApJS..193...25R}
{Rebull}, L.~M., {Guieu}, S., {Stauffer}, J.~R., {et~al.} 2011, \apjs, 193, 25

\bibitem[{{Reipurth}(2008{\natexlab{a}})}]{2008hsf1}
{Reipurth}, B. 2008{\natexlab{a}}, {Handbook of Star Forming Regions, Volume I:
  The Northern Sky}

\bibitem[{{Reipurth}(2008{\natexlab{b}})}]{2008hsf2}
{Reipurth}, B. 2008{\natexlab{b}}, {Handbook of Star Forming Regions, Volume
  II: The Southern Sky}

\bibitem[{{Reipurth} {et~al.}(2007){Reipurth}, {Aspin}, {Beck}, {Brogan},
  {Connelley}, \& {Herbig}}]{2007AJ....133.1000R}
{Reipurth}, B., {Aspin}, C., {Beck}, T., {et~al.} 2007, \aj, 133, 1000

\bibitem[{{Rimoldini et al.}(2022)}]{DR3-DPACP-165}
{Rimoldini et al.} 2022, \aap\ in prep.

\bibitem[{{Rimoldini, L. et al.}(2022)}]{DR3-documentation}
{Rimoldini, L. et al.} 2022, {Gaia DR3 documentation Chapter 10: Variability},
  Gaia DR3 documentation, European Space Agency; Gaia Data Processing and
  Analysis Consortium. Online at
  \url{https://gea.esac.esa.int/archive/documentation/GDR3/index.html}, id. 10

\bibitem[{{Saunders} {et~al.}(2000){Saunders}, {Sutherland}, {Maddox},
  {Keeble}, {Oliver}, {Rowan-Robinson}, {McMahon}, {Efstathiou}, {Tadros},
  {White}, {Frenk}, {Carrami{\~n}ana}, \& {Hawkins}}]{2000MNRAS.317...55S}
{Saunders}, W., {Sutherland}, W.~J., {Maddox}, S.~J., {et~al.} 2000, \mnras,
  317, 55

\bibitem[{{Schneider} {et~al.}(2015){Schneider}, {Ossenkopf}, {Csengeri},
  {Klessen}, {Federrath}, {Tremblin}, {Girichidis}, {Bontemps}, \&
  {Andr{\'e}}}]{2015A&A...575A..79S}
{Schneider}, N., {Ossenkopf}, V., {Csengeri}, T., {et~al.} 2015, \aap, 575, A79

\bibitem[{{Shappee} {et~al.}(2014){Shappee}, {Prieto}, {Grupe}, {Kochanek},
  {Stanek}, {De Rosa}, {Mathur}, {Zu}, {Peterson}, {Pogge}, {Komossa}, {Im},
  {Jencson}, {Holoien}, {Basu}, {Beacom}, {Szczygie{\l}}, {Brimacombe},
  {Adams}, {Campillay}, {Choi}, {Contreras}, {Dietrich}, {Dubberley},
  {Elphick}, {Foale}, {Giustini}, {Gonzalez}, {Hawkins}, {Howell}, {Hsiao},
  {Koss}, {Leighly}, {Morrell}, {Mudd}, {Mullins}, {Nugent}, {Parrent},
  {Phillips}, {Pojmanski}, {Rosing}, {Ross}, {Sand}, {Terndrup}, {Valenti},
  {Walker}, \& {Yoon}}]{2014ApJ...788...48S}
{Shappee}, B.~J., {Prieto}, J.~L., {Grupe}, D., {et~al.} 2014, \apj, 788, 48

\bibitem[{{Skrutskie} {et~al.}(2006){Skrutskie}, {Cutri}, {Stiening},
  {Weinberg}, {Schneider}, {Carpenter}, {Beichman}, {Capps}, {Chester},
  {Elias}, {Huchra}, {Liebert}, {Lonsdale}, {Monet}, {Price}, {Seitzer},
  {Jarrett}, {Kirkpatrick}, {Gizis}, {Howard}, {Evans}, {Fowler}, {Fullmer},
  {Hurt}, {Light}, {Kopan}, {Marsh}, {McCallon}, {Tam}, {Van Dyk}, \&
  {Wheelock}}]{2006AJ....131.1163S}
{Skrutskie}, M.~F., {Cutri}, R.~M., {Stiening}, R., {et~al.} 2006, \aj, 131,
  1163

\bibitem[{{Soszy{\'n}ski} {et~al.}(2016){Soszy{\'n}ski}, {Pawlak},
  {Pietrukowicz}, {Udalski}, {Szyma{\'n}ski}, {Wyrzykowski}, {Ulaczyk},
  {Poleski}, {Koz{\l}owski}, {Skowron}, {Skowron}, {Mr{\'o}z}, \&
  {Hamanowicz}}]{2016AcA....66..405S}
{Soszy{\'n}ski}, I., {Pawlak}, M., {Pietrukowicz}, P., {et~al.} 2016, \actaa,
  66, 405

\bibitem[{{Soszy{\'n}ski} {et~al.}(2017){Soszy{\'n}ski}, {Udalski},
  {Szyma{\'n}ski}, {Wyrzykowski}, {Ulaczyk}, {Poleski}, {Pietrukowicz},
  {Koz{\l}owski}, {Skowron}, {Skowron}, {Mr{\'o}z}, {Pawlak}, {Rybicki}, \&
  {Jacyszyn-Dobrzeniecka}}]{2017AcA....67..297S}
{Soszy{\'n}ski}, I., {Udalski}, A., {Szyma{\'n}ski}, M.~K., {et~al.} 2017,
  \actaa, 67, 297

\bibitem[{{Strai{\v{z}}ys} \& {Kazlauskas}(2010)}]{2010BaltA..19....1S}
{Strai{\v{z}}ys}, V. \& {Kazlauskas}, A. 2010, Baltic Astronomy, 19, 1

\bibitem[{{Stringer} {et~al.}(2019){Stringer}, {Long}, {Macri}, {Marshall},
  {Drlica-Wagner}, {Mart{\'\i}nez-V{\'a}zquez}, {Vivas}, {Bechtol},
  {Morganson}, {Carrasco Kind}, {Pace}, {Walker}, {Nielsen}, {Li}, {Rykoff},
  {Burke}, {Carnero Rosell}, {Neilsen}, {Ferguson}, {Cantu}, {Myron},
  {Strigari}, {Farahi}, {Paz-Chinch{\'o}n}, {Tucker}, {Lin}, {Hatt}, {Maner},
  {Plybon}, {Riley}, {Nadler}, {Abbott}, {Allam}, {Annis}, {Bertin}, {Brooks},
  {Buckley-Geer}, {Carretero}, {Cunha}, {D'Andrea}, {da Costa}, {De Vicente},
  {Desai}, {Doel}, {Eifler}, {Flaugher}, {Frieman}, {Garc{\'\i}a-Bellido},
  {Gaztanaga}, {Gruen}, {Gschwend}, {Gutierrez}, {Hartley}, {Hollowood},
  {Hoyle}, {James}, {Kuehn}, {Kuropatkin}, {Melchior}, {Miquel}, {Ogando},
  {Plazas}, {Sanchez}, {Santiago}, {Scarpine}, {Schubnell}, {Serrano},
  {Sevilla-Noarbe}, {Smith}, {Smith}, {Soares-Santos}, {Sobreira}, {Suchyta},
  {Swanson}, {Tarle}, {Thomas}, {Vikram}, {Yanny}, \& {DES
  Collaboration}}]{2019AJ....158...16S}
{Stringer}, K.~M., {Long}, J.~P., {Macri}, L.~M., {et~al.} 2019, \aj, 158, 16

\bibitem[{{Sung} {et~al.}(2017){Sung}, {Bessell}, {Chun}, {Yi}, {Naz{\'e}},
  {Lim}, {Karimov}, {Rauw}, {Park}, \& {Hur}}]{2017ApJS..230....3S}
{Sung}, H., {Bessell}, M.~S., {Chun}, M.-Y., {et~al.} 2017, \apjs, 230, 3

\bibitem[{{Szegedi-Elek} {et~al.}(2013){Szegedi-Elek}, {Kun}, {Reipurth},
  {P{\'a}l}, {Bal{\'a}zs}, \& {Willman}}]{2013ApJS..208...28S}
{Szegedi-Elek}, E., {Kun}, M., {Reipurth}, B., {et~al.} 2013, \apjs, 208, 28

\bibitem[{{Taylor}(2005)}]{2005ASPC..347...29T}
{Taylor}, M.~B. 2005, in Astronomical Society of the Pacific Conference Series,
  Vol. 347, Astronomical Data Analysis Software and Systems XIV, ed.
  P.~{Shopbell}, M.~{Britton}, \& R.~{Ebert}, 29

\bibitem[{{Th\'e} {et~al.}(1994){Th\'e}, {de Winter}, \&
  {P\'erez}}]{1994A&AS..104..315T}
{Th\'e}, P.~S., {de Winter}, D., \& {P\'erez}, M.~R. 1994, \aaps, 104, 315

\bibitem[{{Usatov} \& {Nosulchik}(2008)}]{2008OEJV...87....1U}
{Usatov}, M. \& {Nosulchik}, A. 2008, Open European Journal on Variable Stars,
  0087, 1

\bibitem[{{Varga-Vereb{\'e}lyi} {et~al.}(2020){Varga-Vereb{\'e}lyi}, {Kun},
  {Szegedi-Elek}, {{\'A}brah{\'a}m}, {Varga}, {Kiss}, {K{\'o}sp{\'a}l},
  {Marton}, \& {Szabados}}]{2020IAUS..345..378V}
{Varga-Vereb{\'e}lyi}, E., {Kun}, M., {Szegedi-Elek}, E., {et~al.} 2020, in
  Origins: From the Protosun to the First Steps of Life, ed. B.~G. {Elmegreen},
  L.~V. {T{\'o}th}, \& M.~{G{\"u}del}, Vol. 345, 378--379

\bibitem[{{Vieira} {et~al.}(2003){Vieira}, {Corradi}, {Alencar}, {Mendes},
  {Torres}, {Quast}, {Guimar{\~a}es}, \& {da Silva}}]{2003AJ....126.2971V}
{Vieira}, S.~L.~A., {Corradi}, W.~J.~B., {Alencar}, S.~H.~P., {et~al.} 2003,
  \aj, 126, 2971

\bibitem[{{Vogt} {et~al.}(2016){Vogt}, {Contreras-Quijada}, {Fuentes-Morales},
  {Vogt-Geisse}, {Arcos}, {Abarca}, {Agurto-Gangas}, {Caviedes}, {DaSilva},
  {Flores}, {Gotta}, {Pe{\~n}aloza}, {Rojas}, \&
  {Villase{\~n}or}}]{2016ApJS..227....6V}
{Vogt}, N., {Contreras-Quijada}, A., {Fuentes-Morales}, I., {et~al.} 2016,
  \apjs, 227, 6

\bibitem[{{Wenger} {et~al.}(2000){Wenger}, {Ochsenbein}, {Egret}, {Dubois},
  {Bonnarel}, {Borde}, {Genova}, {Jasniewicz}, {Lalo{\"e}}, {Lesteven}, \&
  {Monier}}]{2000A&AS..143....9W}
{Wenger}, M., {Ochsenbein}, F., {Egret}, D., {et~al.} 2000, \aaps, 143, 9

\bibitem[{{Wolk} {et~al.}(2008){Wolk}, {Spitzbart}, {Bourke}, {Gutermuth},
  {Vigil}, \& {Comer{\'o}n}}]{2008AJ....135..693W}
{Wolk}, S.~J., {Spitzbart}, B.~D., {Bourke}, T.~L., {et~al.} 2008, \aj, 135,
  693

\bibitem[{{Wright} {et~al.}(2010){Wright}, {Eisenhardt}, {Mainzer}, {Ressler},
  {Cutri}, {Jarrett}, {Kirkpatrick}, {Padgett}, {McMillan}, {Skrutskie},
  {Stanford}, {Cohen}, {Walker}, {Mather}, {Leisawitz}, {Gautier}, {McLean},
  {Benford}, {Lonsdale}, {Blain}, {Mendez}, {Irace}, {Duval}, {Liu}, {Royer},
  {Heinrichsen}, {Howard}, {Shannon}, {Kendall}, {Walsh}, {Larsen}, {Cardon},
  {Schick}, {Schwalm}, {Abid}, {Fabinsky}, {Naes}, \&
  {Tsai}}]{2010AJ....140.1868W}
{Wright}, E.~L., {Eisenhardt}, P. R.~M., {Mainzer}, A.~K., {et~al.} 2010, \aj,
  140, 1868

\bibitem[{{Zucker} {et~al.}(2019){Zucker}, {Speagle}, {Schlafly}, {Green},
  {Finkbeiner}, {Goodman}, \& {Alves}}]{2019ApJ...879..125Z}
{Zucker}, C., {Speagle}, J.~S., {Schlafly}, E.~F., {et~al.} 2019, \apj, 879,
  125

\end{thebibliography}

\begin{appendix}
\section{KYSO - The Konkoly Optical YSO catalogue}\label{kysoappendix}
\begin{figure}
   \centering
   \includegraphics[width=\hsize]{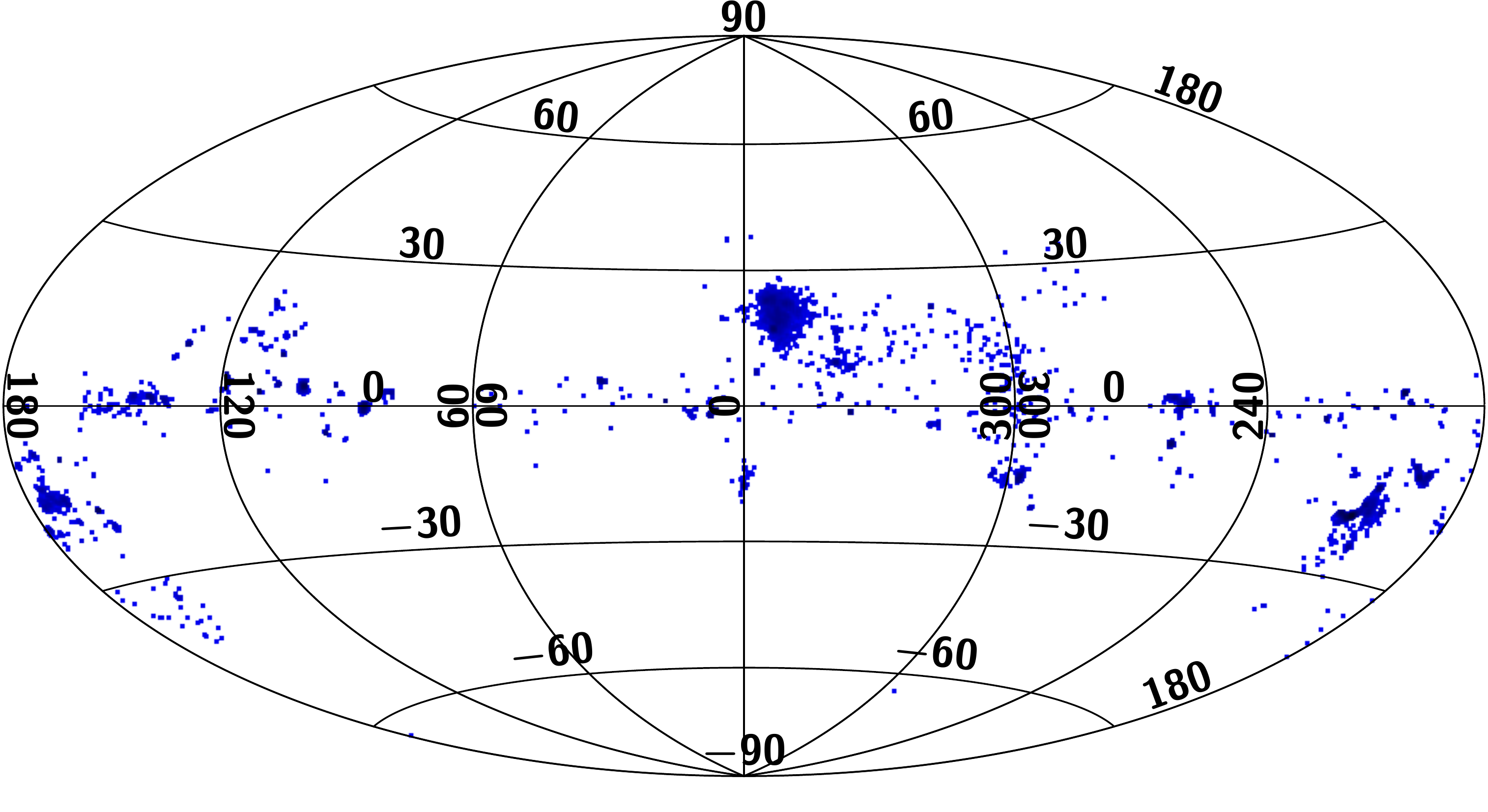}
      \caption{ Konkoly Optical YSO (KYSO) catalogue  of optically selected young stars presented in Aitoff projection in Galactic coordinates. 
              }
         \label{kysosky}
   \end{figure}

We compiled a large catalogue of optically detected, bona fide YSOs.\footnote{The KYSO table is only available in electronic form at the CDS via anonymous ftp to \hyperref[cdsarc.cds.unistra.fr]{cdsarc.cds.unistra.fr} (130.79.128.5) or via \hyperref[https://cdsarc.cds.unistra.fr/cgi-bin/qcat?J/A+A/]{https://cdsarc.cds.unistra.fr/cgi-bin/qcat?J/A+A/}.} The main source of our compilation was the Handbook of Star Forming Regions \citep[Volumes I and II, ][]{2008hsf1,2008hsf2} which  consists of 62~chapters, with each chapter describing a region of the sky (mostly one constellation or a part of it) and not a well-defined star forming region. All of the included regions are located within $\sim$2~kpc of the Sun. Most chapters contain a list of YSOs located in the specific star forming regions, but sometimes only references to the catalogues are given. In the latter case, we took the list of the YSOs from the original (discovery) papers. We also performed an extensive literature search to include YSO catalogues published after the Handbook of Star Forming Regions. Furthermore, we included the stars of the comprehensive Herbig and Bell \citeyearpar{1988cels.book.....H} catalogue of pre-main sequence stars, and Herbig Ae/Be stars from \citet{1994A&AS..104..315T} and \citet{2003AJ....126.2971V}. The celestial distribution of the KYSOs is shown in Fig.~\ref{kysosky}.

Most of the stars of the KYSO catalogue were classified as YSOs based on optical spectra, i.e. they exhibited strong emission lines and/or strong lithium absorption. Moreover, we included H$\alpha$ emission stars located in star forming regions and detected by slitless spectroscopy. In addition to the spectroscopically identified YSOs, we included a few datasets containing optical counterparts of candidate YSOs selected by IR and X-ray observations. These are listed as follows: (i) IR variable stars detected by \citet{2001AJ....121.3160C} in the Orion~A cloud; (ii) optically visible low-mass stars classified as YSOs based on combined X-ray and IR criteria  by \citet{2009ApJ...699.1454G} in Cepheus\,B, and by \citet{2008AJ....135..693W} in RCW\,108; (iii) optical counterparts of IR sources in Barnard~59, classified as YSOs by IR spectroscopy \citep{2010ApJ...722..971C}; (iv) candidate YSOs in the Camelopardalis region, classified by optical--IR SED \citep{2010BaltA..19....1S}.  

The individual YSO catalogues that we collected are very diverse regarding data structure, detection methods, limiting magnitude, and angular resolution. Our aim is to create a unified database from these various sources. As a first step, we extracted the celestial coordinates and names of the objects from the individual catalogues. In the case of older measurements, we converted the epoch from B1950 to J2000 and checked the positions in original finding charts. These data were then loaded into a database using a common scheme. Apart from the name, coordinates, and the detection method, no other information was used from the YSO catalogues.

A major issue with the data is that a source can be present in several catalogues under different names and with slightly different coordinates. 
For the detection of duplicate sources, we applied a semi-automatic approach.
The automatic part of the process was the identification of clusters of objects within a search radius of 1$\farcs$5. Then we manually examined the duplicate candidates. We checked the probability of being a duplicate based on the distribution of positions from the various catalogues, known binarity, and the naming of objects from the SIMBAD database. When a duplicate was found, we removed the duplicate entries from our database and retained the data from the most recent catalogue. Binary stars can also appear as duplicate candidates. Three criteria were used in the identification of multiple stars: (1) When duplicate candidates are present in the same catalogue, this suggests that the sources are indeed distinct, probably binaries. (2) In other cases, the names of the objects indicate binarity. (3) The source is present in the Young Visual Binary Star Database\footnote{\url{http://www2.lowell.edu/users/lprato/YBIN/Binary_Star_Database.html}}.

Next, we consistently associated all of the KYSO objects with their parent star forming regions. Generally, we took the names of the regions from the individual papers describing the catalogues. In some cases, we found inconsistencies in the naming, when, e.g. the same region had different designations. To resolve these issues, we chose the naming convention present in SIMBAD. In the case of some smaller regions, where there is a known larger star forming region to which the given smaller region belongs, we used both names in the following arrangement: `Large star forming region: Small region' (e.g. `Rosette Complex: NGC 2244'). In total, the KYSO catalogue contains 124 star forming regions.

We supplemented the catalogue with the columns of YSO type and variability type of the sources. YSO types are as follows: CTT* -- classical T~Tauri star, K--M~type stars with emission spectra and accretion disc, GTT* -- similar to CTT* with G0--K0 spectral type \citep[see][]{1999AJ....118.1043H}, IMTT* -- intermediate-mass TT* similar to CTT* with F5--F8 spectral type;  WTT* -- weak-line T~Tauri stars identified by G--M spectral types and strong Li~I absorption line at 6707\,\AA. PTT* is for post-T~Tauri stars \citep{2002AJ....124.1670M}. HAeBe stars are B--F5 type pre-main sequence stars. FU~Ori and EX~Lupi type stars are regarded as distinct classes. TT* is assigned when no data were available on the accretion;  Y*O is assigned when this was the only information. We have not included normal, main sequence OB stars, but included a few objects classified as high-mass YSOs (HMY*O).
Variability data were obtained from an extensive search in VizieR. In addition to the published variability types, we inspected the light curves available in various VizieR catalogues. A previous version of the catalogue is described in \citet{2020IAUS..345..378V}.

\section{Parameter distributions}\label{distributions}
\subsection{\bpminrp distributions}

   \begin{figure}
   \centering
   \includegraphics[width=\hsize]{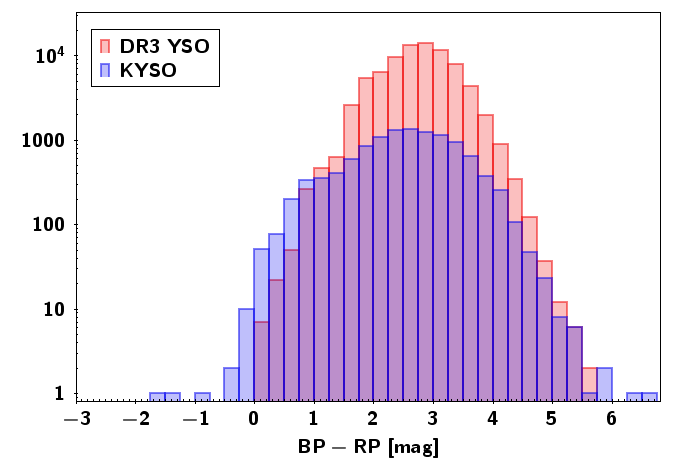}
      \caption{\bpminrp colour distribution of the \textit{Gaia}~DR3 YSOs (red bars) and KYSOs (blue bars) on a logarithmic scale. KYSOs show an excess towards bluer colours.
              }
         \label{bprpkyso}
   \end{figure} 

   \begin{figure}
   \centering
   \includegraphics[width=\hsize]{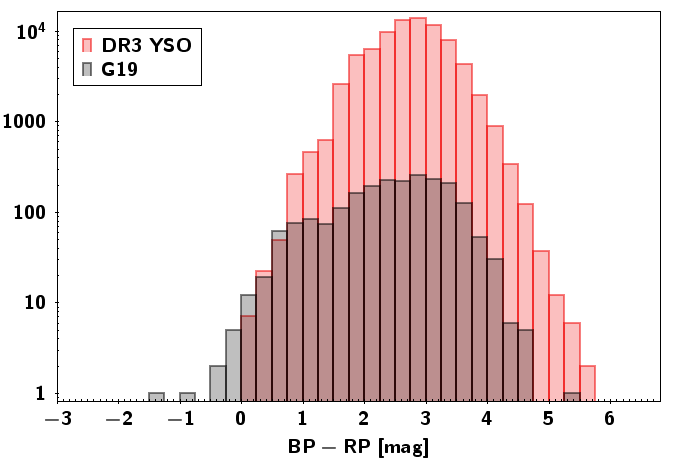}
      \caption{Same as Fig.~\ref{bprpkyso} but for the \textit{Gaia}~DR3 YSOs (red bars) and the \citet{2019A&A...622A.149G} G19 YSOs (grey bars). 
              }
         \label{bprpg19}
   \end{figure} 

   \begin{figure}
   \centering
   \includegraphics[width=\hsize]{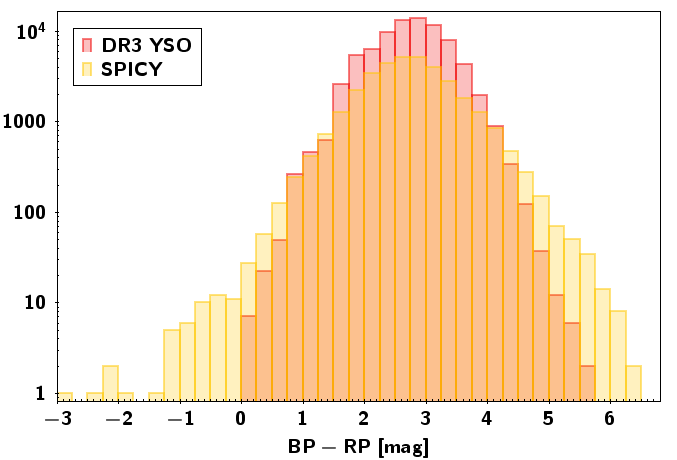}
      \caption{Same as Fig.~\ref{bprpkyso} but for the \textit{Gaia}~DR3 YSOs (red bars) and the \citet{2021ApJS..254...33K} SPICY YSOs (yellow bars).
              }
         \label{bprpspicy}
   \end{figure} 

   \begin{figure}
   \centering
   \includegraphics[width=\hsize]{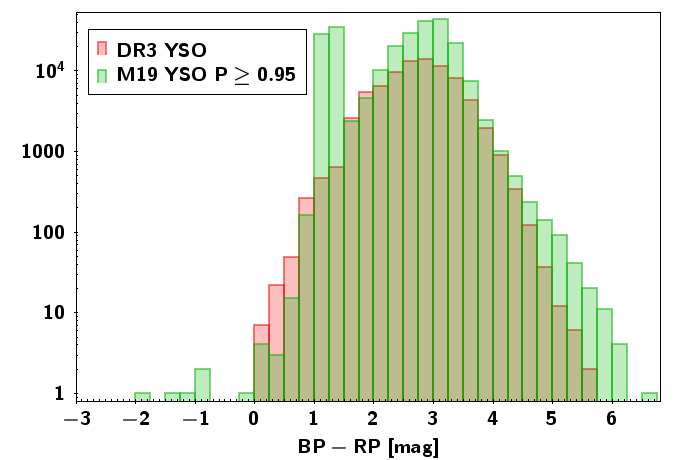}
      \caption{Same as Fig.~\ref{bprpkyso} but for the \textit{Gaia}~DR3 YSOs (red bars) and the \citet{2019MNRAS.487.2522M} M19 YSOs with $R\geq0.5$ and $LY\geq0.95$ or $R<0.5$ and $SY\geq0.95$ (green bars). The M19 YSOs show a significant excess at \bpminrp=1. These are all YSO candidates seen towards the Galactic midplane.
              }
         \label{bprpm19}
   \end{figure} 

\subsection{Absolute median \gmag}

   \begin{figure}
   \centering
   \includegraphics[width=\hsize]{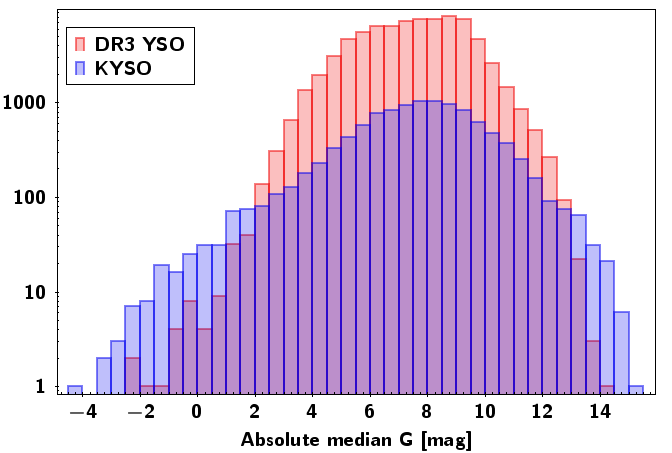}
      \caption{\textit{Gaia} absolute \gmag-band magnitude distribution of the \textit{Gaia}~DR3 YSOs (red bars) and KYSOs (blue bars) on a logarithmic scale. KYSOs show a slight excess towards both ends, showing that very faint and bright sources were not classified as YSOs.
              }
         \label{absgkyso}
   \end{figure}

   \begin{figure}
   \centering
   \includegraphics[width=\hsize]{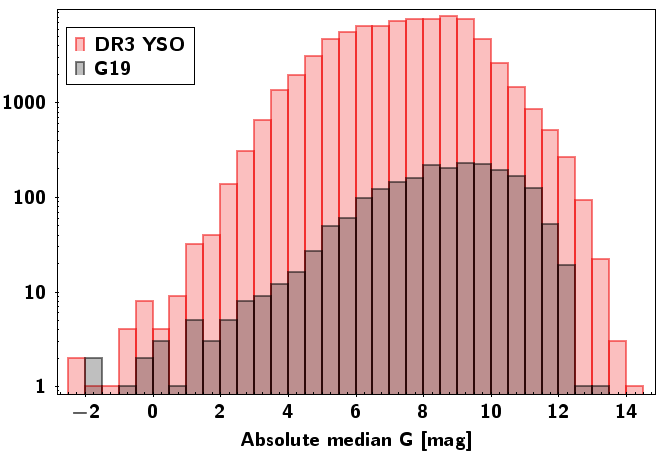}
      \caption{Same as Fig.~\ref{absgkyso}, but for the \textit{Gaia}~DR3 YSOs (red bars) and the G19 YSOs (grey bars).
              }
         \label{absgg19}
   \end{figure} 
   
   \begin{figure}
   \centering
   \includegraphics[width=\hsize]{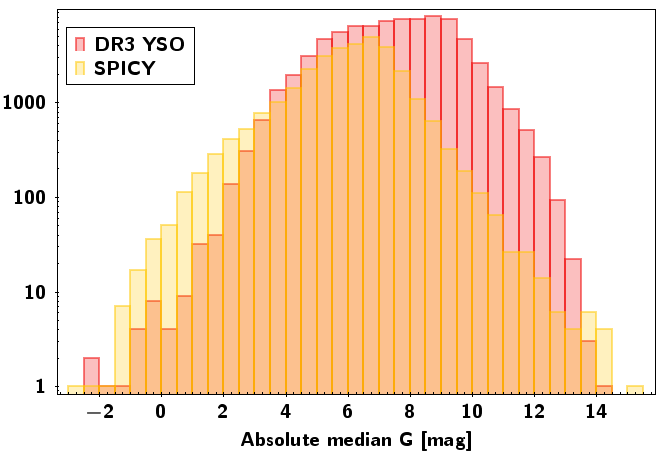}
      \caption{Same as Fig.~\ref{absgkyso}, but for the \textit{Gaia}~DR3 YSOs (red bars) and the SPICY YSOs (yellow bars).
              }
         \label{absgspicy}
   \end{figure} 
   
   \begin{figure}
   \centering
   \includegraphics[width=\hsize]{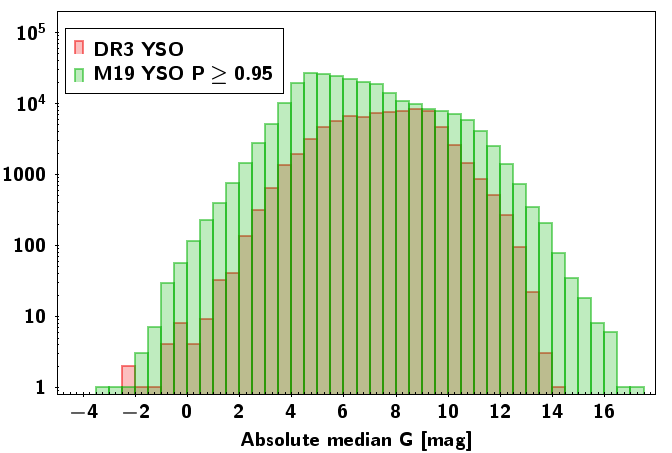}
      \caption{Same as Fig.~\ref{absgkyso}, but for the \textit{Gaia}~DR3 YSOs (red bars) and the M19 YSOs (green bars).
              }
         \label{absgm19}
   \end{figure}    

\subsection{J-H 2MASS colour}
  \begin{figure}
   \centering
   \includegraphics[width=\hsize]{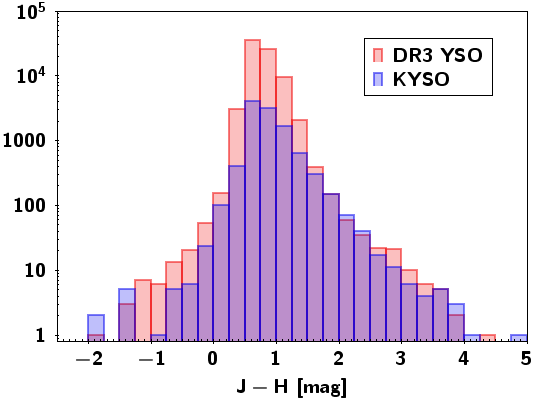}
      \caption{2MASS J$-$H colour distribution of the \textit{Gaia}~DR3 YSOs (red bars) and KYSOs (blue bars) on a logarithmic scale.
              }
         \label{jhkyso}
   \end{figure}

   \begin{figure}
   \centering
   \includegraphics[width=\hsize]{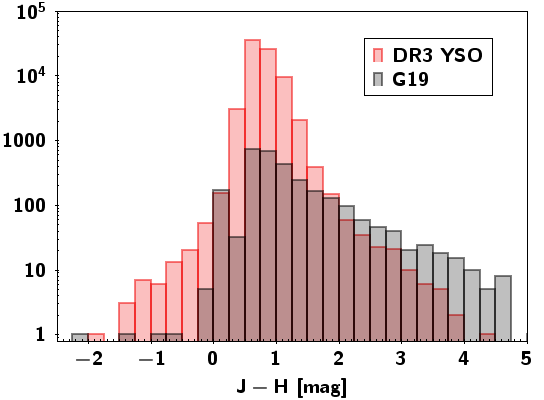}
      \caption{Same as Fig.~\ref{jhkyso}, but for the \textit{Gaia}~DR3 YSOs (red bars) and the G19 YSOs (grey bars).
              }
         \label{jhg19}
   \end{figure} 
   
   \begin{figure}
   \centering
   \includegraphics[width=\hsize]{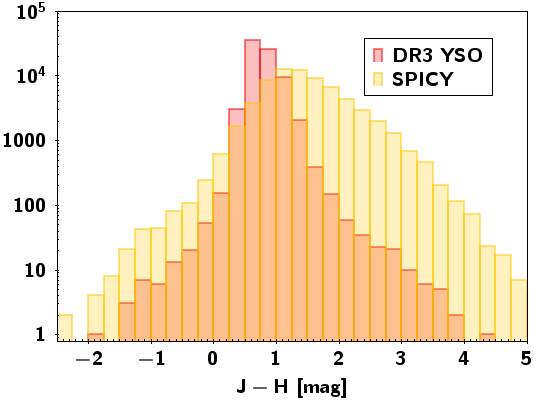}
      \caption{Same as Fig.~\ref{jhkyso}, but for the \textit{Gaia}~DR3 YSOs (red bars) and the SPICY YSOs (yellow bars).
              }
         \label{jhspicy}
   \end{figure} 
   
   \begin{figure}
   \centering
   \includegraphics[width=\hsize]{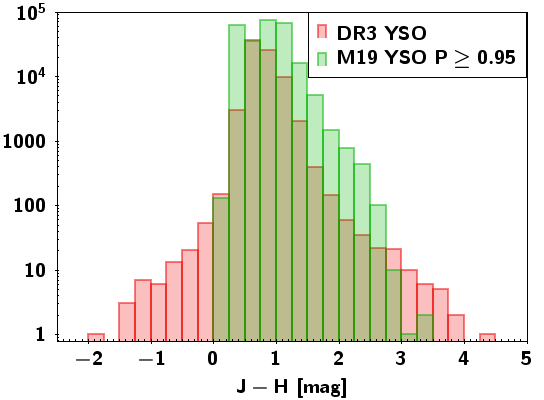}
      \caption{Same as Fig.~\ref{jhkyso}, but for the \textit{Gaia}~DR3 YSOs (red bars) and the M19 YSOs (green bars).
              }
         \label{jhm19}
   \end{figure}

\subsection{H-K$_s$ 2MASS colour}
  \begin{figure}
   \centering
   \includegraphics[width=\hsize]{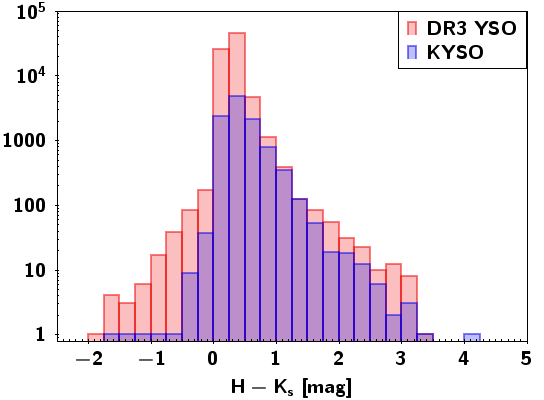}
      \caption{2MASS H$-$K$_s$ colour distribution of the \textit{Gaia}~DR3 YSOs (red bars) and KYSOs (blue bars) on a logarithmic scale.
              }
         \label{hkkyso}
   \end{figure}

   \begin{figure}
   \centering
   \includegraphics[width=\hsize]{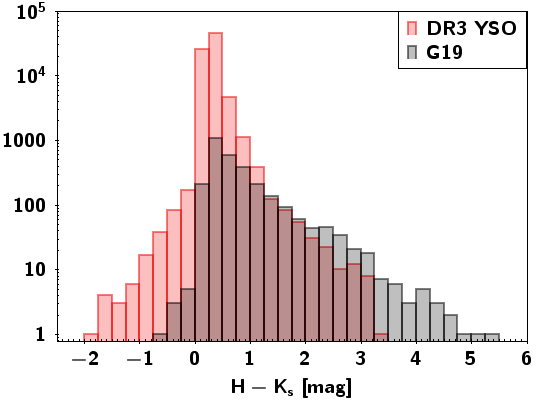}
      \caption{Same as Fig.~\ref{hkkyso}, but for the \textit{Gaia}~DR3 YSOs (red bars) and the G19 YSOs (grey bars).
              }
         \label{hkg19}
   \end{figure} 
   
   \begin{figure}
   \centering
   \includegraphics[width=\hsize]{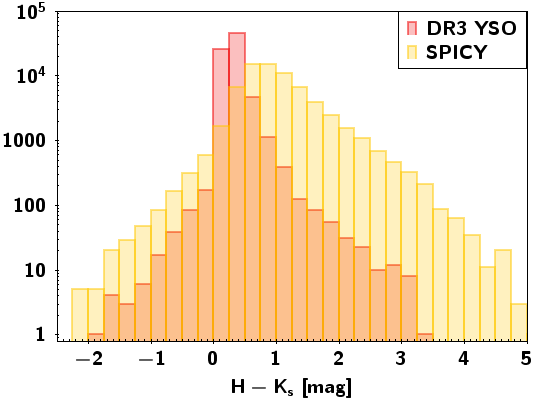}
      \caption{Same as Fig.~\ref{hkkyso}, but for the \textit{Gaia}~DR3 YSOs (red bars) and the SPICY YSOs (yellow bars).
              }
         \label{hkspicy}
   \end{figure} 
   
   \begin{figure}
   \centering
   \includegraphics[width=\hsize]{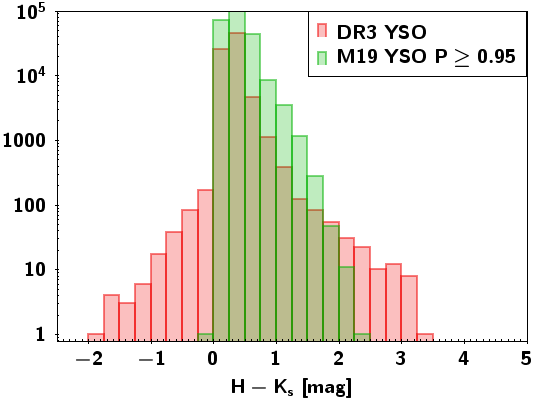}
      \caption{Same as Fig.~\ref{hkkyso}, but for the \textit{Gaia}~DR3 YSOs (red bars) and the M19 YSOs (green bars).
              }
         \label{hkm19}
   \end{figure}     
\end{appendix}

\end{document}